\documentclass{article}

\usepackage{subcaption}
\usepackage{graphicx}
\usepackage[utf8]{inputenc} 
\usepackage[T1]{fontenc}    
\usepackage{hyperref}       
\usepackage{url}            
\usepackage{booktabs}       
\usepackage{amsfonts}       
\usepackage{nicefrac}       
\usepackage{microtype}      
\usepackage{amsmath}
\usepackage[english]{babel}

\usepackage{algorithm,algpseudocode}
\usepackage[preprint]{neurips_2021}
\newcommand{\equal}{\textsuperscript{*}}
\title{Predicting Daily Trading Volume \\via Various Hidden States}

\author{
   Shaojun Ma \thanks{Equal contribution, the work is done while interning at Invesco Ltd}\\
   Department of Mathematics \\
   Georgia Institute of Technology\\
    Atlanta, GA 30332, USA\\
   \texttt{shaojunma@gatech.edu} \\
   \And
  Pengcheng Li \equal \\
  Global Head of Trading Alpha Research\\
  Invesco Ltd\\
  Atlanta, GA 30309, USA \\
}

\begin{document}

\maketitle

\begin{abstract}
Predicting intraday trading volume plays an important role in trading alpha research. Existing methods such as rolling means(RM) and a two-states based Kalman Filtering method have been presented in this topic. We extend two states into various states in Kalman Filter framework to improve the accuracy of prediction. Specifically, for different stocks we utilize cross validation and determine best states number by minimizing mean squared error of the trading volume. We demonstrate the effectivity of our method through a series of comparison experiments and numerical analysis.
\end{abstract}

\section{Introduction}
Trading volume is the total quantity of shares or contracts traded for specified securities such as stocks, bonds, options contracts, futures contracts and all types of commodities. It can be measured on any type of security traded during a trading day or a specified time period. In our case, daily volume of trade is measured on stocks. The volume of trading is an essential component in trading alpha research since it tells investors about the market's activity and liquidity. Over the past decade, along with the improved accessibility of ultra-high-frequency financial data, evolving data-based computational technologies has attracted many attentions on the financial industry. Meanwhile, the development of algorithmic and electronic trading has shows great potential of trading volume since many trading models require intraday volume forecasts as an key input. As a result, there is growing interest in developing models for precisely predicting intraday trading volume.

Researchers aims to propose various strategies to accomplish trading efficiently in the electronic financial markets, meanwhile they wish to minimize transaction costs and market impact. The study of trading volume generally falls into two lines to achieve the goals. One line of work is focused on giving optimal trading sequence and amount, while another line is investigating the relationships among trading volume and other financial variables or market activities such as bid-ask spread, return volatility and liquidity, etc. Thus a precise model that provides insights of trading volume can be regarded as a basis for two lines of work.

There are several existing methods to estimate future trading volume. As a fundamental approach, rolling means(RM) predict intraday volume during a time interval by averaging volume traded within the same interval over the past days. The concept of RM model is straightforward, but it fails to adequately capture the intraday regularities. One classical publicly available intraday volume prediction model decomposes trading volume into three components, namely, a daily average component, an intraday periodic component, and an intraday dynamic component, then adopts the Component Multiplicative Error Model (CMEM) to estimate the three terms \citep{cmem}. Though this model outperforms RM, the limitations such as high sensitivity to noise and initial parameters complicate its practical implementation. \citet{kffortrading} propose a new model to deal with the logarithm of intraday volume to simplify the multiplicative model into an additive one. The model is constructed within the scope of a two-state (intraday and overday features) Kalman Filter \citep{kforiginal} framework, the authors adopt the expectation-maximization (EM) algorithm for the parameter estimation. Though the model provides a novel view to study intraday and overday factors, the flexibility is not satisfied since the model treat the number of hidden states of all stocks as two, thus there may be information loss. Moreover, from experiment we see that the dominant term in the model is actually daily seasonality, the learning process of parameters is not robust.

As an extension of two-state Kalman Fiter, our new model has advantages such as higher prediction precision, stability and simple structure. In general our contributions are:
\begin{itemize}
  \item Firstly, we develop a new way that combines cubic spline and statistical process to determine the best degrees of freedom (DOFs) for different stocks.
  \item Secondly, by choosing suitable DOFs, we provide a smoothing prediction of traded volume.
  \item Finally, we demonstrate that our model outperforms RM and two-state Kalman Fiter through experiments on 978 stocks.
\end{itemize}

\section{Methodologies}
We denote the $i$-th observation on day $t$ as $vol_{t,i}$, the local indices $i \in \{0,1,2,...,I\}$ and $t\in \{0,1,2,...,T\}$, we set global index $\tau$=$t*I+i$ thus $vol_{t,i}$ = $vol_{\tau}$. 
\subsection{two-state Kalman Filter Model}
Before introducing our method, we would like to review two-state Kalman Fiter model. Within the model, the volume is defined as the number of shares traded normalized by daily outstanding shares:
\begin{align}
    vol_{t,i} = \frac{\text{shares traded}_{t,i}}{\text{daily outstanding shares}_t}.
\end{align}

\begin{figure}[ht!]
\centering
\includegraphics[scale=0.15]{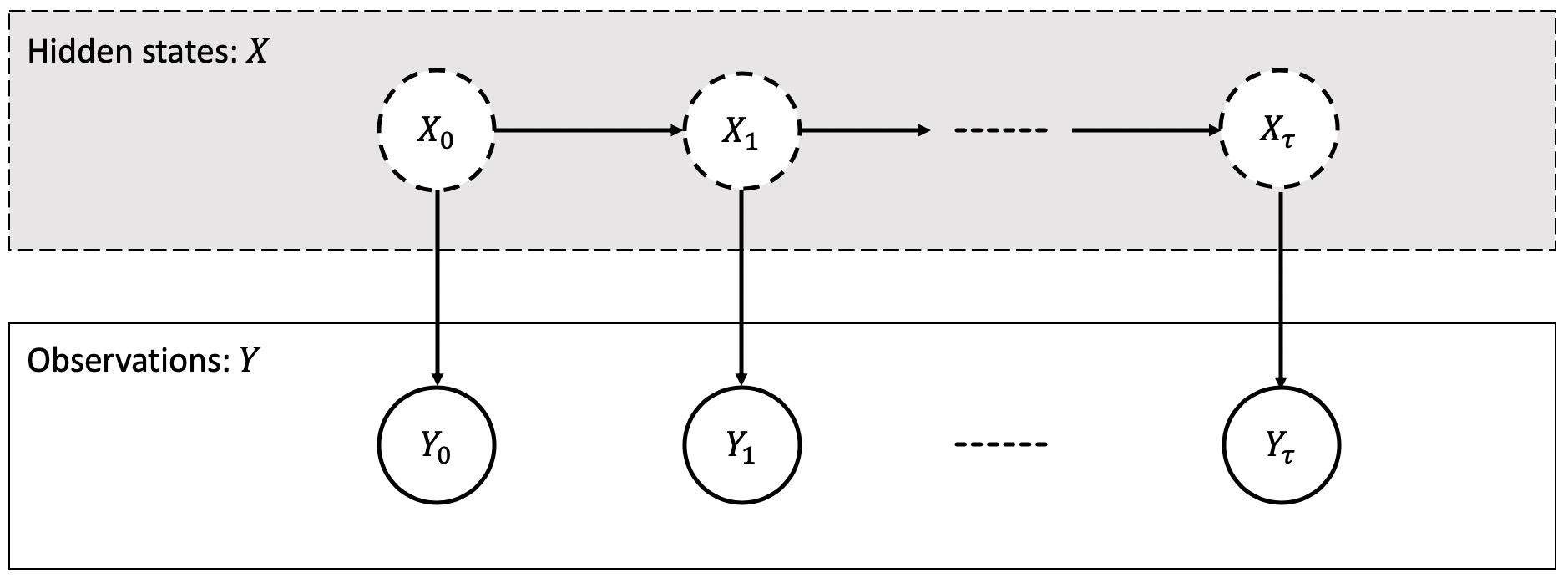}
\caption{A graphical representation of the Kalman Filter model: each
vertical slice represents a time instance; the top node $X$ in each slice
is the hidden state variable corresponding to the underlying volume
components; and the bottom node $Y$ in each slice is the observed volume
in the market}
\label{fig:model}
\end{figure}

This ratio is one way of normalization \citep{kffortrading} since the normalization helps to correct low-frequency variation caused by change of traded volume. Log-volume refers to the natural logarithm of traded volume. The researchers train their model with log-volume and evaluate the predictive performance based on volume. The reason for using log-volume is that it converts the multiplicative terms \citep{cmem} to additive relationships and makes it naturally fit Kalman Fiter framework, moreover, the logarithmic transformation facilitate to reduce the skewness property of volume data \citep{behaviour}.
\citet{kffortrading}'s model is built within Kalman Fiter framework as shown in Figure \ref{fig:model}. $X$ represents hidden state that is not observable, $Y$ represents logarithm of observed traded volume. The mathematical updates are:
\begin{align}
    X_{\tau + 1} &= A_{\tau}X_{\tau} + q_{\tau},\nonumber \\
    Y_{\tau} &= Cx_{\tau} + \phi_{\tau} + v_{\tau},
    \label{kf1}
\end{align}
for $\tau={1,2,...,T*I}$, where $X_{\tau}=[\eta_{\tau}\ \mu_{\tau}]^T$ is the hidden state vector containing two parameters, namely, the daily average part and the intraday dynamic part; $A_{\tau}=diag(a_{\tau}^{\eta}, a_{\tau}^{\mu})$ is the state transition matrix; observation matrix $C$ is fixed as $[1 \ 1]$; $q_{\tau}=[q_{\tau}^{\eta}\ a_{\tau}^{\mu}]^T \sim N(0,Q_{\tau})$ where $Q_{\tau} = diag((\sigma_{\tau}^{\eta})^2, (\sigma_{\tau}^{\mu})^2)$; $v_{\tau} \sim N(0,r)$, and $\phi=[\phi_1,\phi_2,...,\phi_I]$ is treated the seasonality; initial state $X_1 \sim N(\pi_1, \Sigma_1)$. The unknown system parameters $\theta(\pi_1,\Sigma_1,a^{\eta}, a^{\mu}, \sigma^{\eta}, \sigma^{\mu}, r, \phi)$ are estimated by closed form equations, which are derived from expectation-maximization(EM) algorithm. For more details of two-state model, we suggest readers review the original paper.

\subsection{Our Model: Various-states Kalman Filter}
In two-state model mentioned above, the DOF of hidden state variable is two since it has intra-day and over-day two factors. Since there is no systematic way to determine a correct DOF of hidden state variable, especially for various stocks. Our concern is that how to find a better DOF for each stock and predict more precisely. Thus our new method still falls into the Kalman Fiter framework shown in Figure \ref{fig:model}, however, we change equation \ref{kf1} to the most common Kalman Fiter update equation:
\begin{align}
    X_{t + 1} &= BX_{t} + \gamma,\nonumber \\
    Y_{t} &= DX_{t} + \psi.
    \label{kf2}
\end{align}
The differences among Equation \ref{kf1} and Equation \ref{kf2} are as follows: $X_t$ represents hidden state whose dimension is $n\times 1$ that $n$, as the DOF of hidden state variable, will be determined in Section 2.2.1; state transition matrix $B$ is a $n\times n$ matrix while observation matrix $D$ is a $I\times n$ matrix; $\gamma \sim N(0, \Gamma)$ where transition covariance matrix $\Gamma$ is a $n\times n$ matrix; $\psi \sim N(0, \Psi)$ where observation covariance matrix $\Psi$ is a $I\times I$ matrix; initial state $X_1 \sim N(\pi_1, \Sigma_1)$ and observation $Y_t$ is a $I\times 1$ vector. Notice that $B$, $D$, $\Gamma$ and $\Psi$ are uniquely determined by training data. The reason that we use $t$ as subscript is that every time we predict one day's traded volume.

Within the framework of our model, the data we use is historical daily trading volume. We define observation as multiplication of traded volume and olume Weighted Average Price.
\begin{align}
    vol_{t,i}^{\text{new}} = \text{Volume} \times \text{VWAP}.
    \label{v1}
\end{align}
Furthermore we model and evaluate performance with the percentage ratio:
\begin{align}
    p_{t,i} = \frac{vol_{t,i}^{\text{new}}}{\sum_i^I vol_{t,i}^{\text{new}}} \times 100.
    \label{p1}
\end{align}

\begin{algorithm}[ht!]
\caption{Find DOF by Cross Validation}
\begin{algorithmic}[1]
\Require Training data $Y$, shuffle times $N_{s}$ and cross validation times $N_{cv}$
\For{DOF = 1,2,...,I}
\For{i = 1,2,...,$N_s$}
\State shuffle $Y$
\For{j = 1,2,...,$N_{cv}$}
\State shuffle and split $Y$ into training set and test set
\State compute cubic smoothing spline on training set
\State compute and store MSE on test set
\State compute and store standard error(SE) of each MSE
\EndFor
\State find best DOF by  ‘‘one standard error rule''
\EndFor
\EndFor
\end{algorithmic}
\label{alg1}
\end{algorithm}

\subsubsection{DOF of State Space}
\begin{figure}[t]
\begin{subfigure}{0.3\textwidth}
  \includegraphics[width=\linewidth]{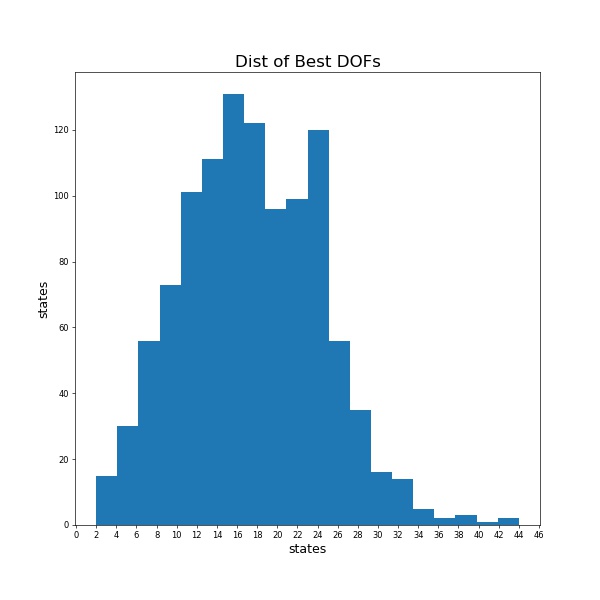}  
  \caption{Distribution of best DOFs}
  \label{fig:dof}
\end{subfigure}\hfill
\begin{subfigure}{0.3\textwidth}
  \includegraphics[width=\linewidth]{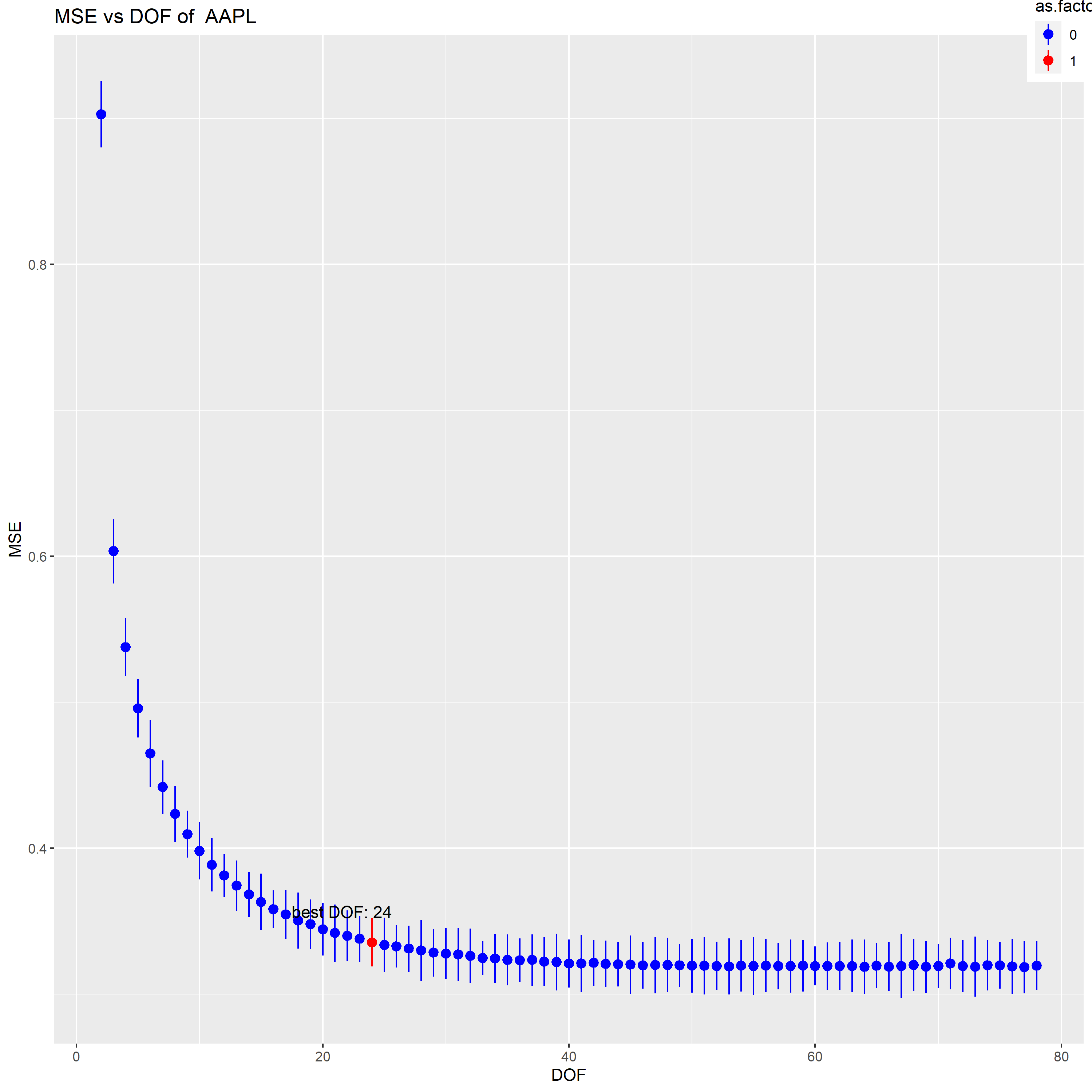}  
  \caption{Best DOF of AAPL}
  \label{fig:dof-aapl}
\end{subfigure}\hfill
\begin{subfigure}{0.3\textwidth}
  \includegraphics[width=\linewidth]{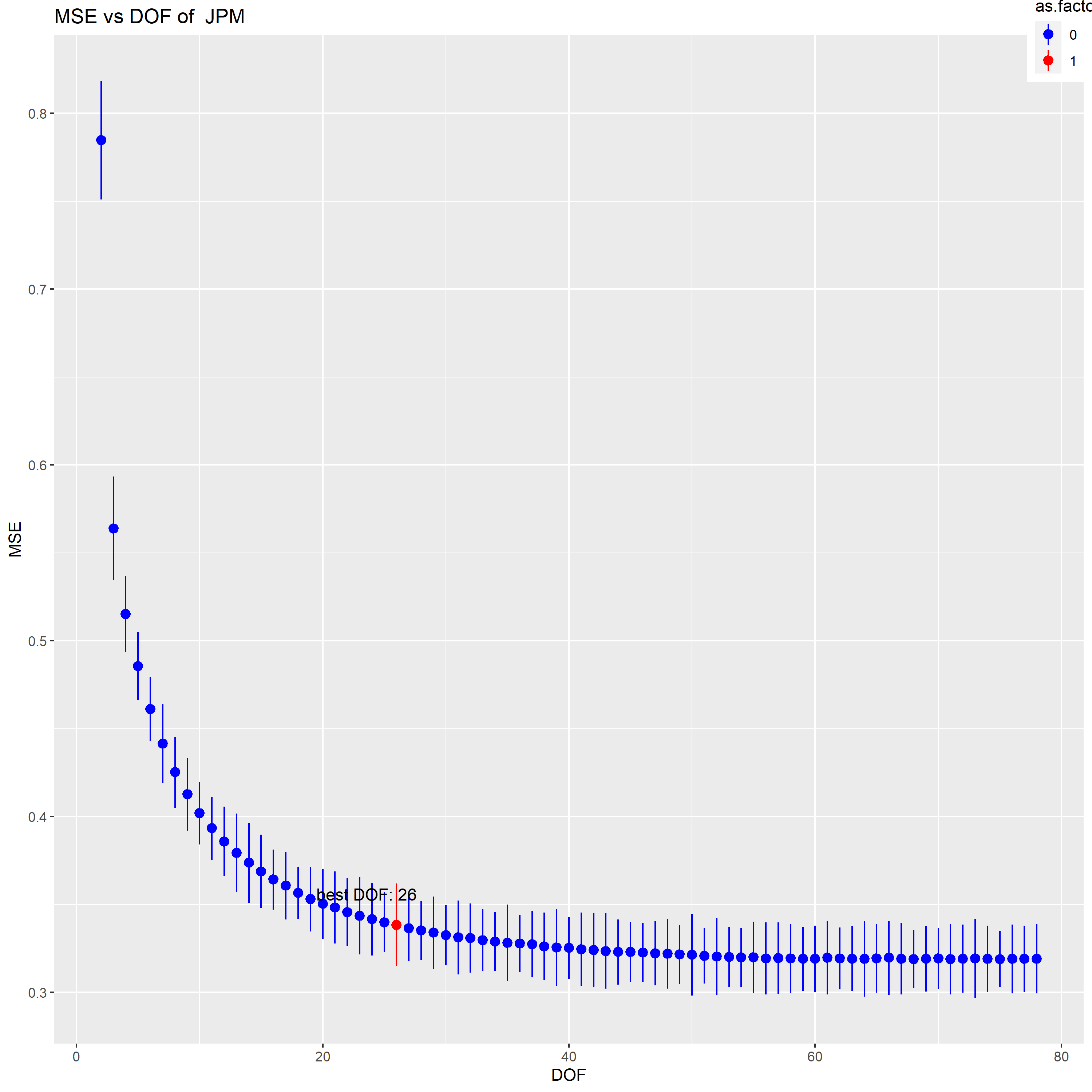}  
  \caption{Best DOF of JPM}
  \label{fig:dof-jpm}
\end{subfigure}
\caption{DOFs of states}
\label{fig:dofdist}
\end{figure}

In our assumption, different stocks will have distinct number of parameters in hidden state $X_{\tau}$. For a specific stock, we call the number of elements in $X_{\tau}$ as degrees of freedom(DOFs), thus for stock with index $s$, $X_{\tau}^s = [x_1^s,x_2^s,...,x_{dof}^s]$. The key concern is how to determine DOFs for each stock. By experiment, we find that seasonality $\phi$ dominates the prediction of traded volume in two-state model, and in both of two-state model and our model, $I$=77, which means each day for each stock we have 78 observations and $\phi$ has 78 parameters.  We look for including seasonality in the hidden states and drop $\phi$ to avoid dominant term. In terms of avoiding overfitting and reduce computations, we use cubic spline to fit $78$ observations smoothly in each day. Given a series of observations $y_i, i\in {1,2,...,I}$, cubic spline is to find the function $g$ that minimizes:
\begin{align}
    \sum_{i=1}^I(y_i-g(x_i))^2 + \lambda \int g''(x)^2dx,
    \label{ncs}
\end{align}
where $x=1,2,...,I$ in our case, $\lambda$ is a nonnegative tuning parameter. The function $g$ that minimizes \ref{ncs} is known as a smoothing spline. The term $\sum_{i=1}^I(y_i-g(x_i))^2$ encourages $g$ to fit the data well, and the term $\lambda \int g''(x)^2dx$ is regarded as a penalty that controls smoothness of the spline. By solving \ref{ncs} we have:
\begin{align}
    \boldsymbol{g_{\lambda}} = \boldsymbol{S_{\lambda}}\boldsymbol{y},
    \label{ncs1}
\end{align}
where $\boldsymbol{g_{\lambda}}$, as the solution to \ref{ncs} for a particular choice of $\lambda$, is a $n$-vector containing the fitted values of the smoothing spline at the training points $x_1, x_2,..., x_I$. Equation \ref{ncs1} indicates that the vector of fitted values can be written as a $n\times n$ matrix $S_{\lambda}$ times the response vector y. Then the DOF is defined to be the trace of the matrix $S_{\lambda}$. 
Our first purpose is to give a specific DOF and get a corresponding spline. Thanks to the work of B. D. Ripley and Martin Maechler (\url{https://rdrr.io/r/stats/smooth.spline.html}), we are able to get fitting splines when given reasonable DOFs. After fitting process then we use cross validation to find DOF that achieves lowest mean squared error(MSE). Algorithm \ref{alg1} outlines the mechanism of finding DOF of each stock. We analyze the DOFs of 978 stocks, examples of the distribution of DOFs and best DOFs of some stocks are shown in Figure \ref{fig:dofdist}.

\subsubsection{Kalman Filter}
Given the best DOF from our method, then we use Kalman Filter to do predictions. Kalman Filter is an online algorithm to precisely estimate the mean and covariance matrix of hidden states. Suppose parameters in Equation \ref{kf2} are known, Algorithm 1 outlines the mechanism of the Kalman Filtering. We model the distribution of hidden state $X_{t+1}$ conditional on all the percentage observations up to time $t$. Since we suppose $\gamma$ and $\psi$ in Equation \ref{kf2} are Gaussian noise, thus all hidden states will follow a Gaussian distribution and it is only necessary to characterize the conditional mean and the conditional covariance as shown in line 3 and line 4. Then given new observation we correct the mean and covariance in line 7 and line 8.
\begin{algorithm}[ht!]
\caption{Kalman Filtering}
\begin{algorithmic}[1]
\Require Parameters $B, D, \Gamma, \Psi, \pi_0, \Sigma_0$
\State initialize $X_0\sim N(\pi_0, \Sigma_0)$
\For{t=0,1,2,...,T-1}
\State predict next state mean: $X_{t+1|t}=BX_{t|t}$
\State predict next state covariance: $S_{t+1|t}=B_{t}S_{t|t}B_t\top+\Gamma_t$
\State obtain measurement $Y_t$
\State compute Kalman gain: $K_{t+1} = S_{t+1|t} D^\top (D S_{t+1|t} D^\top + \Psi)^{-1}$
\State update current state mean: $X_{t+1|t+1} = X_{t+1|t} + K_{t+1} (Y_t - DX_{t+1|t})$
\State update current state covariance: $S_{t+1|t+1} = (I - K_{t+1}D) S_{t+1|t}$
\EndFor
\end{algorithmic}
\label{alg2}
\end{algorithm}

Our ultimate goal is to make predictions of $X_t$ and $Y_t$ by Algorithm \ref{alg1} and  $Y_t^{\text{pre}}=DX_{t}^{\text{pre}}$, respectively. Thus we need to estimate parameters precisely. The method to calibrate parameters is expectation-maximization(EM) algorithm. Smoothing process infer past states before $X_T$ conditional on all the observations in the training set, which is a necessary step in model calibration because it provides more accurate information of the unobservable states. We outlines Kalman smoothing process as Algorithm \ref{alg3}.

\begin{algorithm}[ht!]
\caption{Kalman Smoothing}
\begin{algorithmic}[1]
\Require Parameters $B, D, \Gamma, \Psi, \pi_1, \Sigma_1$ and $X_{T|T}, S_{T|T}$ from Kalman Filtering
\For{t=T-1,T-2,...,0}
\State compute: $J_t=S_{t|t}B_t^TS_{t+1|t}^{-1}$
\State compute mean: $X_{t|T}=X_{t|t}+J_t(X_{t+1|T}-BX_{t+1|t})$
\State compute covariance:$S_{t|T}=S_{t|t}+J_t(S_{t+1|T}-S_{t+1|t})J_t^\top$
\State compute joint covariance:$S_{t,t-1|T}=S_{t|t}J_{t-1}^\top+J_t(S_{t+1,t|T}-BS_{t|t})J_{t-1}^\top$
\EndFor
\State specially, $S_{T,T-1|T}=(I-K_TD)BS_{T-1|T-1}$
\end{algorithmic}
\label{alg3}
\end{algorithm}

After performing Algorithm \ref{alg1}, \ref{alg2} and \ref{alg3}, we need to estimate parameters by EM method, as shown in algorithm \ref{alg4}. EM algorithm is one common way to estimate parameters of Kalman Filter problem. It extends the maximum likelihood estimation to cases where hidden states are involved \citep{emalg}. The EM iteration alternates between performing an E-step (i.e., Expectation step), which constructs a global convex lower bound of the expectation of log-likelihood using the current estimation of parameters, and an M-step (i.e., Maximization step), which computes parameters to maximize the lower bound found in E-step. Two advantages of EM algorithm are fast convergence and existence of closed-form solution. The derivations of Kalman Filter and EM algorithm beyond the scope of this paper, we refer interested readers to \citet{kforiginal}'s work for more details.

\begin{algorithm}[ht!]
\caption{EM algorithm}
\begin{algorithmic}[1]
\Require Initial $B, D, \Gamma, \Psi, \pi_1, \Sigma_1$, results from Kalman Filtering and Kalman smoothing
\While{not converge}
\For{t=T-1,T-2,...,0}
\State $P_{t+1}=S_{t+1|T}+X_{t+1|T}X_{t+1|T}^\top$
\State $P_{t+1,t}=S_{t+1,t|T}+X_{t+1|T}X_{t|T}^\top$
\EndFor
\State $\pi_1 = X_{1|T}$
\State $\Sigma_1 = P_1 - X_{1|T}X_{1|T}^\top$
\State $B=(\sum_{t=0}^{T-1}P_{t+1,t})(\sum_{t=2}^{T}P_{t})^{-1}$
\State $\Gamma = \frac{1}{T}\sum_{t=0}^{T-1}(P_{t+1}+P_{t}-P_{t+1,t}B^\top - BP_{t+1,t}^\top)$
\State $\Psi = \frac{1}{T+1}\sum_{t=0}^T(Y_t Y_t^\top+DP_t D^\top-Y_t X_{t|T}^\top D^\top- DX_{t|T}Y_t^\top)$
\EndWhile
\end{algorithmic}
\label{alg4}
\end{algorithm}

\section{Experiment}
\subsection{Data Introduction}
Our collect empirically analyze intraday volume of 978 stocks traded on major U.S. markets.
For example, a glance of the information of stock "AAPL" is summarized in Table \ref{tab:stock}. The data covers the period from January 3rd 2017 to May 29th 2020, excluding none trading days. Each trading day consists of 78 5-minute bins. Volume and percentage are computed as Equation \ref{v1} and \ref{p1} respectively. All historical data used in the experiment are obtained from the Invesco Inc.

\begin{table}[t]
\centering
\caption{Historical intrady trading volume of stock "AAPL"}
\begin{tabular}{ c c c c c c c c} 
\hline
 Time & LAST & FIRST & HIGH & LOW & VOLUME & VWAP & time bin\\ 
\hline
 1/3/2017 & 106.24 & 105.9 & 106.42 & 105.59 & 1139228 & 106.11 & 14:30\\ 
 1/3/2017 & 105.28 & 106.24 & 106.34 & 105.19 & 1245847 & 105.77 & 14:35 \\
 1/3/2017 & 105.51 & 105.29 & 105.53 & 104.85 & 1289865 & 105.23 & 14:40 \\ 
 $\cdots$ & $\cdots$ & $\cdots$ & $\cdots$ & $\cdots$ & $\cdots$ & $\cdots$ & $\cdots$\\
 1/3/2017 & 106.23 & 106.04 & 106.23 & 106 & 1675070 & 106.09 & 20:55 \\
 $\cdots$ & $\cdots$ & $\cdots$ & $\cdots$ & $\cdots$ & $\cdots$ & $\cdots$ & $\cdots$\\
 5/29/2020 & 318.25 & 319.25 & 320 & 318.22 & 747433 & 319.21 & 13:30\\
 $\cdots$ & $\cdots$ & $\cdots$ & $\cdots$ & $\cdots$ & $\cdots$ & $\cdots$ & $\cdots$\\
 5/29/2020 & 317.92 & 319.29 & 319.62 & 317.46 & 1969841 & 318.92 & 19:55 \\
 \hline
\end{tabular}
\label{tab:stock}
\end{table}

\subsection{Experiment Set-up}
We choose two-state model and RM model mentioned before as baselines. Data from January 3rd 2017 to June 30th 2017 is considered as training set while data from July 5th 2017 to January 2nd 2018 is treated as test set. Both of training set and test set contain $T_{\text{train}}$=125 trading days(from day $0$ to day $124$). We initialize $B, \Gamma, \Psi$ as identity matrices, and $D$ as smooth matrix, then perform Algorithms \ref{alg1} to \ref{alg4} on traning set to obtain model parameters, finally make predictions on next $T_{\text{test}}$=125 days(from day $125$ to day $249$). We evaluate performances of three models by mean absolute percentage error(MAPE):
\begin{align}
    \text{MAPE} = \frac{1}{M}\sum_{\tau=1}^M|p_{\tau}^{\text{true}} - p_{\tau}^{\text{predicted}}|,
\end{align}
where $\tau = t*I + i$.

\subsection{Results}
In this section we compare our model with two state-of-the-art baselines and perform some analysis of our results.

\subsubsection{MAPE Distribution}
\begin{figure}[ht!]
\begin{subfigure}{0.33\textwidth}
  \includegraphics[width=\linewidth]{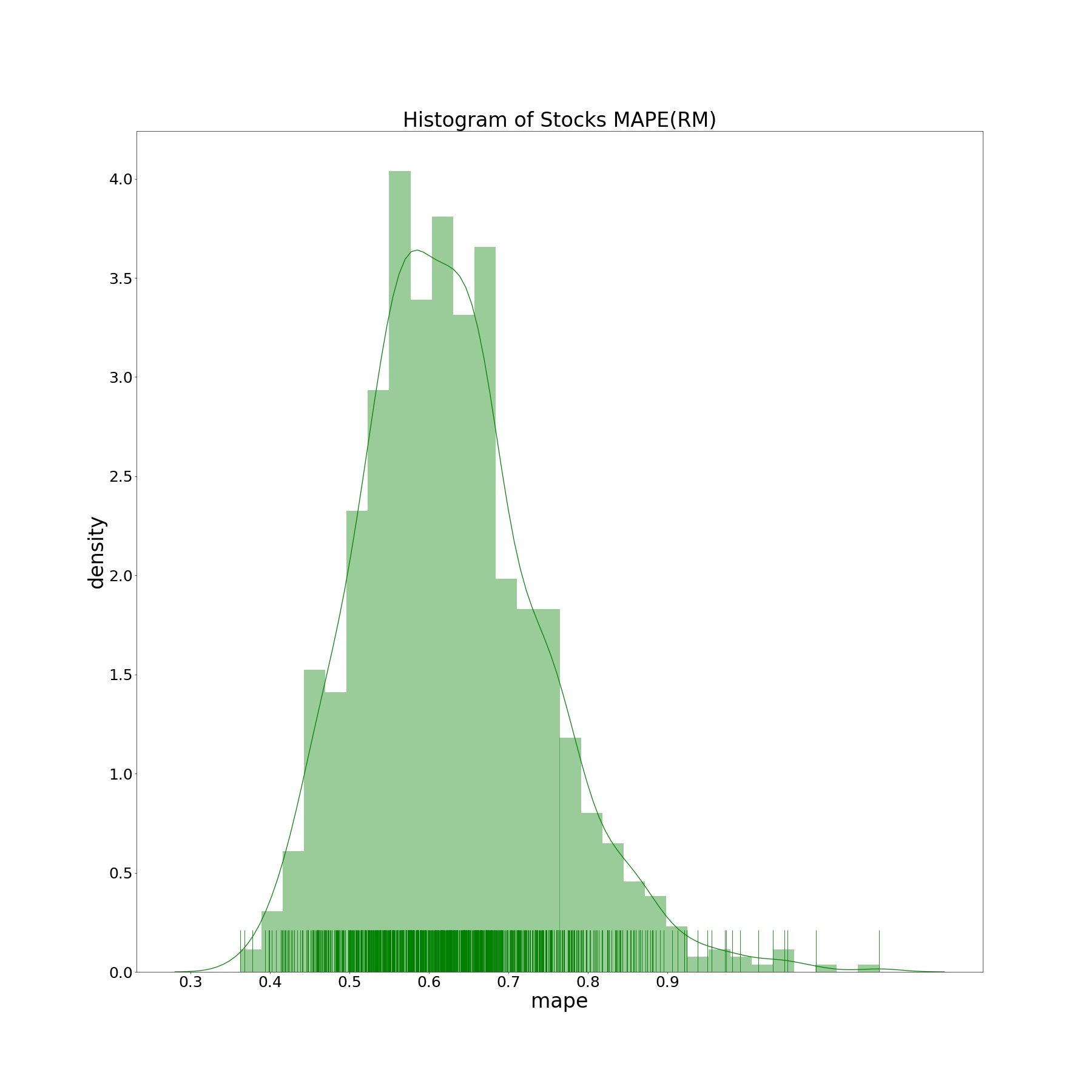}  
  \caption{RM}
  \label{fig:mape-rm}
\end{subfigure}\hfill
\begin{subfigure}{0.33\textwidth}
  \includegraphics[width=\linewidth]{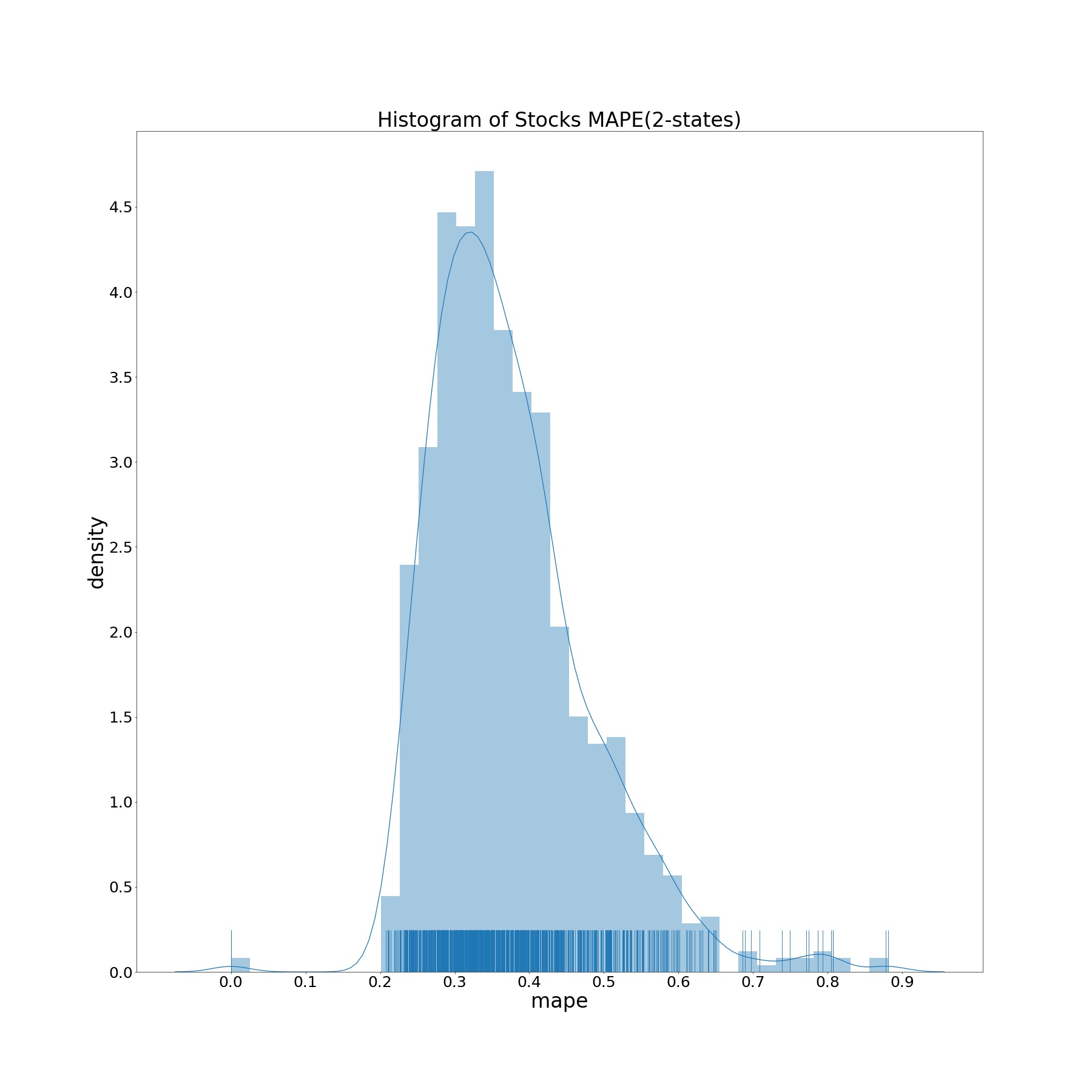}  
  \caption{Two-state}
  \label{fig:mape-2s}
\end{subfigure}\hfill
\begin{subfigure}{0.33\textwidth}
  \includegraphics[width=\linewidth]{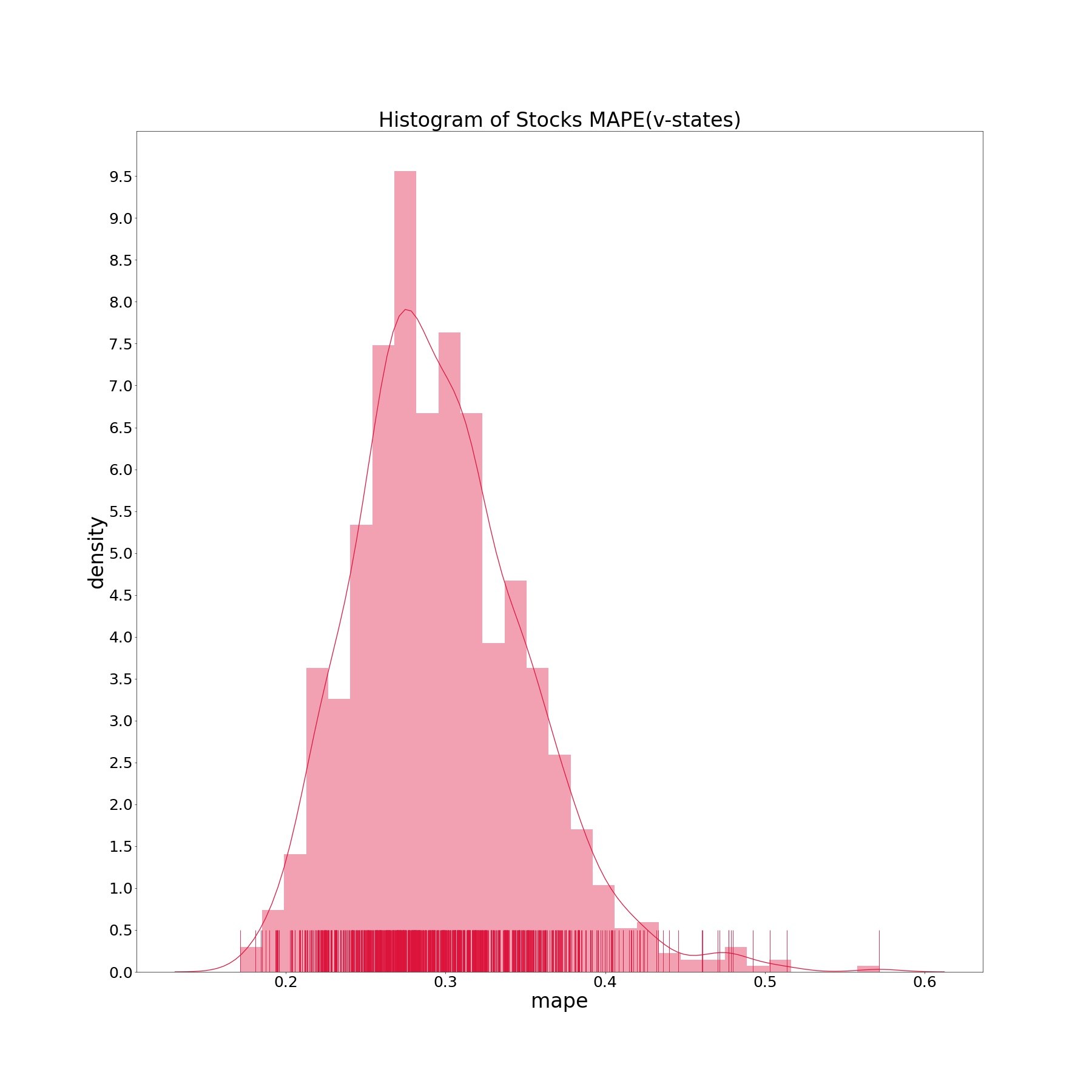}  
  \caption{our model: v-state}
  \label{fig:mape-vs}
\end{subfigure}
\caption{Comparison of MAPE}
\label{fig:mapedist}
\end{figure}

We obtain the distributions of MAPE of 978 stocks from three models and show them in Figure \ref{fig:mapedist}. We see that our v-state model outperforms baselines by giving smaller MAPEs.

\begin{figure}[!ht]
\begin{subfigure}{0.25\textwidth}
  \includegraphics[width=\linewidth]{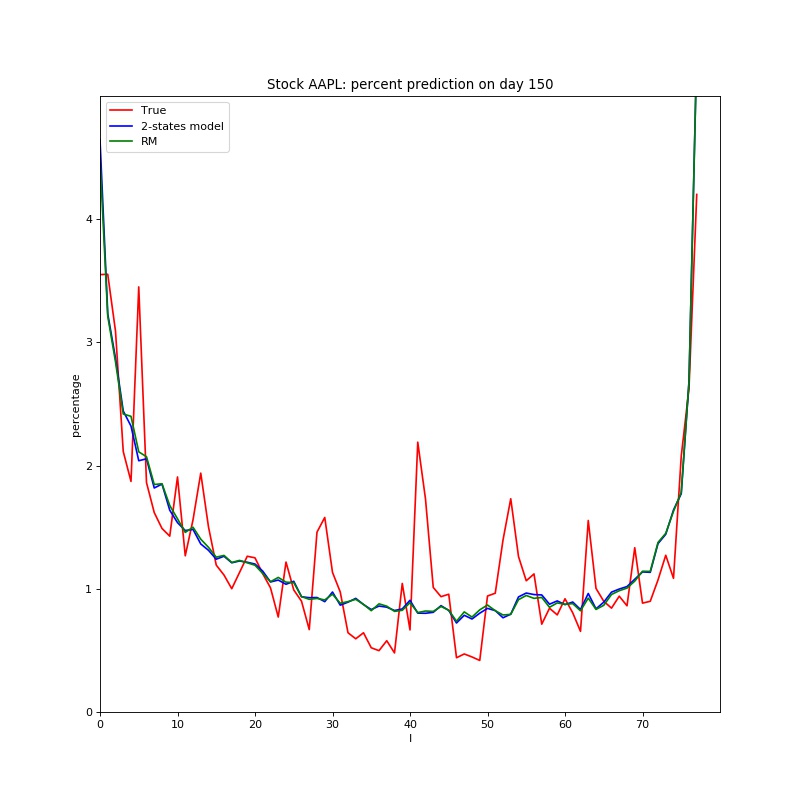}
  \caption{}
  \label{fig:com-aapl1}
\end{subfigure}\hfill
\begin{subfigure}{0.25\textwidth}
  \includegraphics[width=\linewidth]{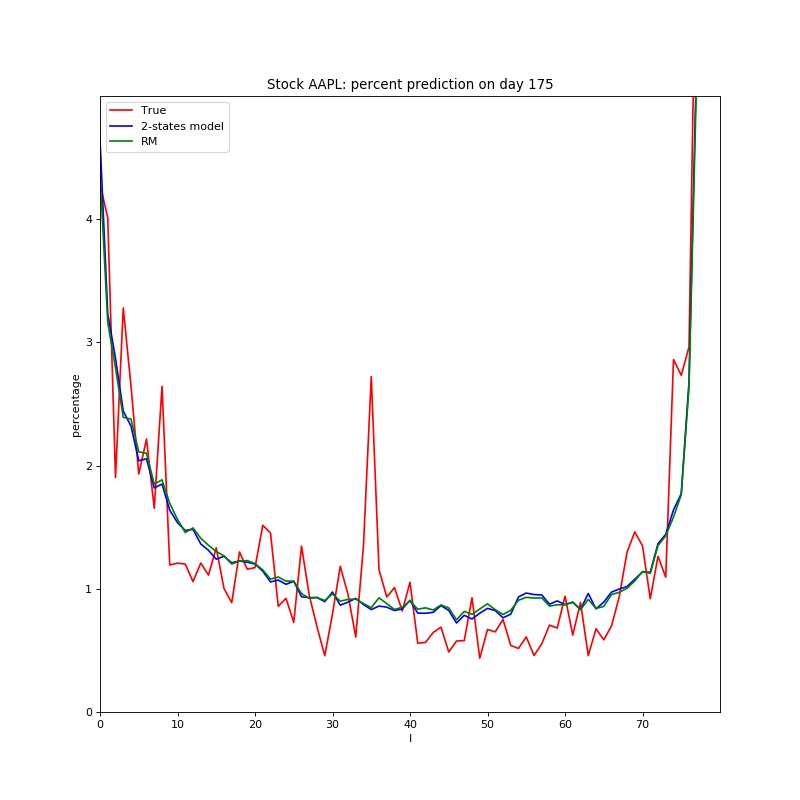}
  \caption{}
  \label{fig:com-aapl2}
\end{subfigure}\hfill
\begin{subfigure}{0.25\textwidth}
  \includegraphics[width=\linewidth]{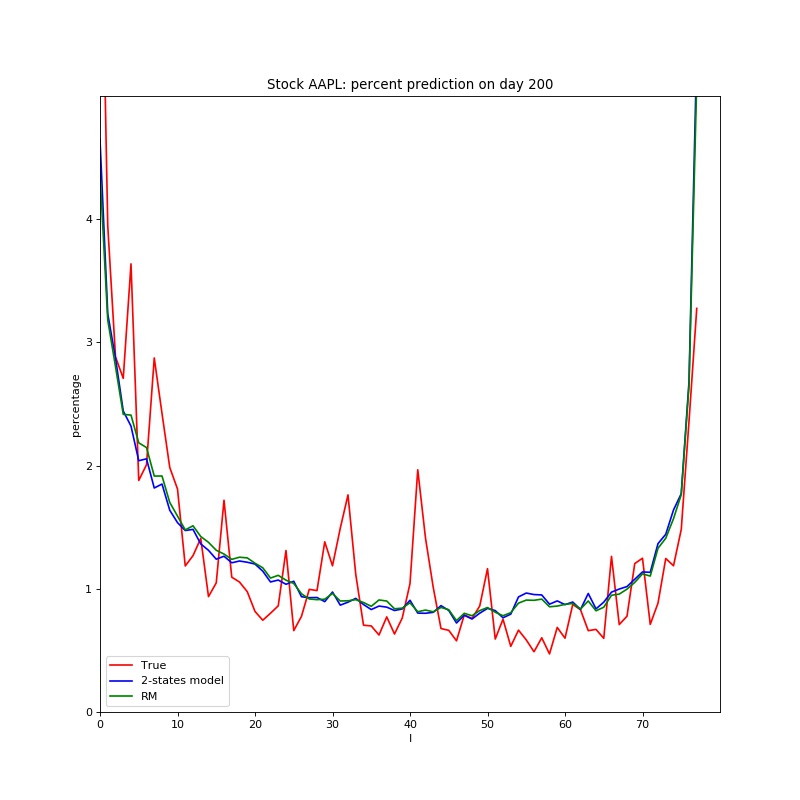}
  \caption{}
  \label{fig:com-aapl3}
\end{subfigure}\hfill
\begin{subfigure}{0.25\textwidth}
  \includegraphics[width=\linewidth]{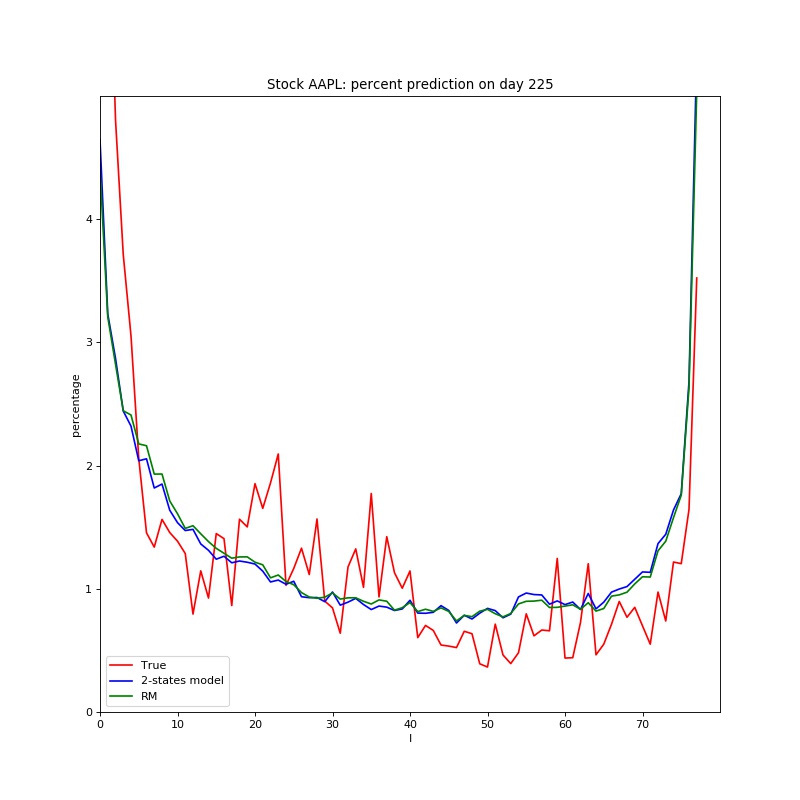}
  \caption{}
  \label{fig:com-aapl4}
\end{subfigure}\\
\begin{subfigure}{0.25\textwidth}
  \includegraphics[width=\linewidth]{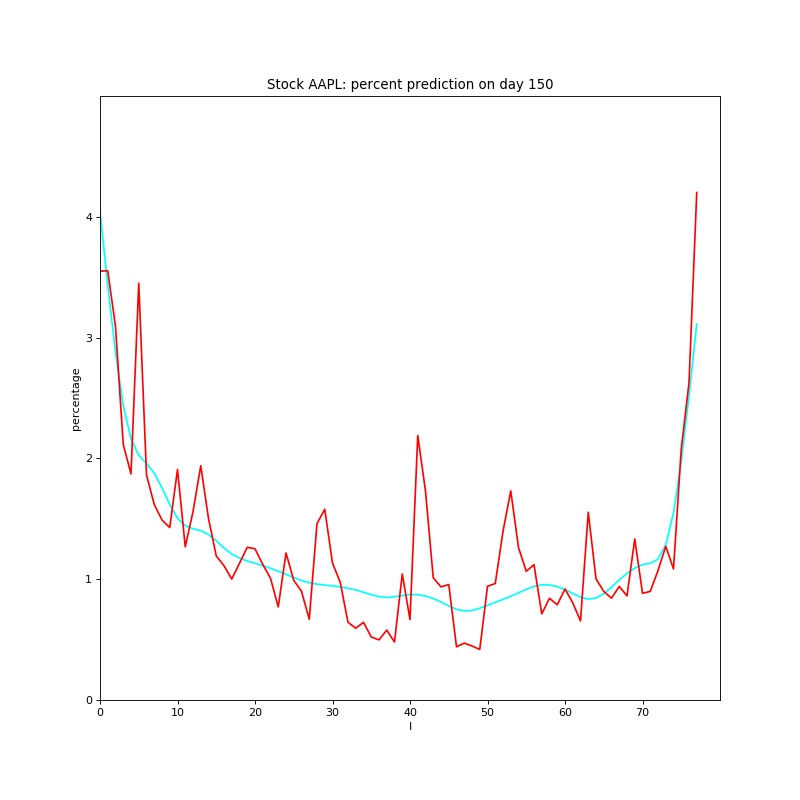}  
  \caption{}
  \label{fig:com-aapl5}
\end{subfigure}\hfill
\begin{subfigure}{0.25\textwidth}
  \includegraphics[width=\linewidth]{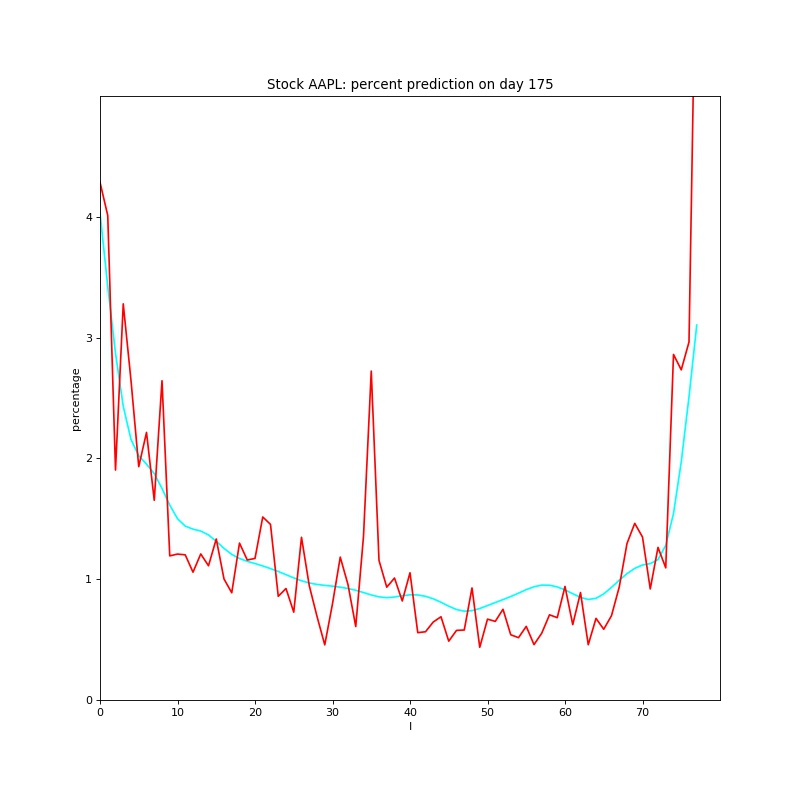}
  \caption{}
  \label{fig:com-aapl6}
\end{subfigure}\hfill
\begin{subfigure}{0.25\textwidth}
  \includegraphics[width=\linewidth]{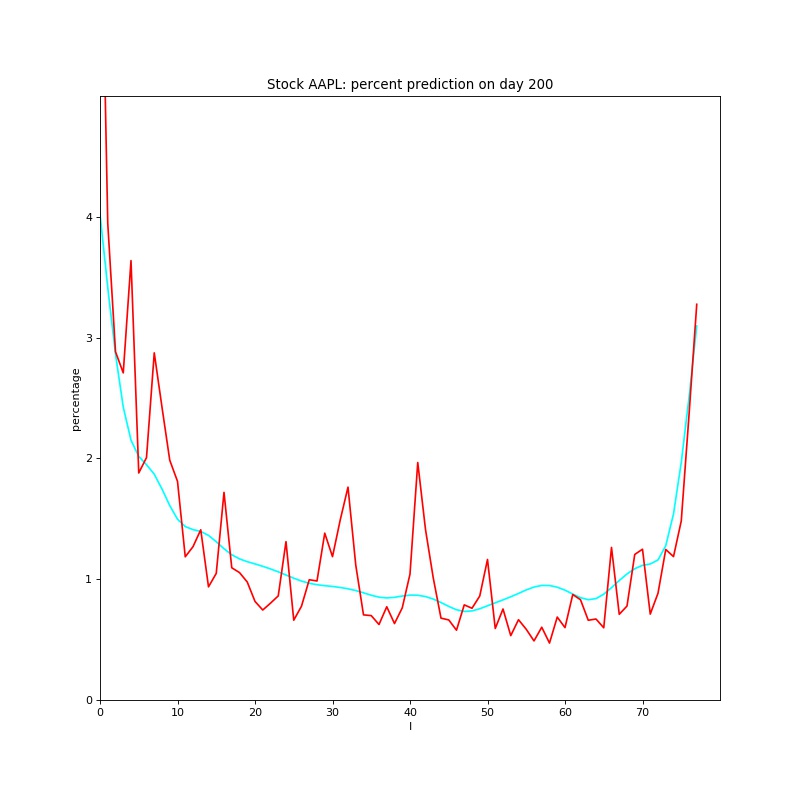}
  \caption{}
  \label{fig:com-aapl7}
\end{subfigure}\hfill
\begin{subfigure}{0.25\textwidth}
  \includegraphics[width=\linewidth]{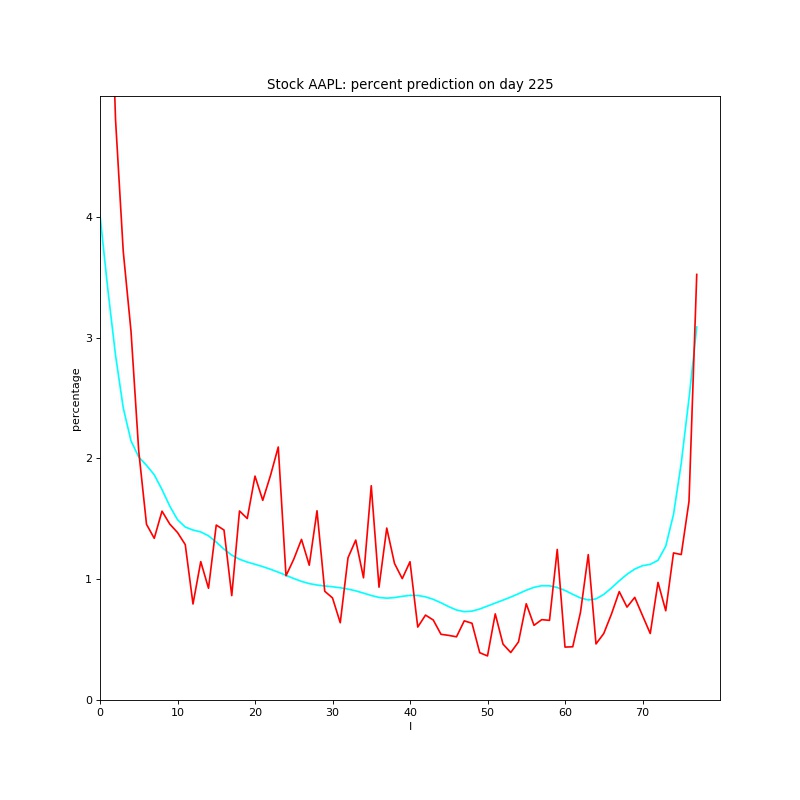}
  \caption{}
  \label{fig:com-aapl8}
\end{subfigure}
\caption{comparison of prediction: (a) to (d):baseline models on AAPL, (e) to (h):our v-state model on stock "AAPL"}
\label{fig:pre}
\end{figure}

\subsubsection{Predictions on Specific Days}
To better visualize comparisons, we pick ten stocks out of the dataset and show their predictions on day 150, 175, 200 and 225. Due to space limitation, we only show one stock "AAPL" here in Figure \ref{fig:pre} and show other nine stocks in Appendix. We see that two-state model almost overlaps RM model. Our v-state model provides a smoother prediction.

\subsubsection{Analysis of v-state Model}
\begin{figure}[!ht]
  \centering
  \includegraphics[width=0.4\linewidth]{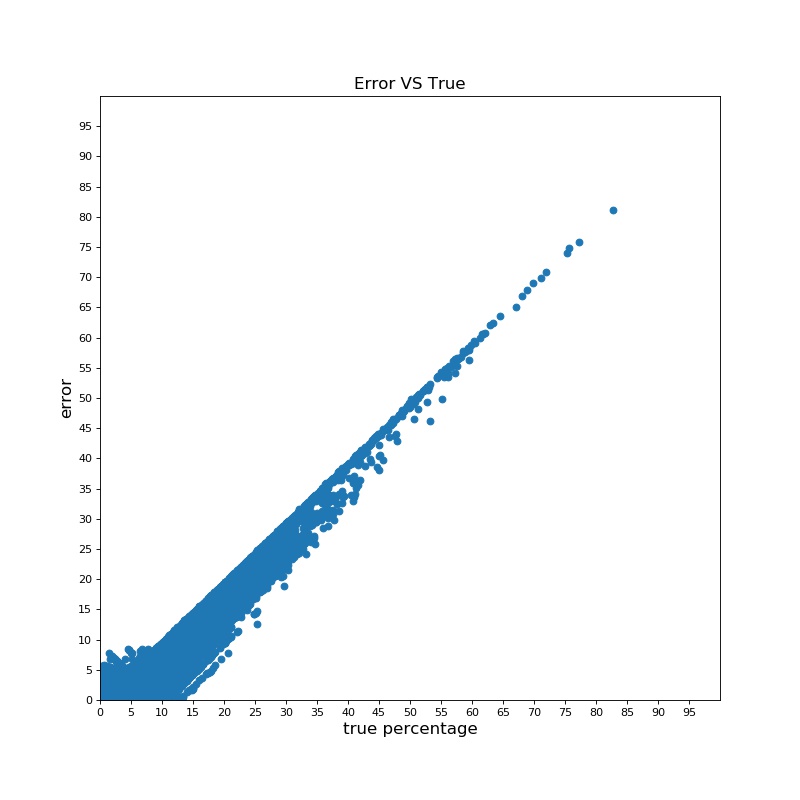}  
  \label{fig:sub-first}
\caption{Relationship between error and true percentage}
\label{fig:errorvstrue}
\end{figure}
We investigate the relationship between absolute error and true percentage for all stocks to further test the precision of our model. The absolute error of stock $s$ on $\tau$-th bin is defined as:
\begin{align}
    \text{error} = |p_{s,\tau}^{\text{predicted}} - p_{s,\tau}^{\text{true}}|.
\end{align}

As illustrated in Figure \ref{fig:errorvstrue}, we plot error versus $p_{s,\tau}^{\text{true}}$ for all 978$\times$125$\times$78 samples. We see there is a nearly linear relationship between absolute error and true percentage, when $p_{s,\tau}^{\text{true}}$ gets larger, the slope gets closer to 1. Moreover, we observe that for 95\% samples that fall into the corner around the original point, we have:
\begin{align}
    &0\leq|p_{s,\tau}^{\text{predicted}} - p_{s,\tau}^{\text{true}}| \leq p_{s,\tau}^{\text{true}}, \nonumber\\
    &0\leq{p_{s,\tau}^{\text{predicted}}}\leq 2{p_{s,\tau}^{\text{true}}}.
    \label{eq:bound1}
\end{align}
And for those samples outside of the corner we plug in the linear equation with slope $1$:
\begin{align}
    &|p_{s,\tau}^{\text{predicted}} - p_{s,\tau}^{\text{true}}| \approx p_{s,\tau}^{\text{true}}, \nonumber\\
    &{p_{s,\tau}^{\text{predicted}}}\ll {p_{s,\tau}^{\text{true}}} \quad  \text{or}\quad   {p_{s,\tau}^{\text{predicted}}} \approx 2{p_{s,\tau}^{\text{true}}}.
    \label{eq:bound2}
\end{align}
From Equation \ref{eq:bound1} and Equation \ref{eq:bound2}, our model provides a lower bound as well as a upper bound for the prediction precision. Due to the high-noisy property of original data, it is still not a trivial task to fully capture the movements of trading volume. It could be one potential direction in the future.

\begin{figure}[!ht]
\begin{subfigure}{0.23\textwidth}
  \includegraphics[width=\linewidth]{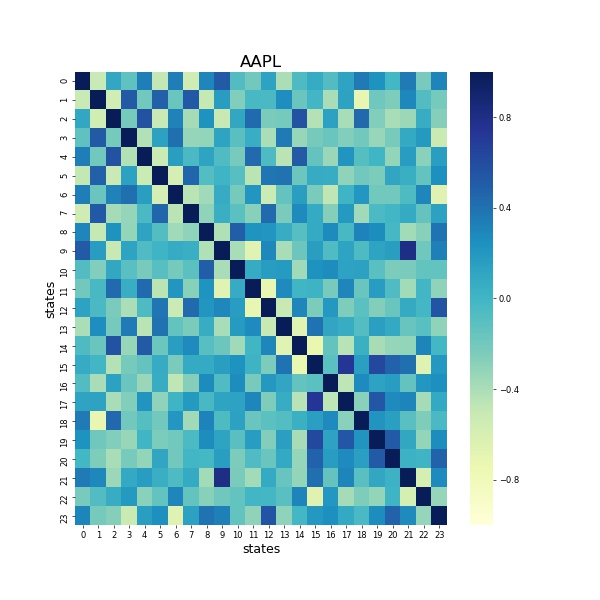}
  \caption{AAPL}
  \label{fig:corr-aapl}
\end{subfigure} \hfill
\begin{subfigure}{0.23\textwidth}
  \includegraphics[width=\linewidth]{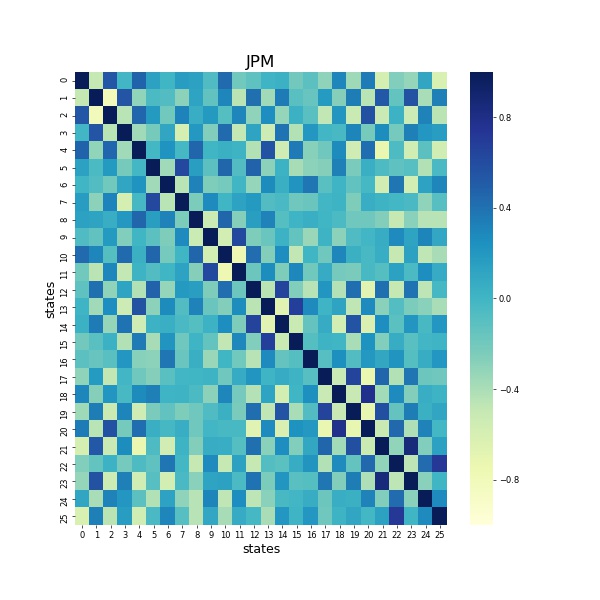}
  \caption{JPM}
  \label{fig:corr-jpm}
\end{subfigure} \hfill
\begin{subfigure}{0.23\textwidth}
  \includegraphics[width=\linewidth]{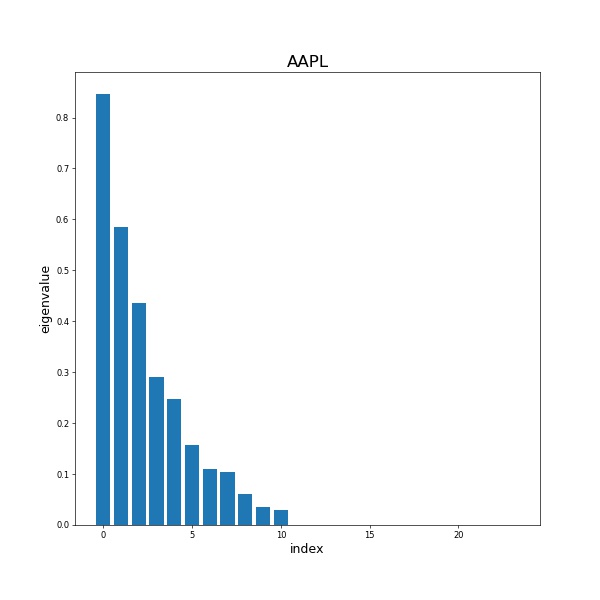}
  \caption{AAPL}
  \label{fig:eig-aapl}
\end{subfigure} \hfill
\begin{subfigure}{0.23\textwidth}
  \includegraphics[width=\linewidth]{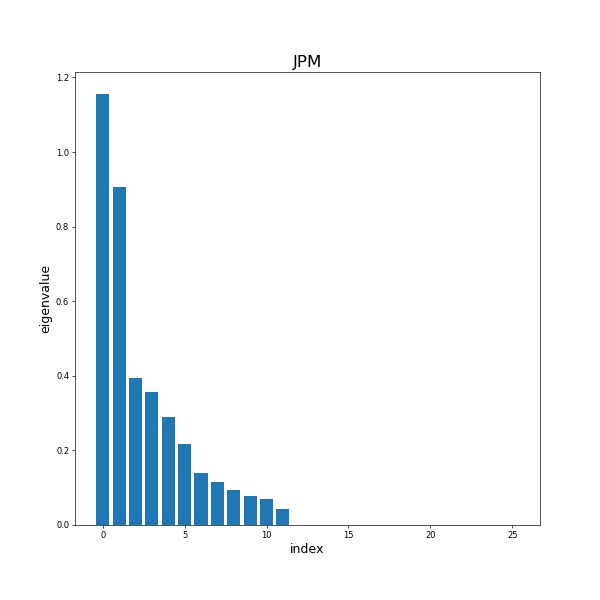}
  \caption{JPM}
  \label{fig:eig-jpm}
\end{subfigure}
\caption{Correlation matrix of hidden states, eigenvalues of transition covariance matrix, "AAPL" and "JPM"}
\label{fig:transcorr}
\end{figure}


We also show examples of correlation matrix of hidden states, eigenvalues of transition covariance matrix in Figure \ref{fig:transcorr} and appendix. It suggests that most of states don't highly correlate to each other, meanwhile a few states that share linear relationships. We find that half of the eigenvalues are very close to 0. Notice that the eigenvalues from a covariance matrix inform us the direction that the data has the maximum spread, the eigenvectors of the covariance matrix will be the direction with the most information. The information of correlation matrix and eigenvalues of covariance matrix could help us further simplify the number of states since more states in the model brings more computations. We also leave this as one of future directions.

\section{Conclusion}
On the basis of Kalman Filter, we develop a new method to determine the dimension of hidden state for each stock. Our methods provides smoothing predictions of intraday traded volume, shows an potential of using gradient based method for further analysis. Through experiments we demonstrate that v-state model gains better prediction precision than two-state model and RM model. Inspired by a series of model analysis, in the future we will conduct further research on reducing and number of DOFs and selecting states.

\bibliography{paper}

\begin{thebibliography}{5}
\providecommand{\natexlab}[1]{#1}
\providecommand{\url}[1]{\texttt{#1}}
\expandafter\ifx\csname urlstyle\endcsname\relax
  \providecommand{\doi}[1]{doi: #1}\else
  \providecommand{\doi}{doi: \begingroup \urlstyle{rm}\Url}\fi

\bibitem[Ajinkya \& Jain(1989)Ajinkya and Jain]{behaviour}
Ajinkya, B. and Jain, P.
\newblock The behavior of daily stock market trading volume.
\newblock \emph{Journal of accounting and economics}, 11\penalty0 (4):\penalty0
  331--359, 1989.

\bibitem[Brownlees et~al.(2011)Brownlees, Cipollini, and Gallo]{cmem}
Brownlees, C., Cipollini, F., and Gallo, G.
\newblock Intra-daily volume modeling and prediction for algorithmic trading.
\newblock \emph{Journal of Financial Econometrics}, 9\penalty0 (3):\penalty0
  489--518, 2011.

\bibitem[Chen et~al.(2016)Chen, Feng, and Palomar]{kffortrading}
Chen, R., Feng, Y., and Palomar, D.
\newblock Forecasting intraday trading volume: a kalman filter approach.
\newblock In \emph{SSRN 3101695}, 2016.

\bibitem[Kalman(1960)]{kforiginal}
Kalman, R.
\newblock A new approach to linear filtering and prediction problems.
\newblock 1960.

\bibitem[Shumway \& Stoffer(1982)Shumway and Stoffer]{emalg}
Shumway, R. and Stoffer, D.
\newblock An approach to time series smoothing and forecasting using the em
  algorithm.
\newblock \emph{Journal of time series analysis}, 3\penalty0 (4):\penalty0
  253--264, 1982.

\end{thebibliography}
\bibliographystyle{nips2021}

\appendix
\begin{figure}[t!]
\begin{subfigure}{0.23\textwidth}
  \includegraphics[width=\linewidth]{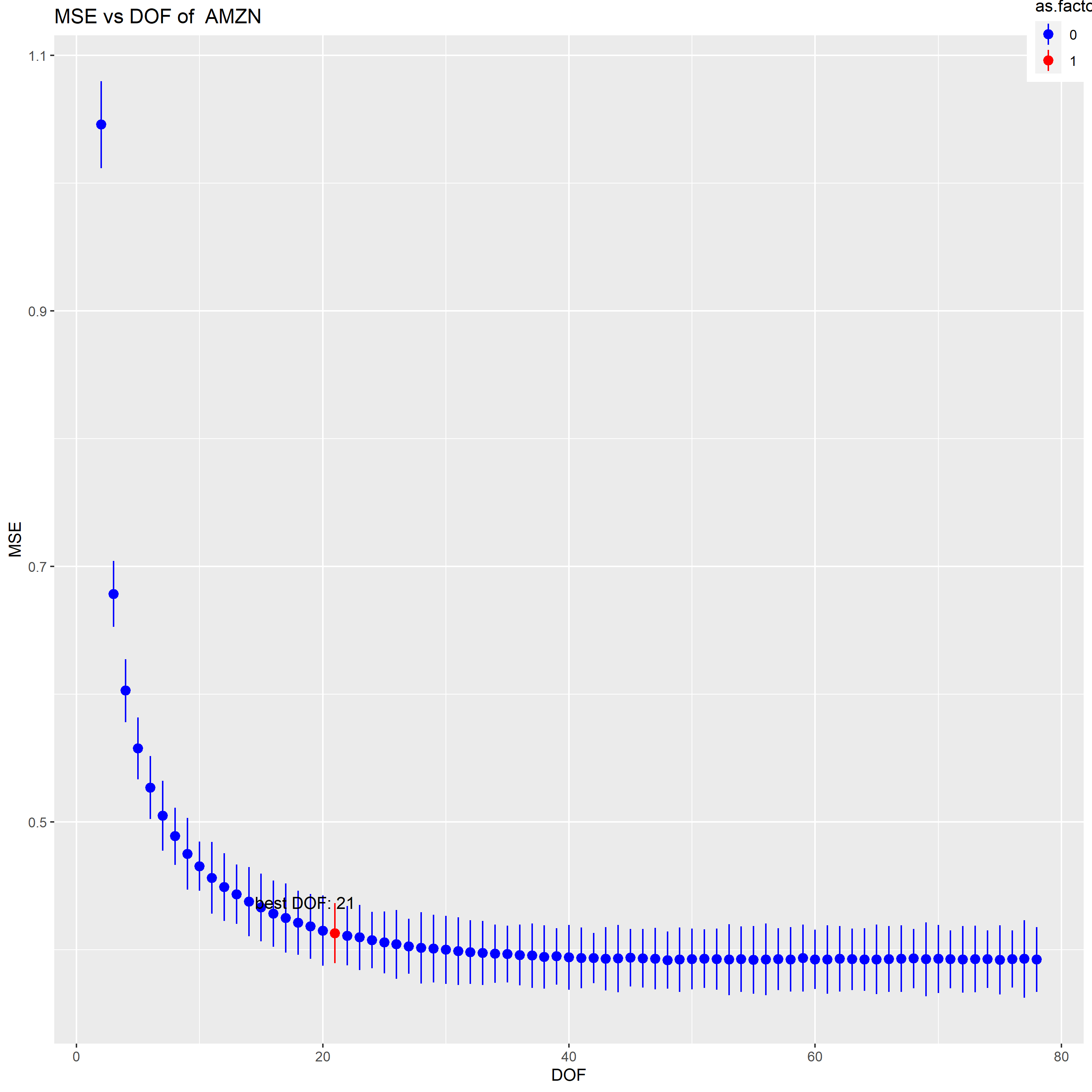}  
  \caption{Best DOF of AMZN}
\end{subfigure}
\begin{subfigure}{0.23\textwidth}
  \includegraphics[width=\linewidth]{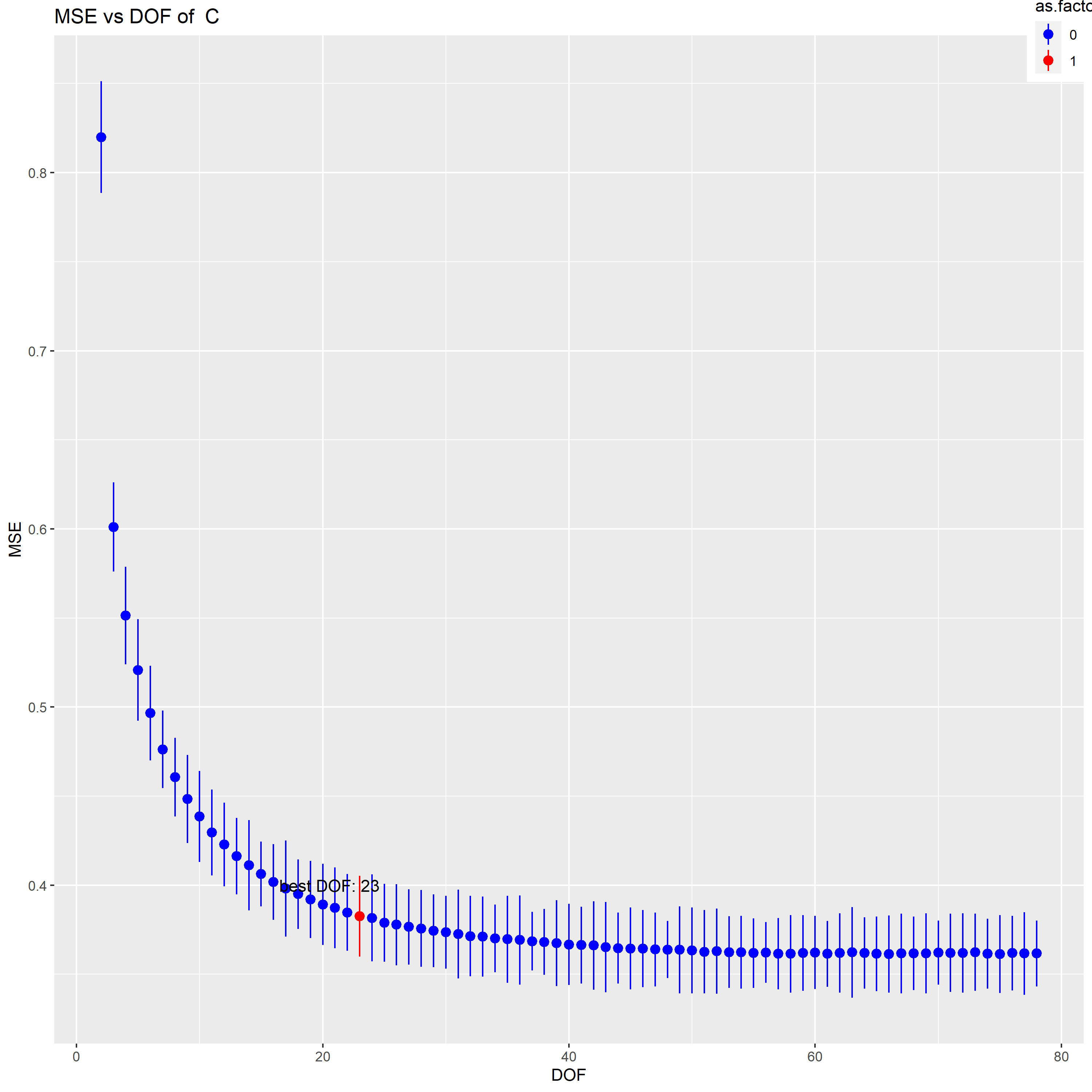}  
  \caption{Best DOF of C}
\end{subfigure}
\begin{subfigure}{0.23\textwidth}
  \includegraphics[width=\linewidth]{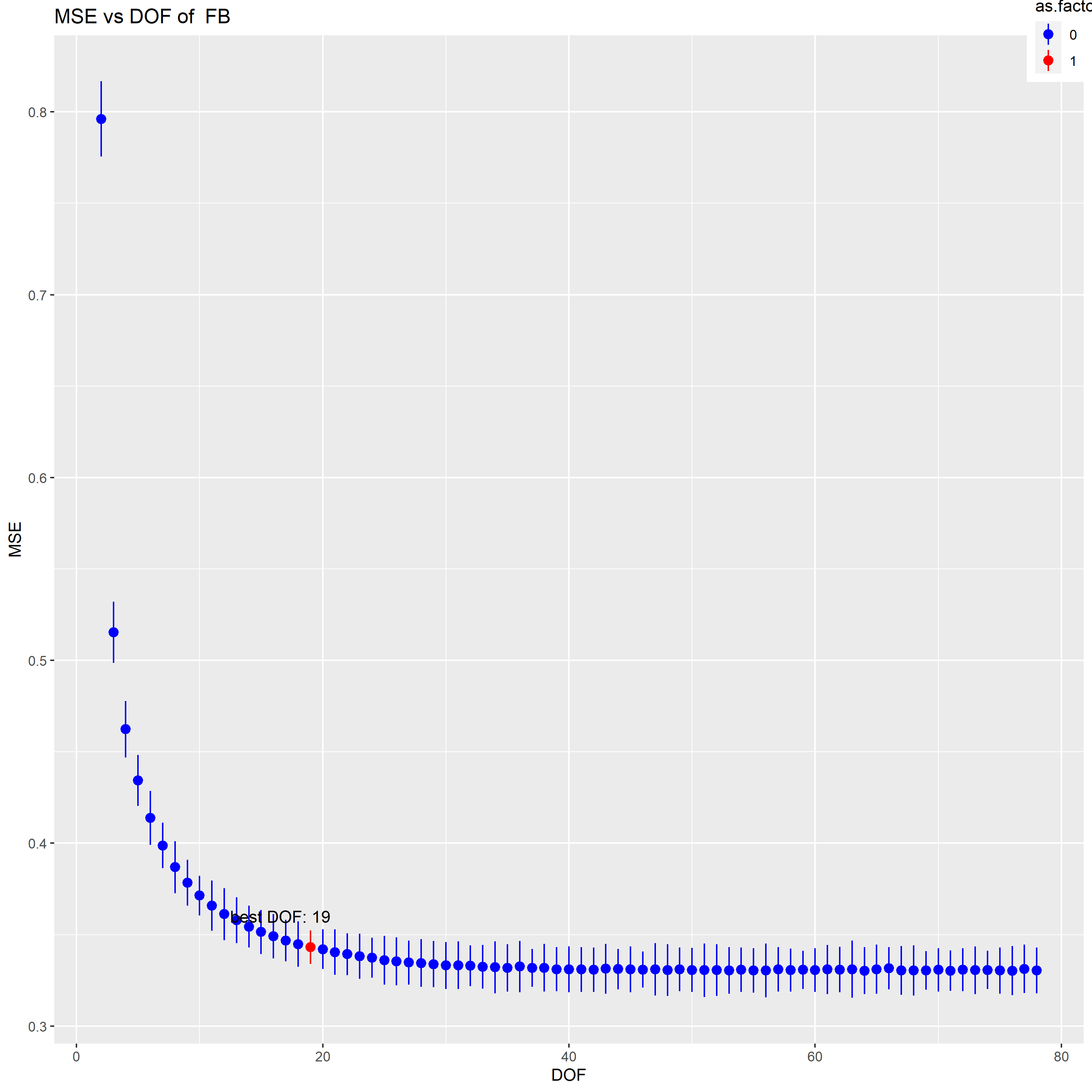}  
  \caption{Best DOF of FB}
\end{subfigure}
\begin{subfigure}{0.23\textwidth}
  \includegraphics[width=\linewidth]{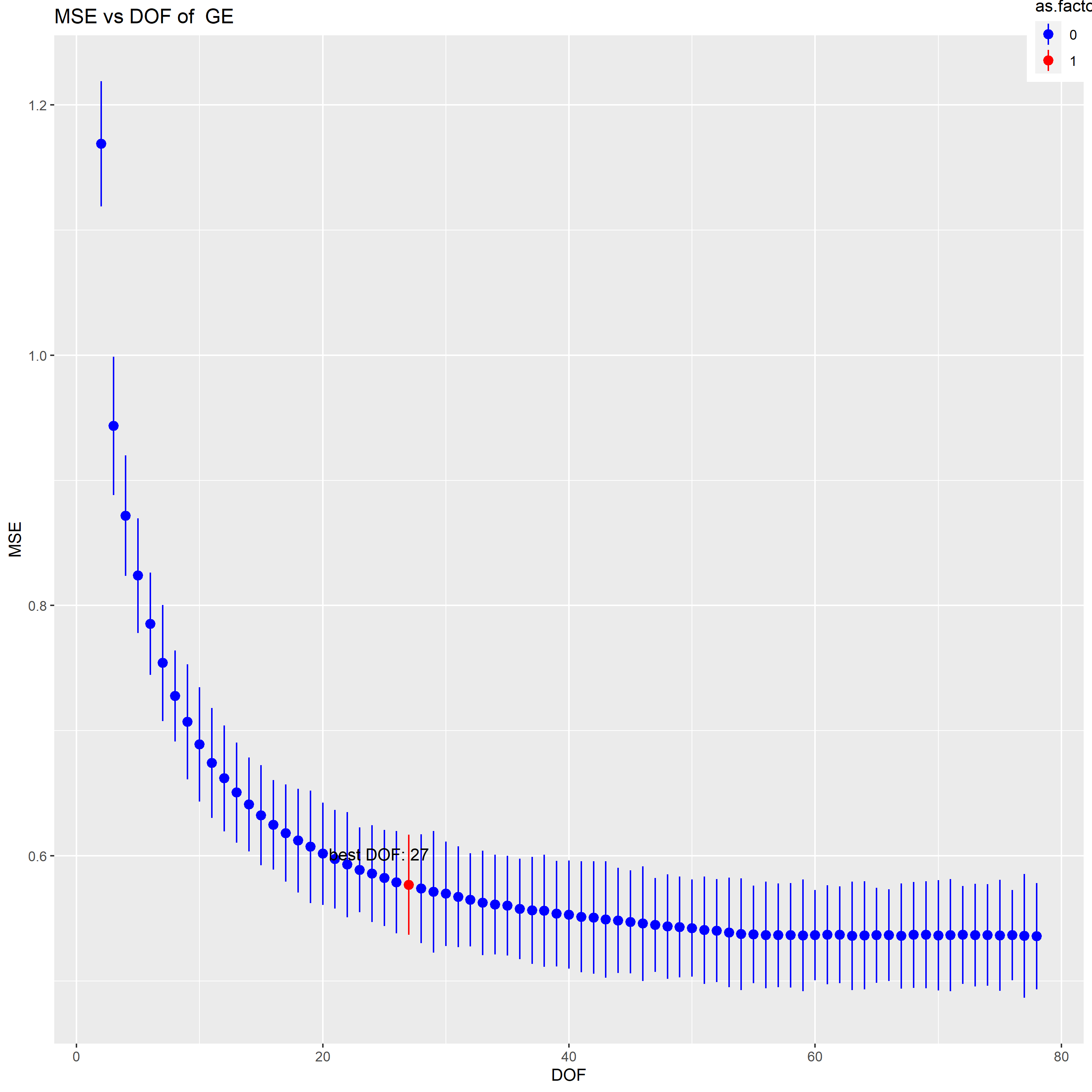}  
  \caption{Best DOF of GE}
\end{subfigure}\\
\begin{subfigure}{0.23\textwidth}
  \includegraphics[width=\linewidth]{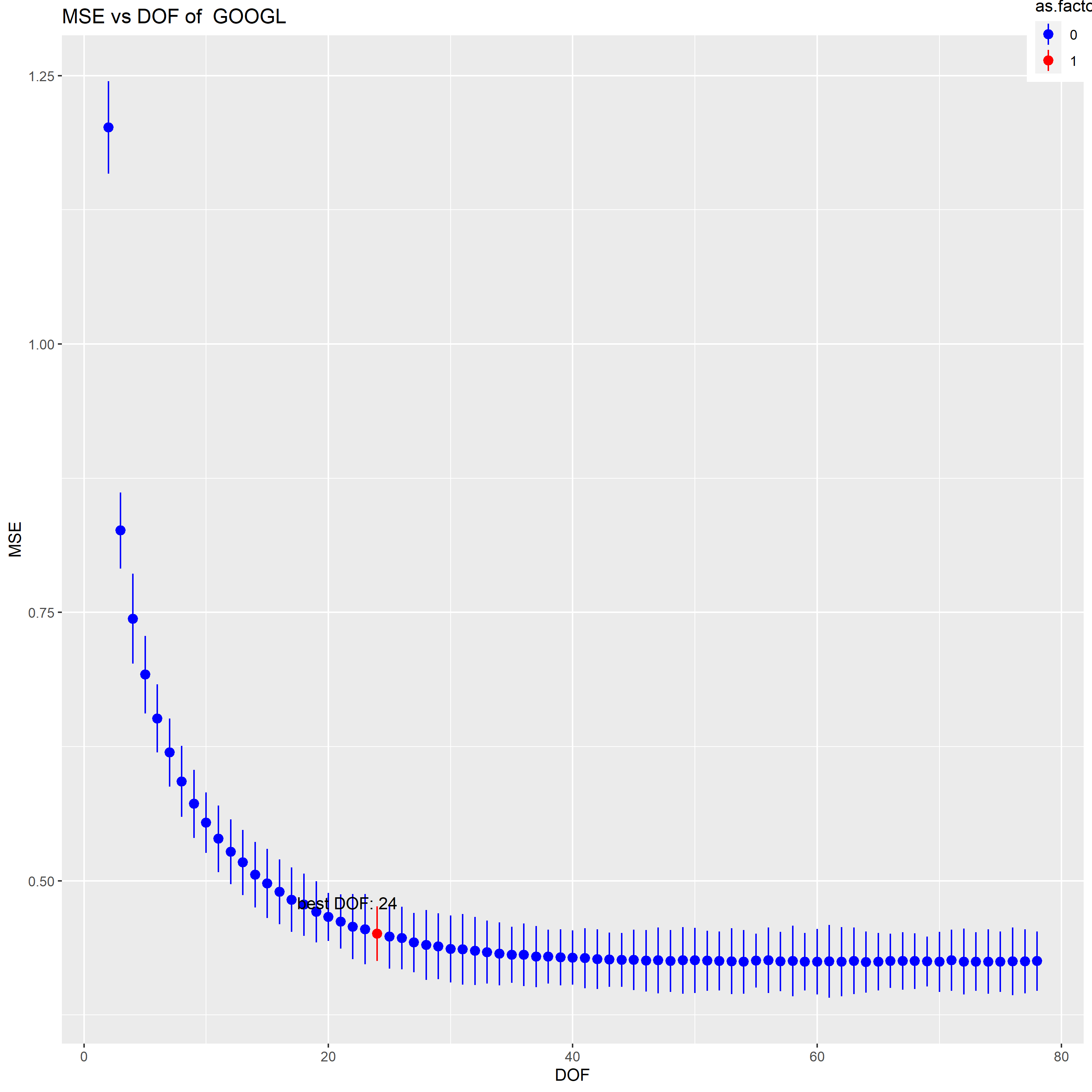}  
  \caption{Best DOF of GOOGL}
\end{subfigure}
\begin{subfigure}{0.23\textwidth}
  \includegraphics[width=\linewidth]{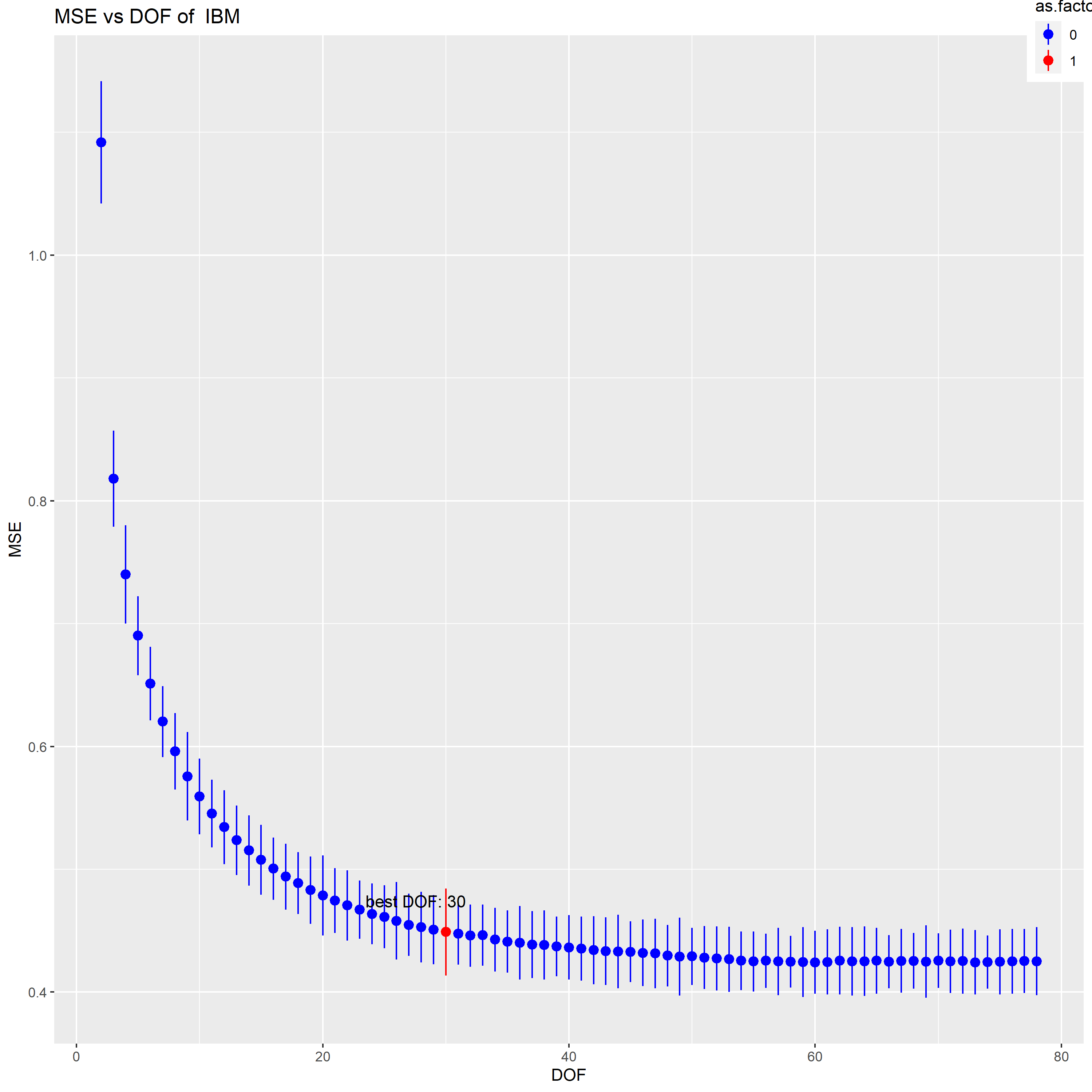}  
  \caption{Best DOF of IBM}
\end{subfigure}
\begin{subfigure}{0.23\textwidth}
  \includegraphics[width=\linewidth]{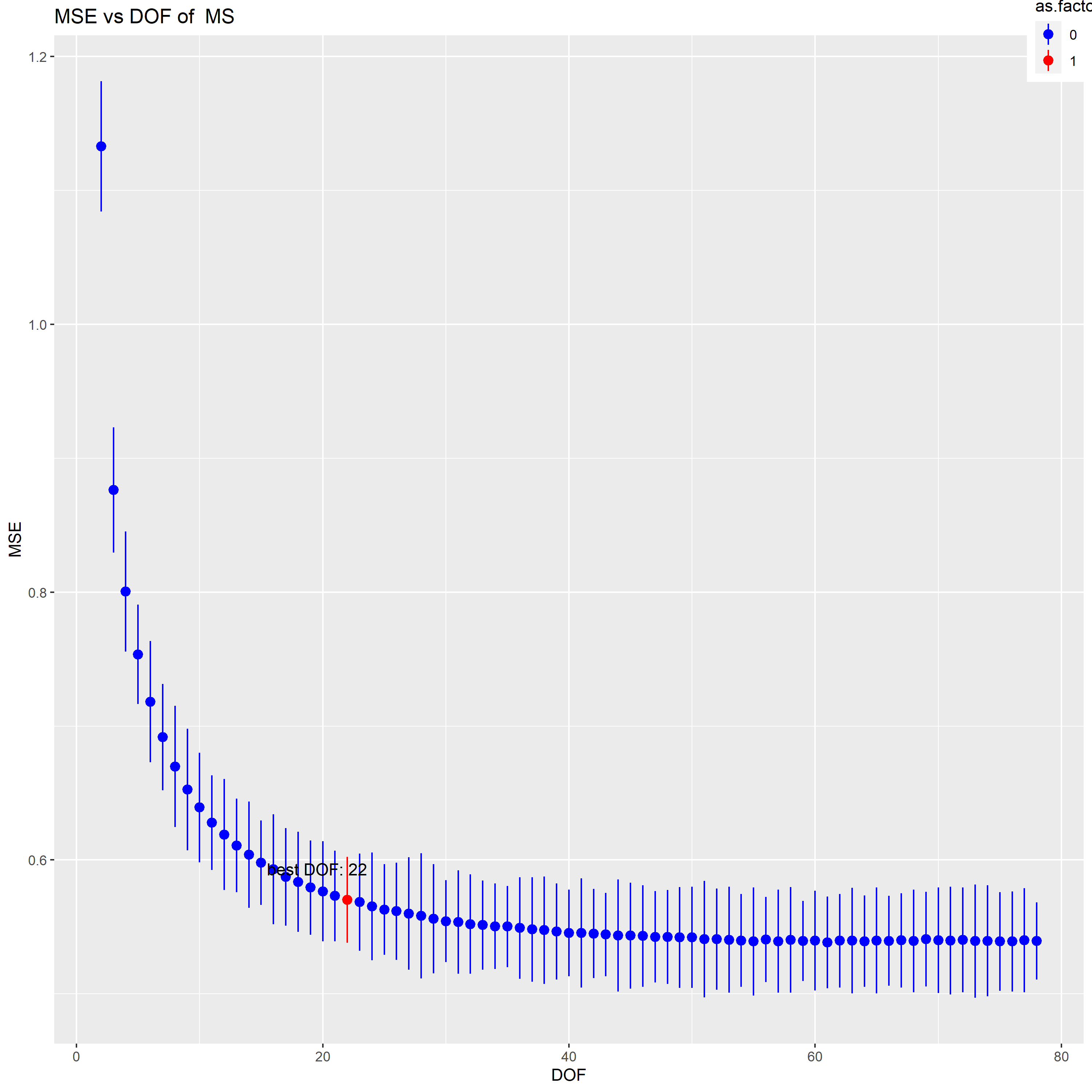}  
  \caption{Best DOF of MS}
\end{subfigure}
\begin{subfigure}{0.23\textwidth}
  \includegraphics[width=\linewidth]{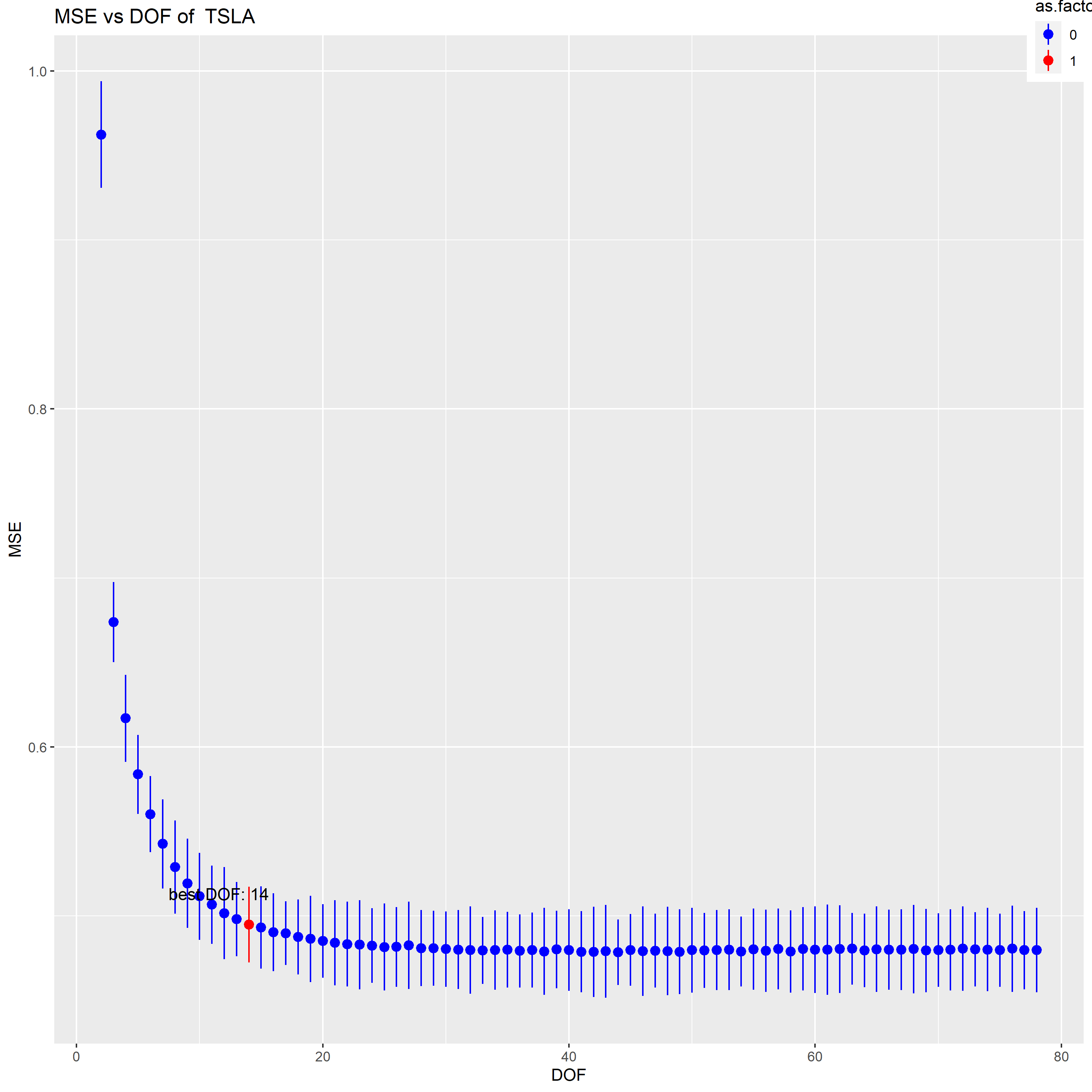}  
  \caption{Best DOF of TSLA}
\end{subfigure}
\caption{DOFs of states}
\label{fig:appen-dofdist}
\end{figure}

\begin{figure}[!ht]
\begin{subfigure}{0.25\textwidth}
  \includegraphics[width=\linewidth]{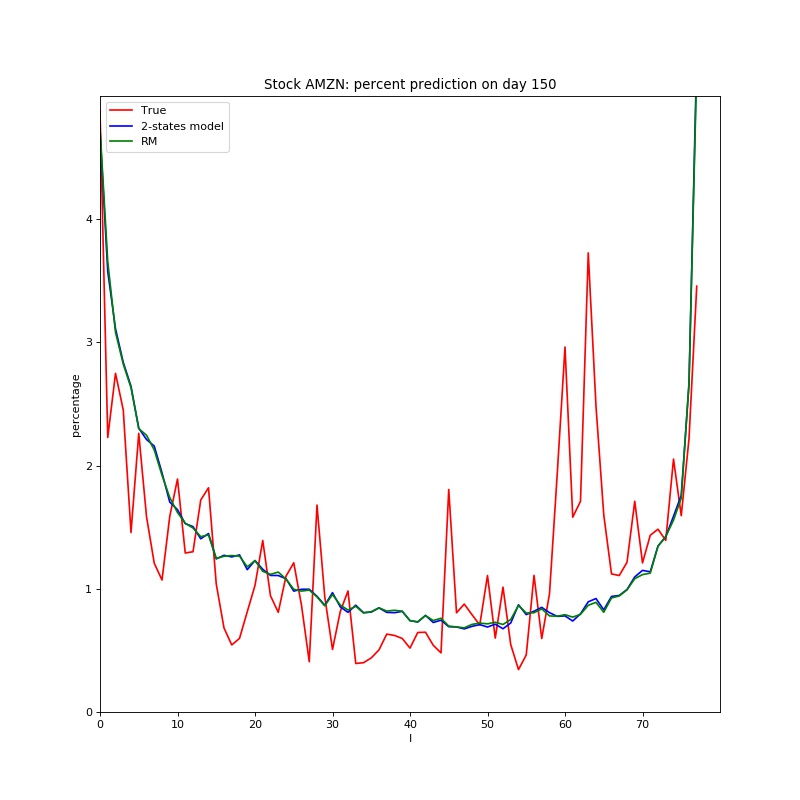}
  \caption{}
  \label{fig:com-AMZN1}
\end{subfigure}\hfill
\begin{subfigure}{0.25\textwidth}
  \includegraphics[width=\linewidth]{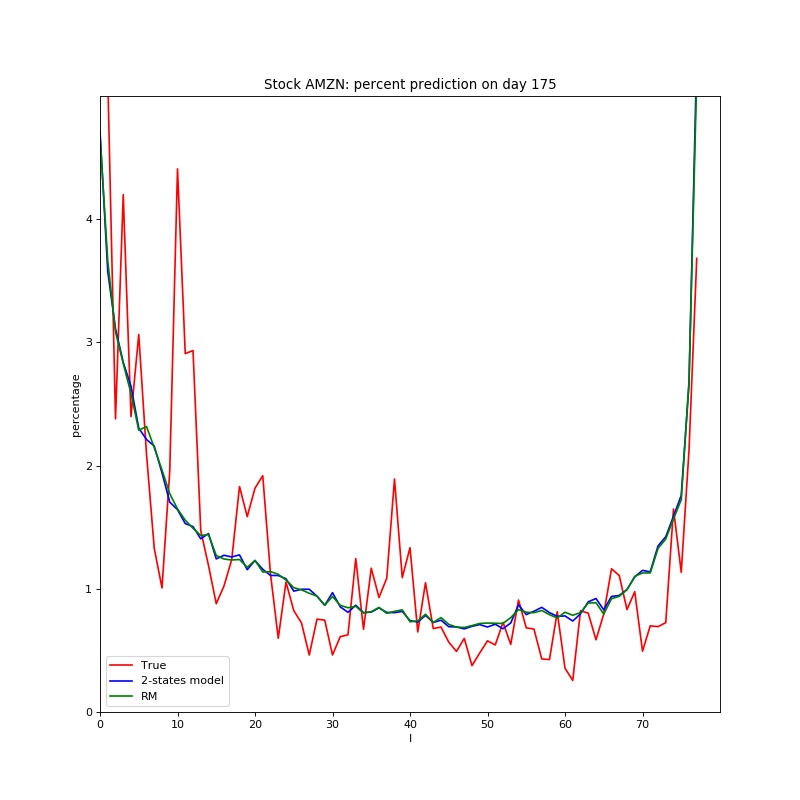}
  \caption{}
  \label{fig:com-AMZN2}
\end{subfigure}\hfill
\begin{subfigure}{0.25\textwidth}
  \includegraphics[width=\linewidth]{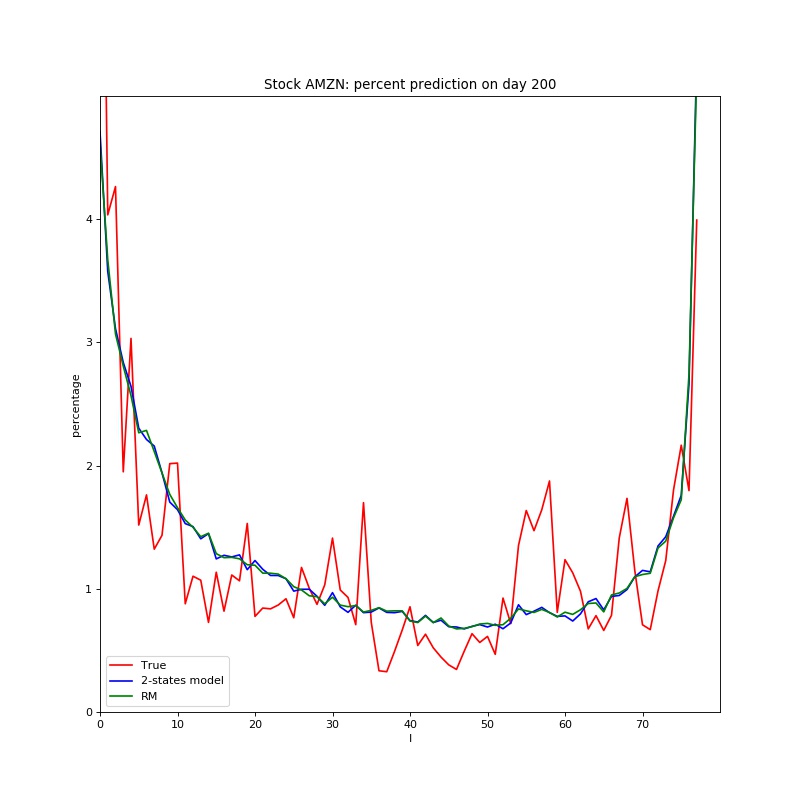}
  \caption{}
  \label{fig:com-AMZN3}
\end{subfigure}\hfill
\begin{subfigure}{0.25\textwidth}
  \includegraphics[width=\linewidth]{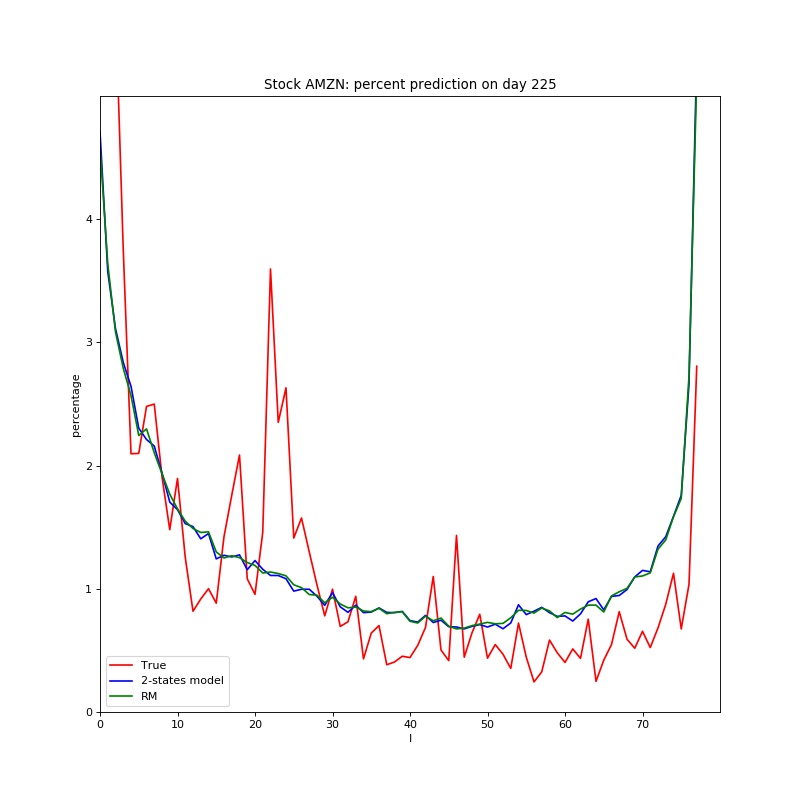}
  \caption{}
  \label{fig:com-AMZN4}
\end{subfigure}\\
\begin{subfigure}{0.25\textwidth}
  \includegraphics[width=\linewidth]{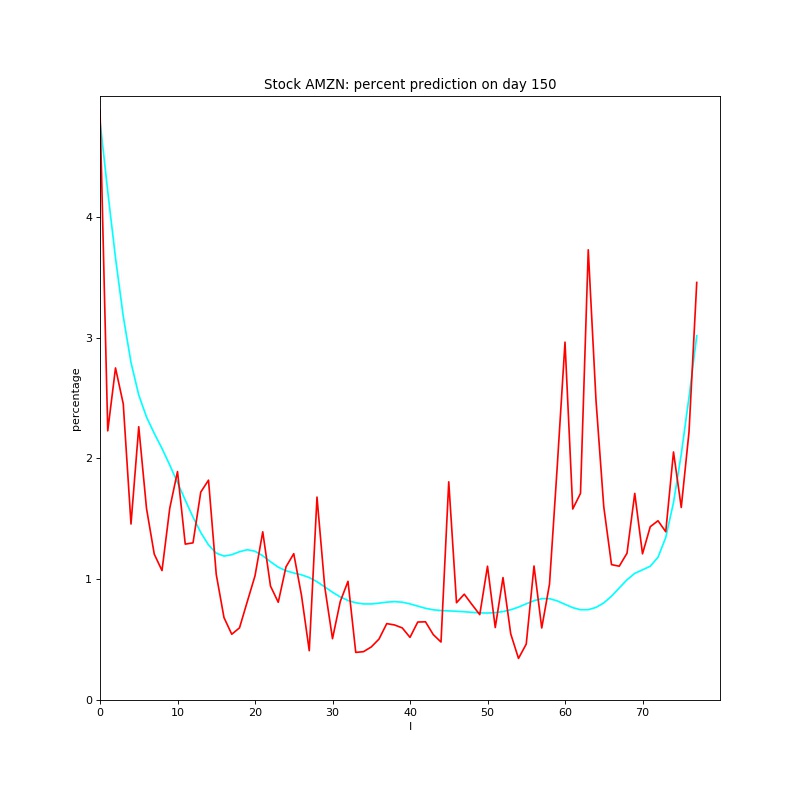}  
  \caption{}
  \label{fig:com-AMZN5}
\end{subfigure}\hfill
\begin{subfigure}{0.25\textwidth}
  \includegraphics[width=\linewidth]{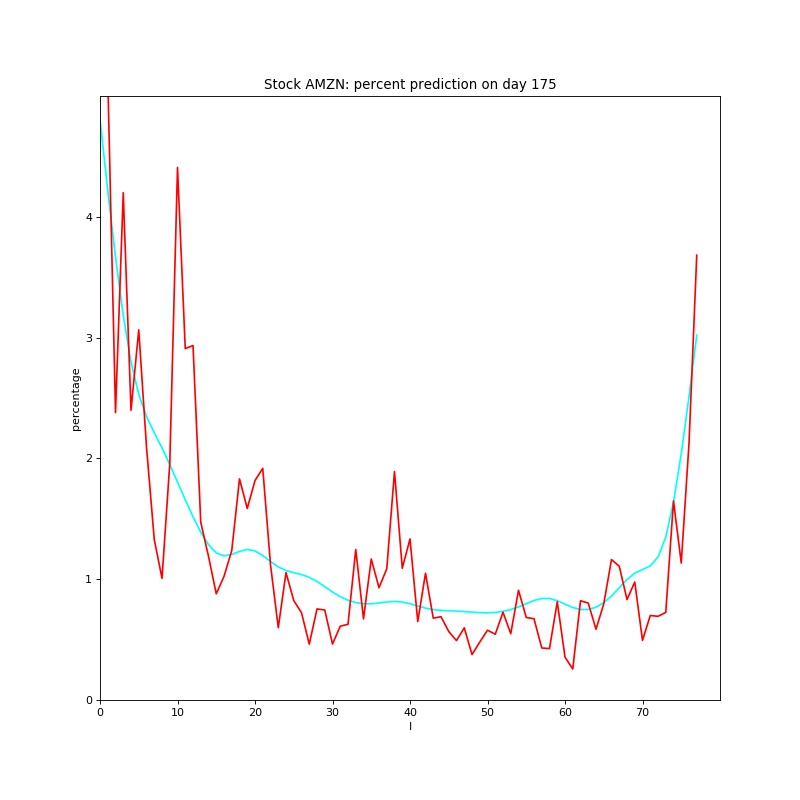}
  \caption{}
  \label{fig:com-AMZN6}
\end{subfigure}\hfill
\begin{subfigure}{0.25\textwidth}
  \includegraphics[width=\linewidth]{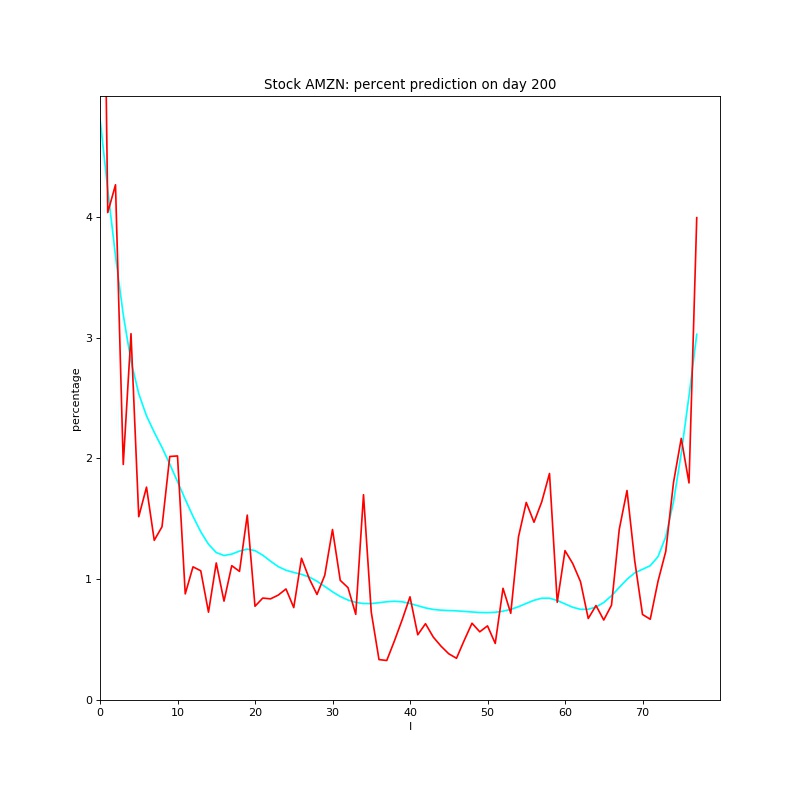}
  \caption{}
  \label{fig:com-AMZN7}
\end{subfigure}\hfill
\begin{subfigure}{0.25\textwidth}
  \includegraphics[width=\linewidth]{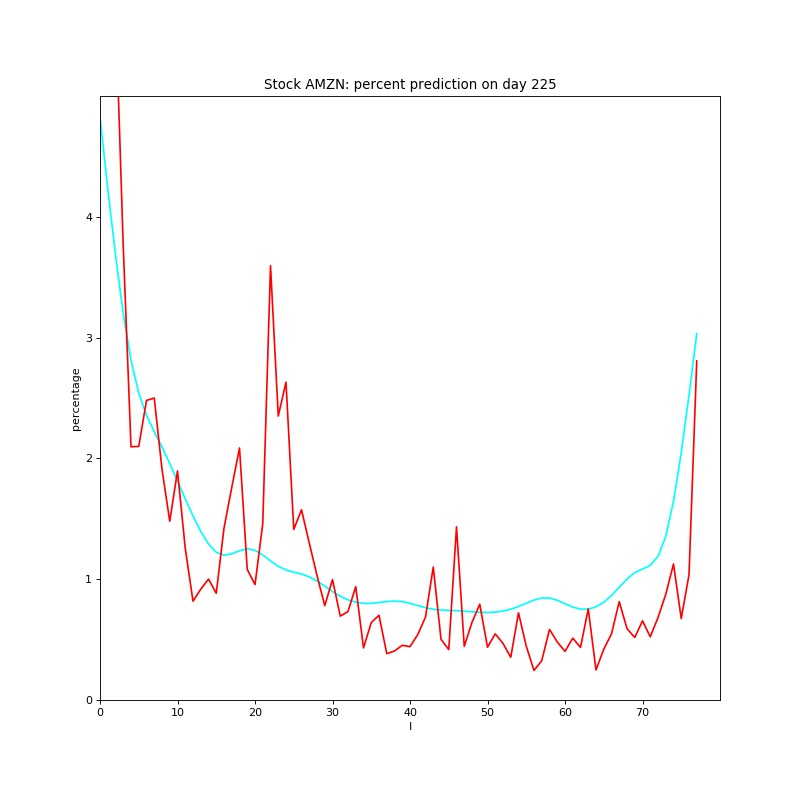}
  \caption{}
  \label{fig:com-AMZN8}
\end{subfigure}
\caption{comparison of prediction: (a) to (d):baseline models on stock "AMZN", (e) to (h):our v-state model on stock "AMZN"}
\label{fig:pre-appen1}
\end{figure}

\begin{figure}[!ht]
\begin{subfigure}{0.25\textwidth}
  \includegraphics[width=\linewidth]{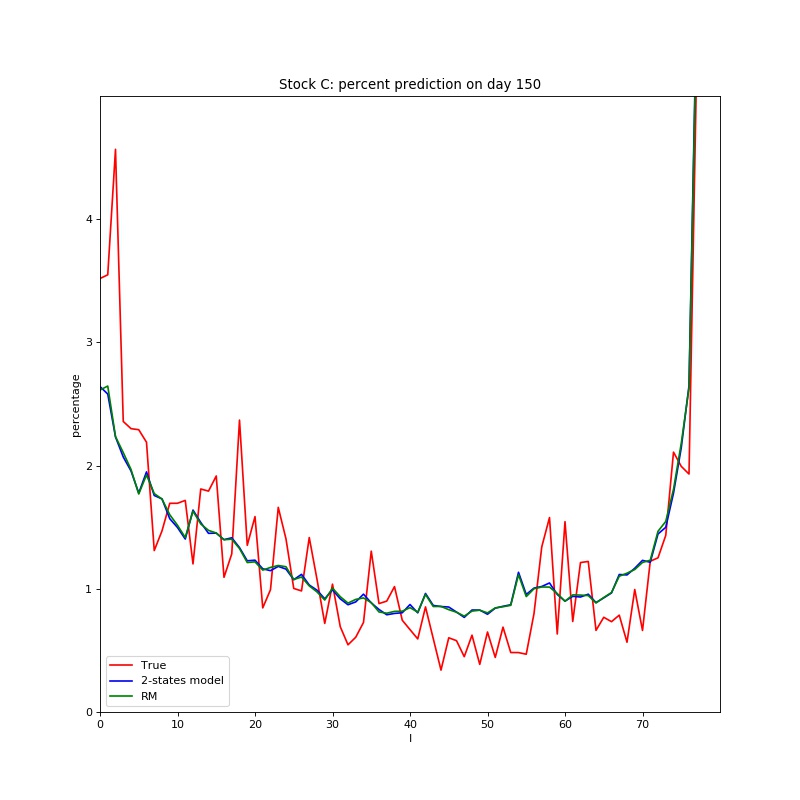}
  \caption{}
  \label{fig:com-C1}
\end{subfigure}\hfill
\begin{subfigure}{0.25\textwidth}
  \includegraphics[width=\linewidth]{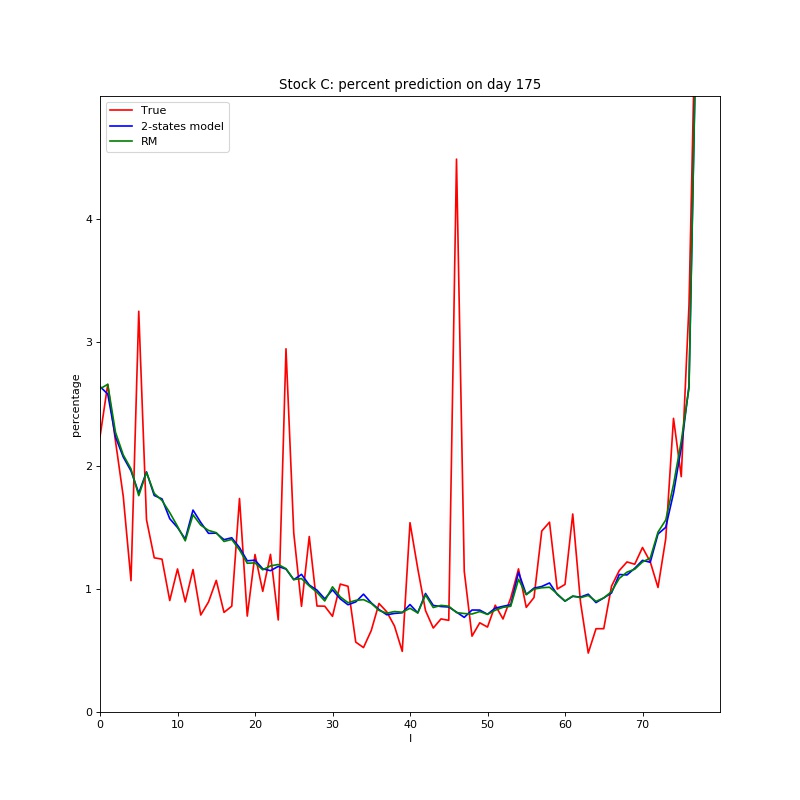}
  \caption{}
  \label{fig:com-C2}
\end{subfigure}\hfill
\begin{subfigure}{0.25\textwidth}
  \includegraphics[width=\linewidth]{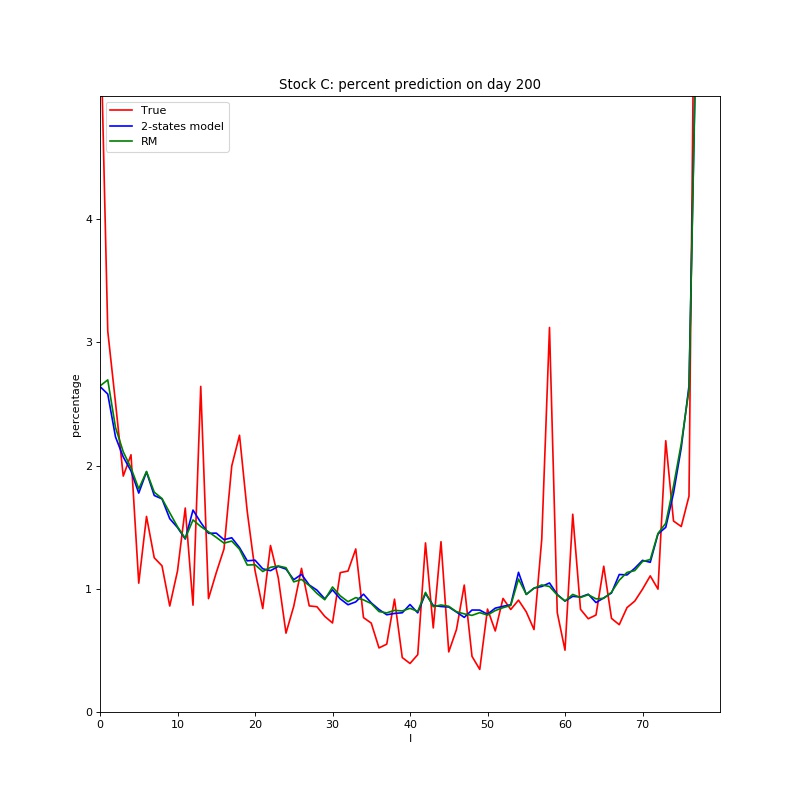}
  \caption{}
  \label{fig:com-C3}
\end{subfigure}\hfill
\begin{subfigure}{0.25\textwidth}
  \includegraphics[width=\linewidth]{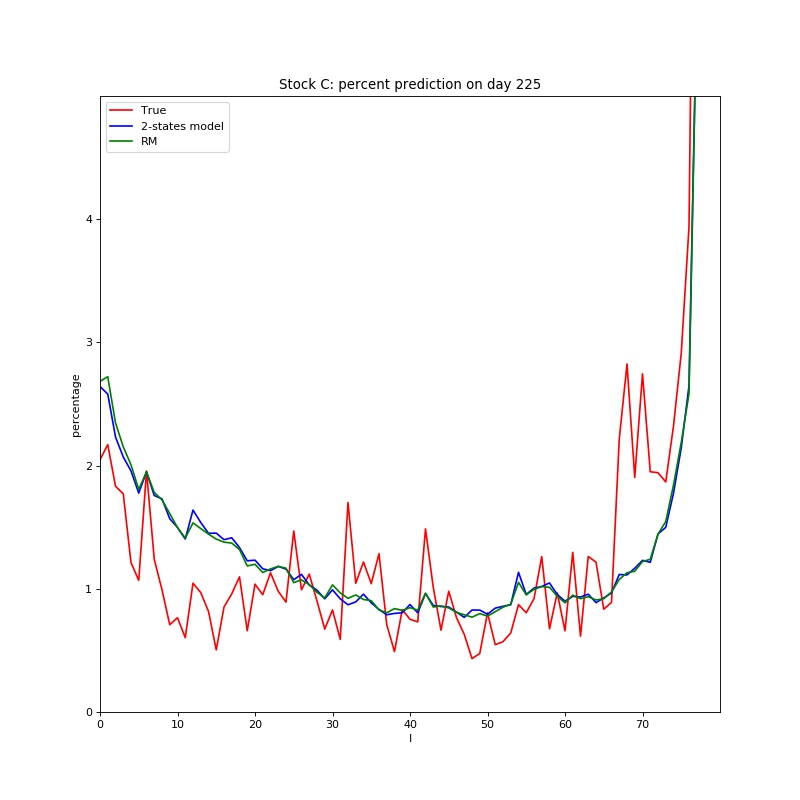}
  \caption{}
  \label{fig:com-C4}
\end{subfigure}\\
\begin{subfigure}{0.25\textwidth}
  \includegraphics[width=\linewidth]{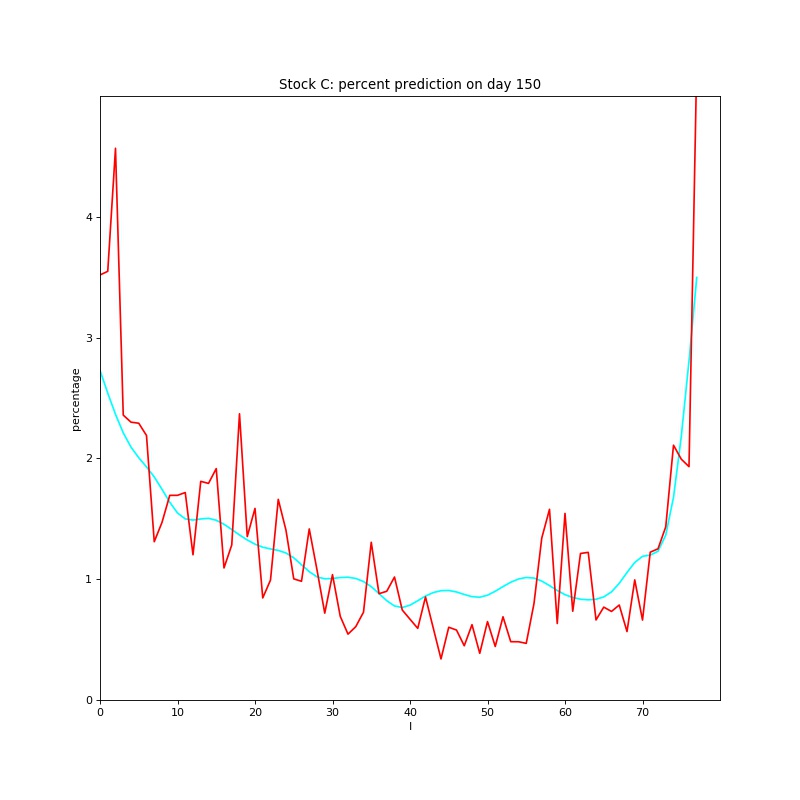}  
  \caption{}
  \label{fig:com-C5}
\end{subfigure}\hfill
\begin{subfigure}{0.25\textwidth}
  \includegraphics[width=\linewidth]{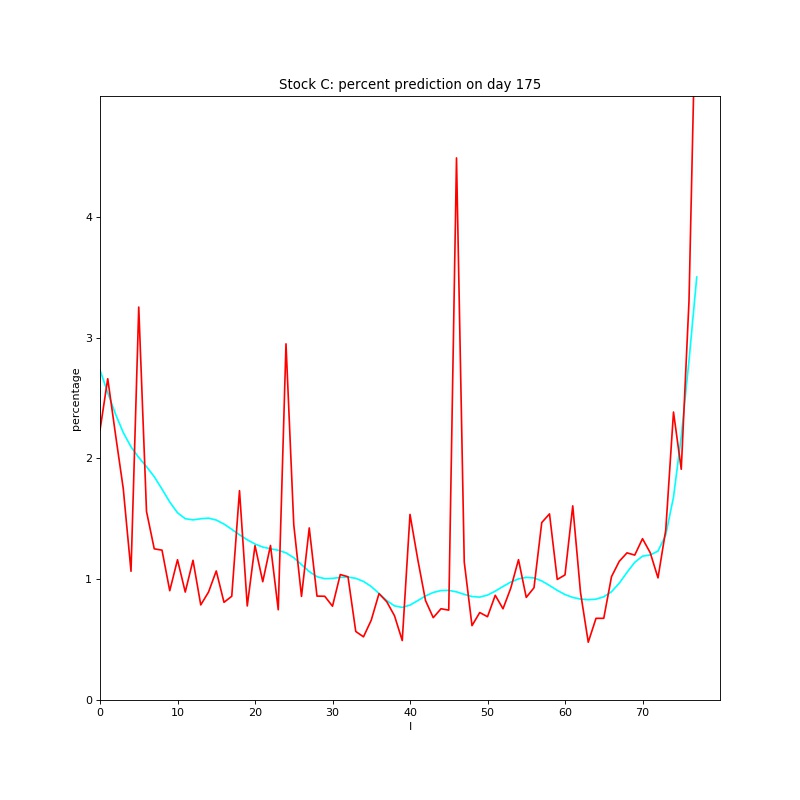}
  \caption{}
  \label{fig:com-C6}
\end{subfigure}\hfill
\begin{subfigure}{0.25\textwidth}
  \includegraphics[width=\linewidth]{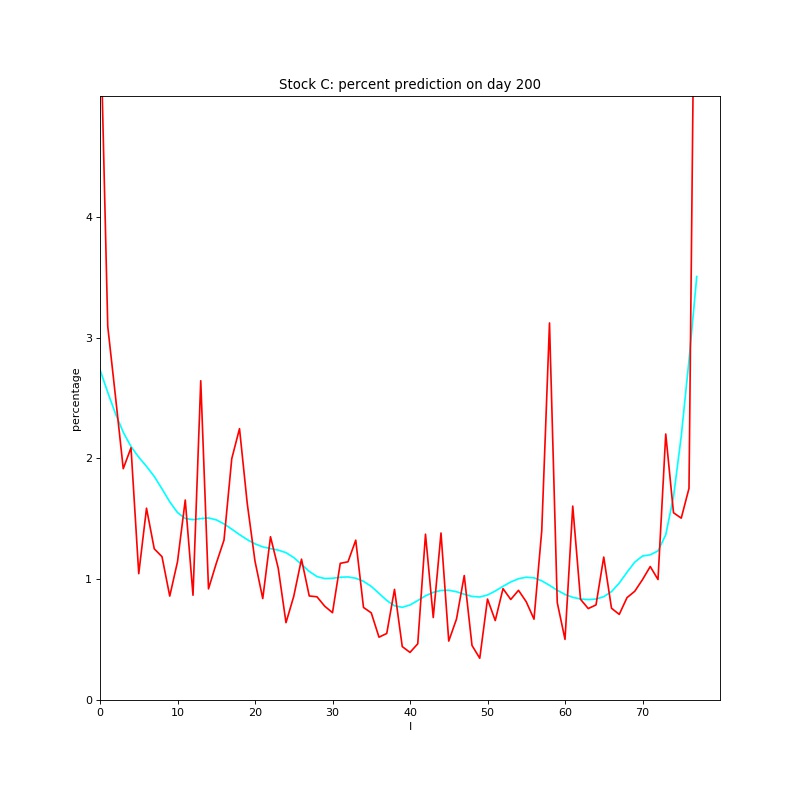}
  \caption{}
  \label{fig:com-C7}
\end{subfigure}\hfill
\begin{subfigure}{0.25\textwidth}
  \includegraphics[width=\linewidth]{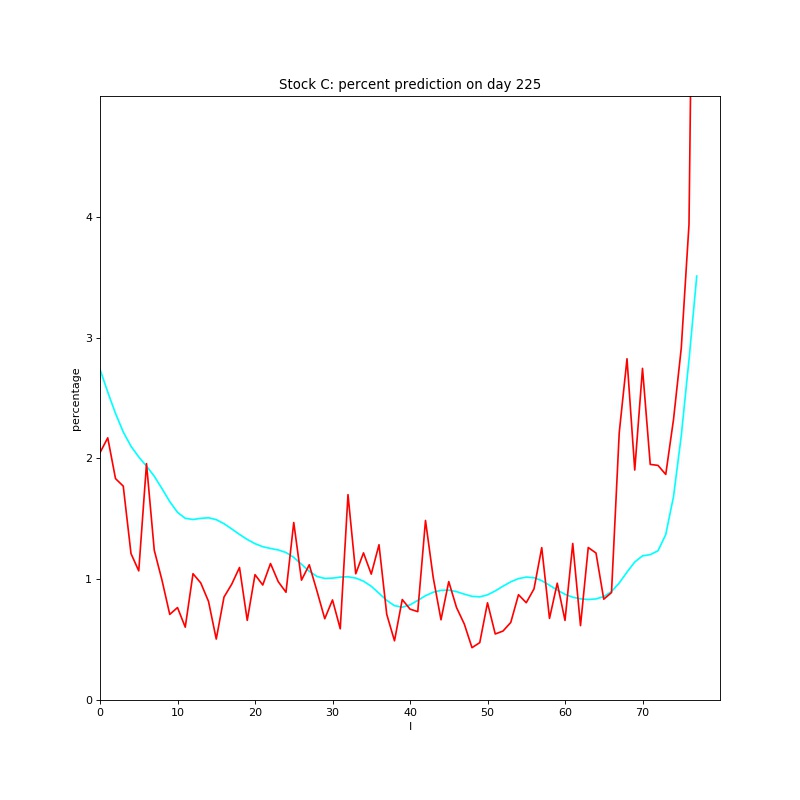}
  \caption{}
  \label{fig:com-C8}
\end{subfigure}
\caption{comparison of prediction: (a) to (d):baseline models on stock "C", (e) to (h):our v-state model on stock "C"}
\label{fig:pre-appen2}
\end{figure}

\begin{figure}[!ht]
\begin{subfigure}{0.25\textwidth}
  \includegraphics[width=\linewidth]{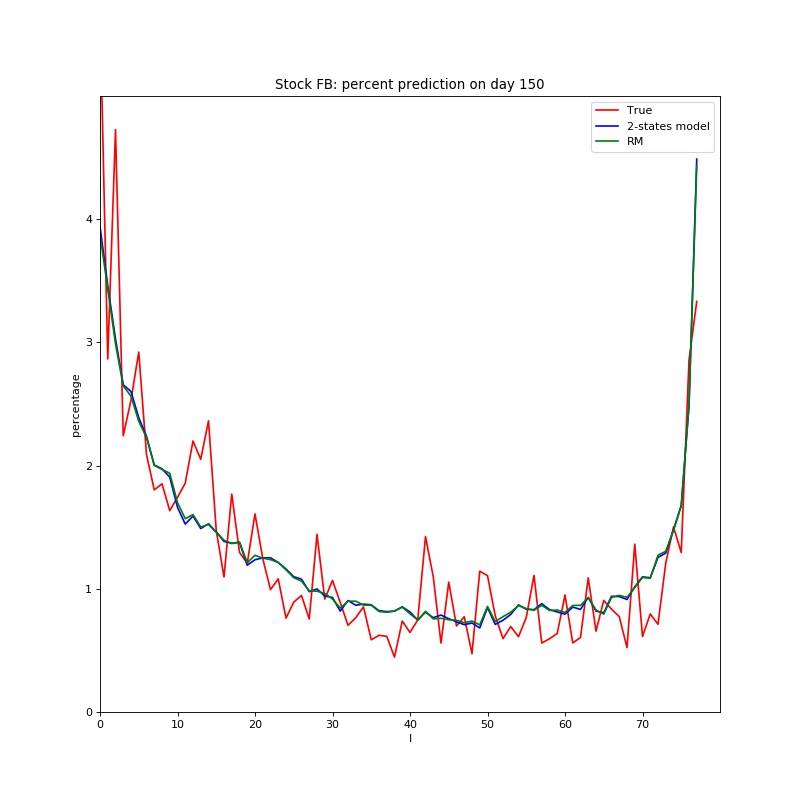}
  \caption{}
  \label{fig:com-FB1}
\end{subfigure}\hfill
\begin{subfigure}{0.25\textwidth}
  \includegraphics[width=\linewidth]{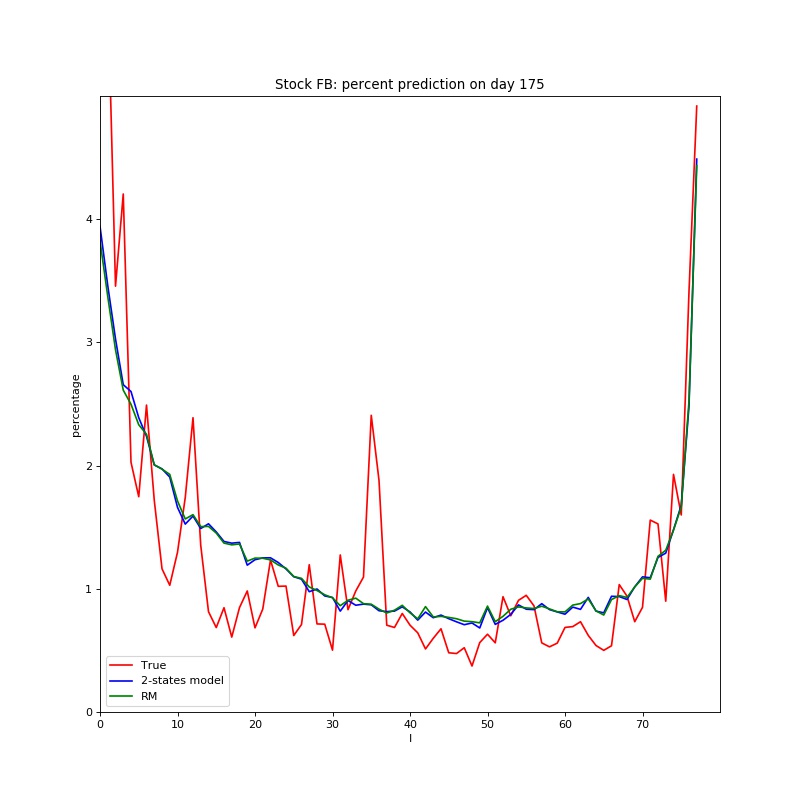}
  \caption{}
  \label{fig:com-FB2}
\end{subfigure}\hfill
\begin{subfigure}{0.25\textwidth}
  \includegraphics[width=\linewidth]{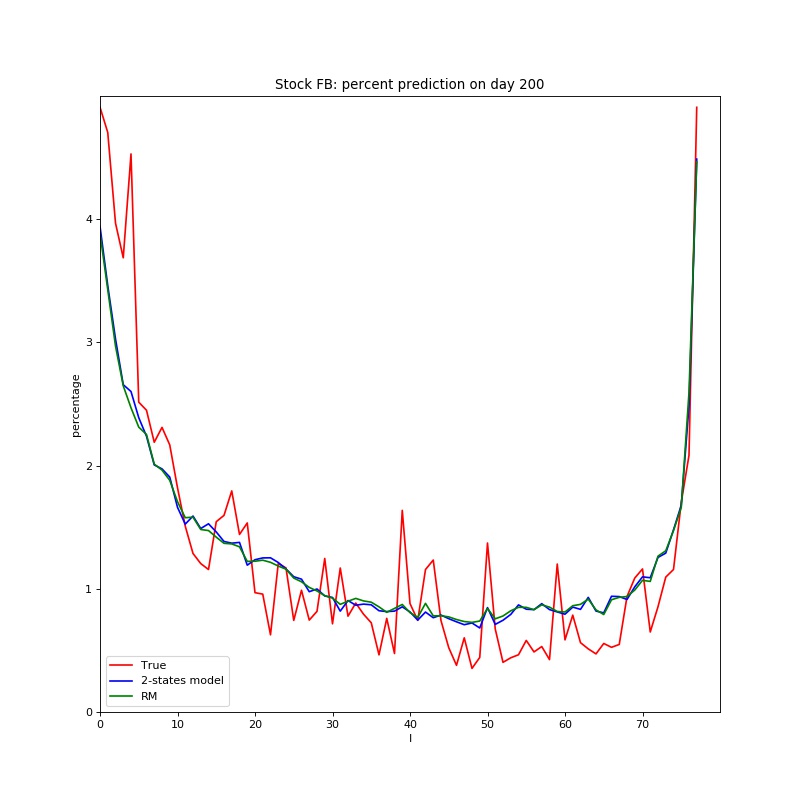}
  \caption{}
  \label{fig:com-FB3}
\end{subfigure}\hfill
\begin{subfigure}{0.25\textwidth}
  \includegraphics[width=\linewidth]{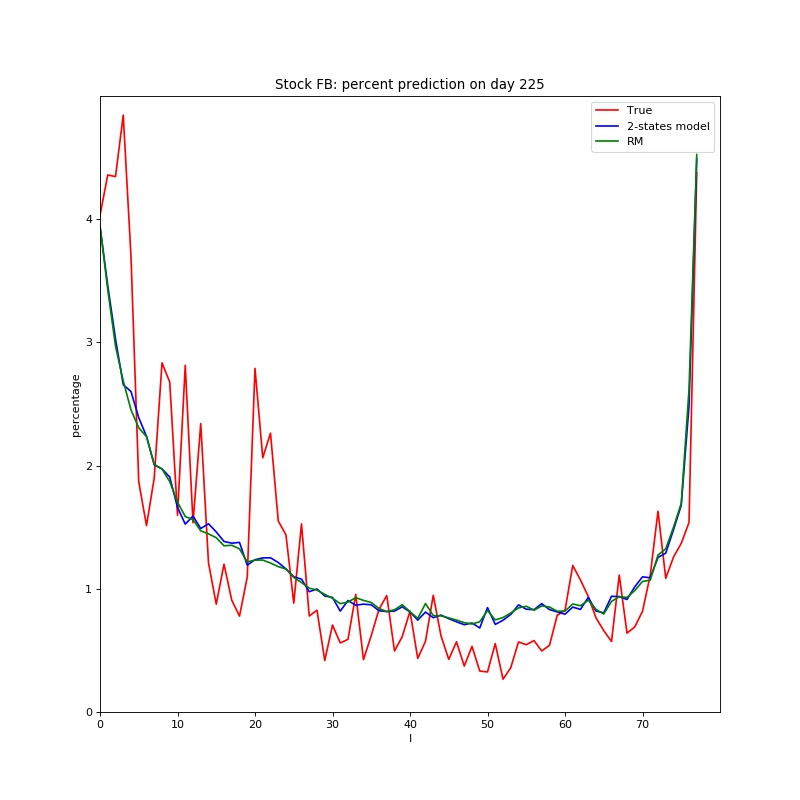}
  \caption{}
  \label{fig:com-FB4}
\end{subfigure}\\
\begin{subfigure}{0.25\textwidth}
  \includegraphics[width=\linewidth]{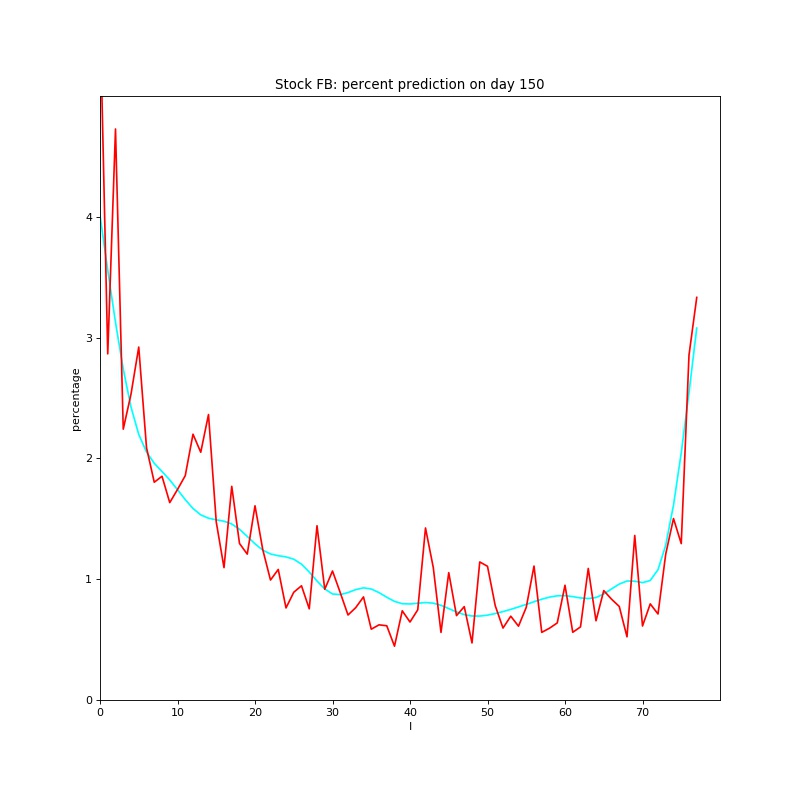}  
  \caption{}
  \label{fig:com-FB5}
\end{subfigure}\hfill
\begin{subfigure}{0.25\textwidth}
  \includegraphics[width=\linewidth]{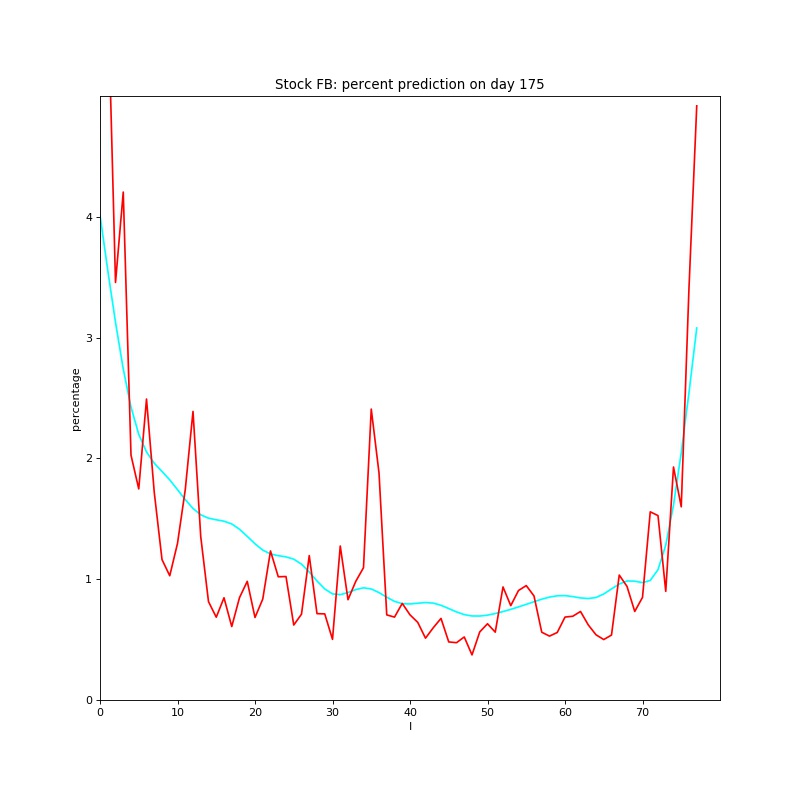}
  \caption{}
  \label{fig:com-FB6}
\end{subfigure}\hfill
\begin{subfigure}{0.25\textwidth}
  \includegraphics[width=\linewidth]{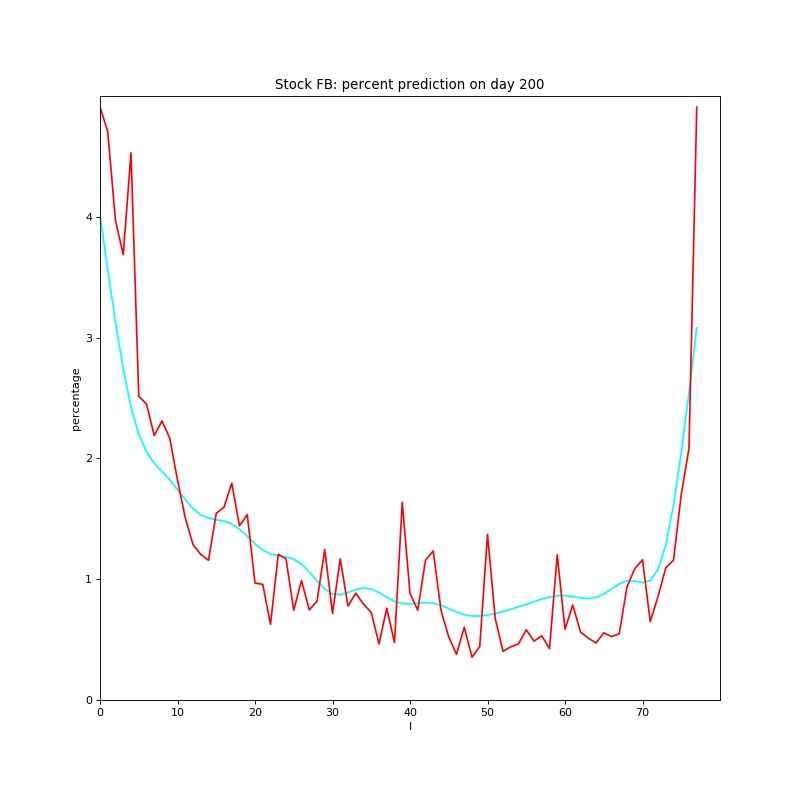}
  \caption{}
  \label{fig:com-FB7}
\end{subfigure}\hfill
\begin{subfigure}{0.25\textwidth}
  \includegraphics[width=\linewidth]{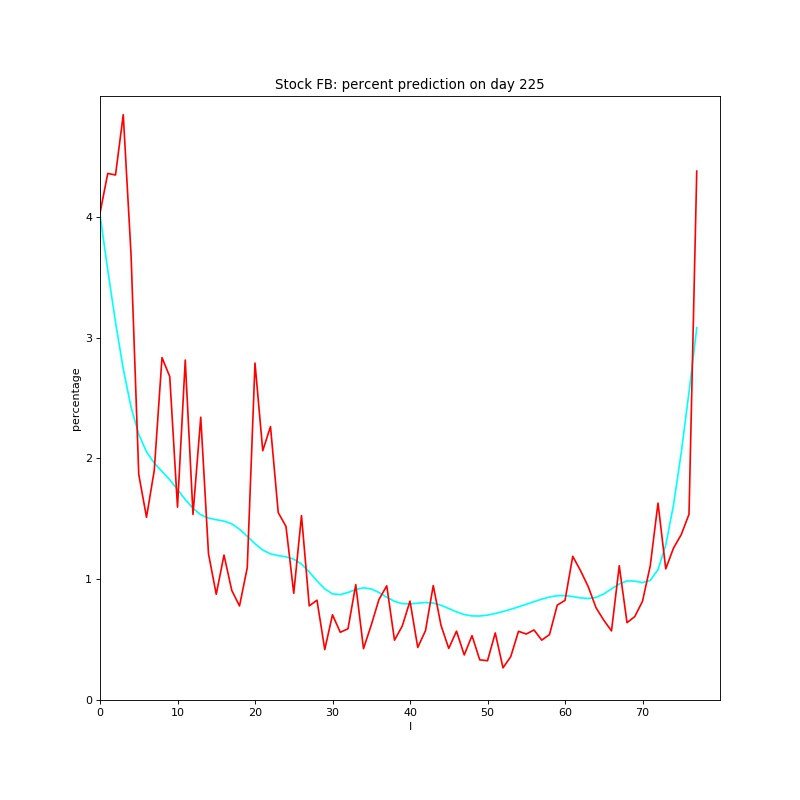}
  \caption{}
  \label{fig:com-FB8}
\end{subfigure}
\caption{comparison of prediction: (a) to (d):baseline models on stock "FB", (e) to (h):our v-state model on stock "FB"}
\label{fig:pre-appen3}
\end{figure}

\begin{figure}[!ht]
\begin{subfigure}{0.25\textwidth}
  \includegraphics[width=\linewidth]{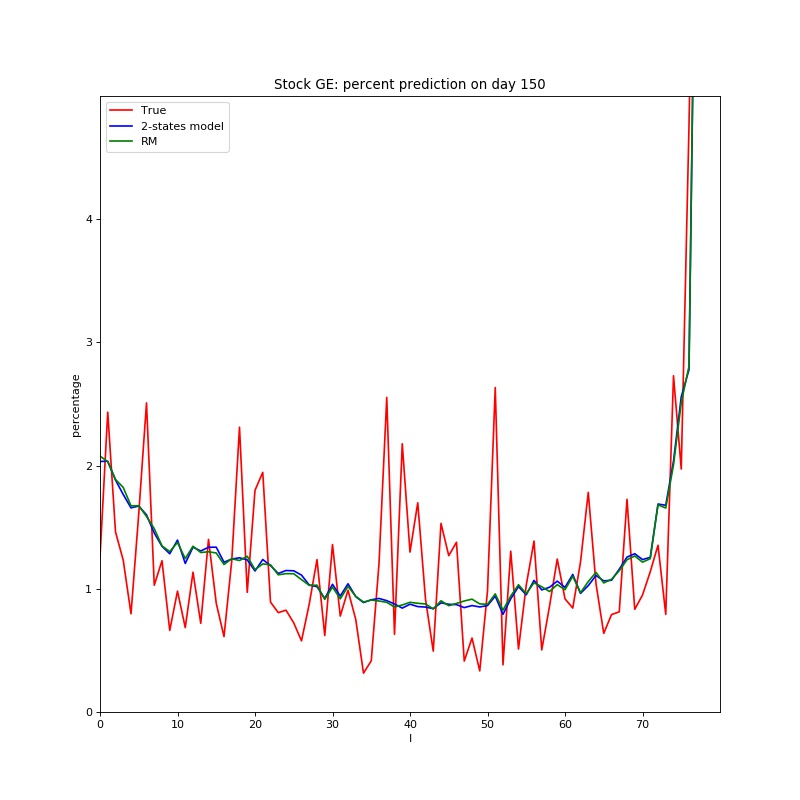}
  \caption{}
  \label{fig:com-GE1}
\end{subfigure}\hfill
\begin{subfigure}{0.25\textwidth}
  \includegraphics[width=\linewidth]{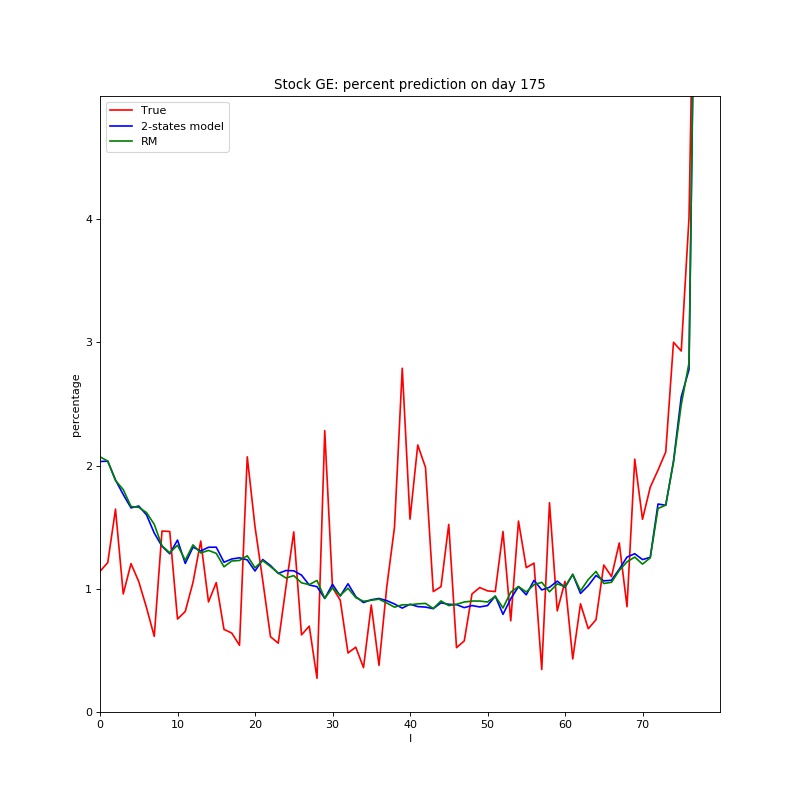}
  \caption{}
  \label{fig:com-GE2}
\end{subfigure}\hfill
\begin{subfigure}{0.25\textwidth}
  \includegraphics[width=\linewidth]{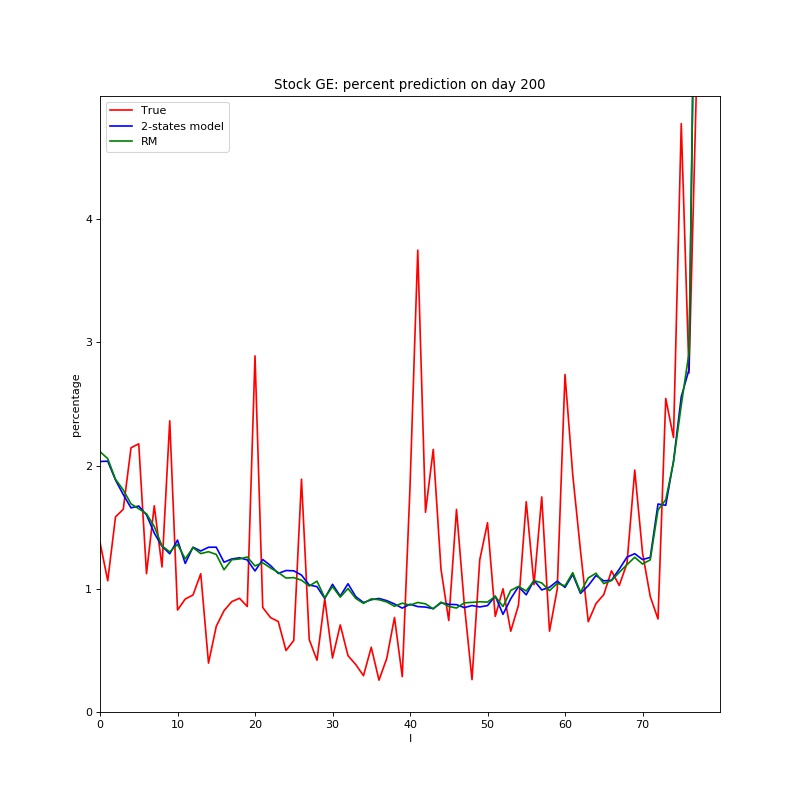}
  \caption{}
  \label{fig:com-GE3}
\end{subfigure}\hfill
\begin{subfigure}{0.25\textwidth}
  \includegraphics[width=\linewidth]{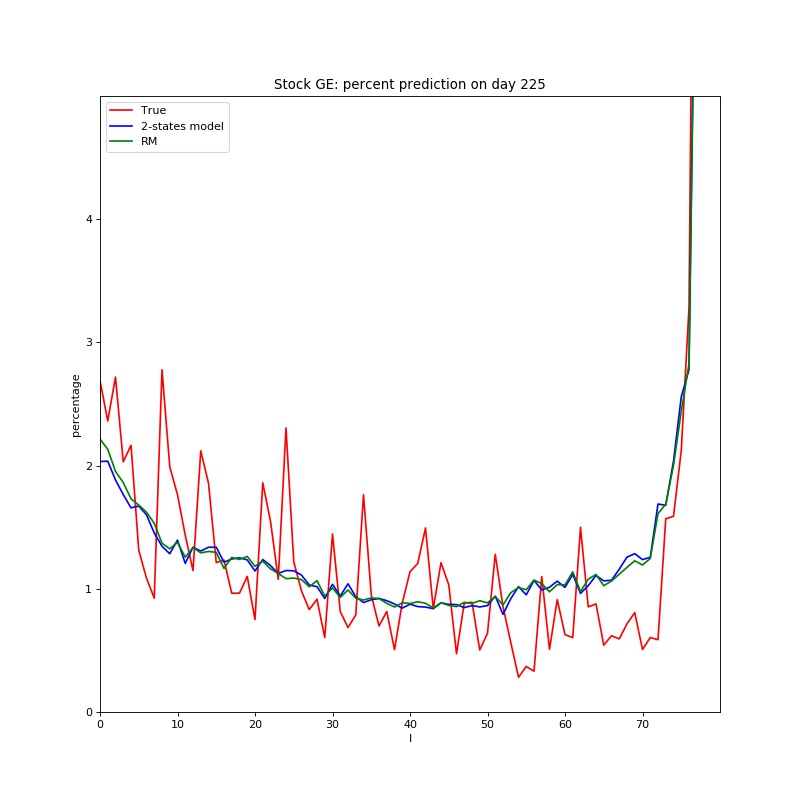}
  \caption{}
  \label{fig:com-GE4}
\end{subfigure}\\
\begin{subfigure}{0.25\textwidth}
  \includegraphics[width=\linewidth]{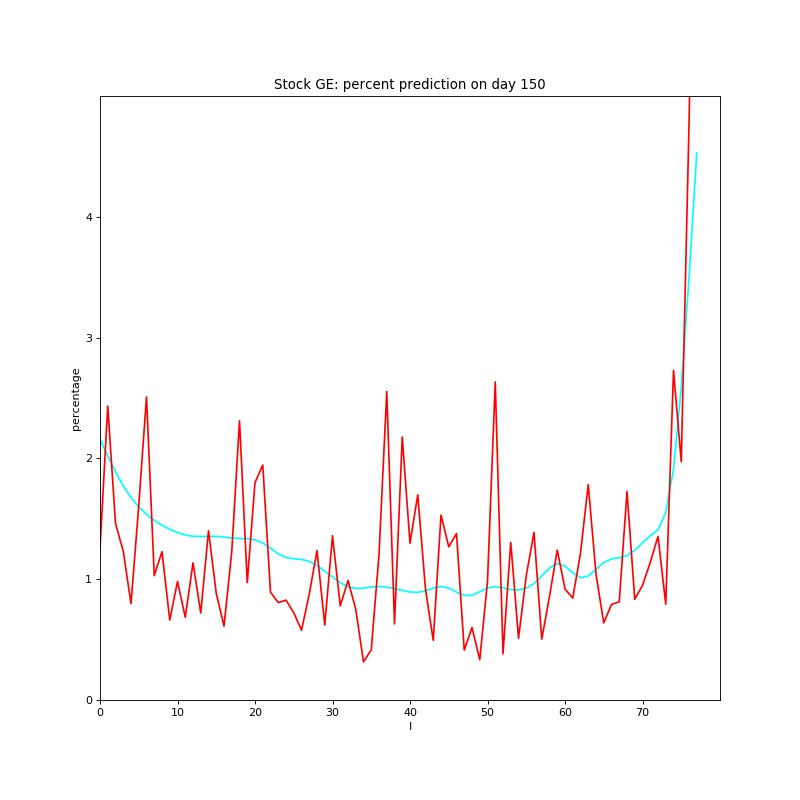}  
  \caption{}
  \label{fig:com-GE5}
\end{subfigure}\hfill
\begin{subfigure}{0.25\textwidth}
  \includegraphics[width=\linewidth]{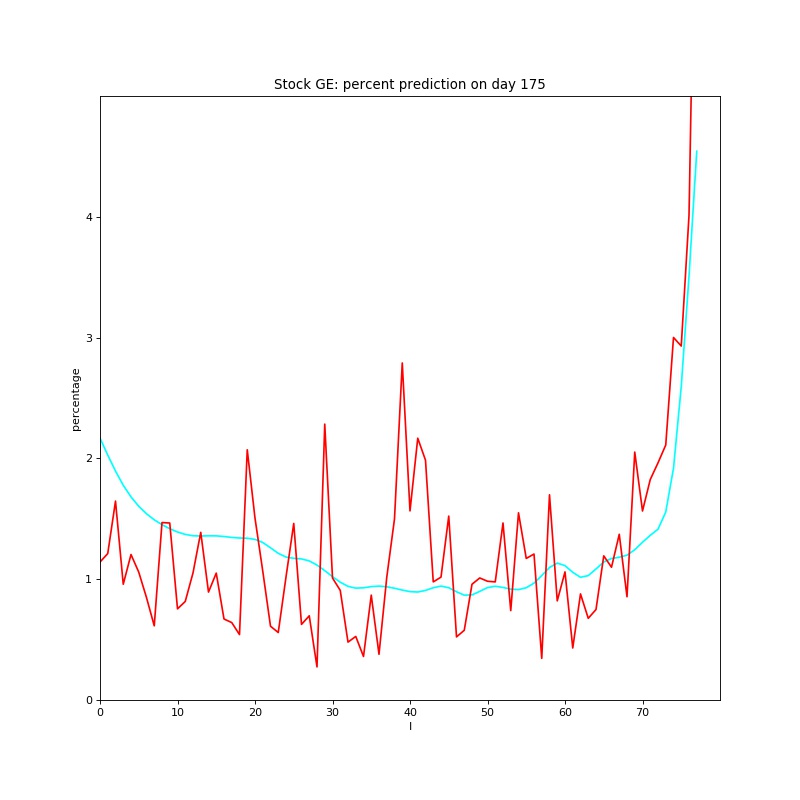}
  \caption{}
  \label{fig:com-GE6}
\end{subfigure}\hfill
\begin{subfigure}{0.25\textwidth}
  \includegraphics[width=\linewidth]{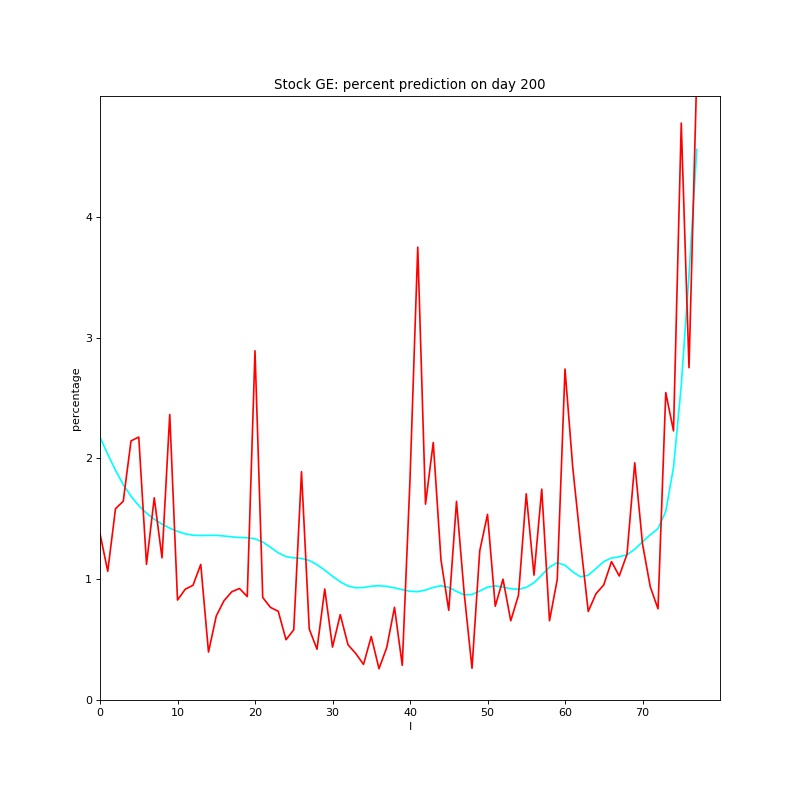}
  \caption{}
  \label{fig:com-GE7}
\end{subfigure}\hfill
\begin{subfigure}{0.25\textwidth}
  \includegraphics[width=\linewidth]{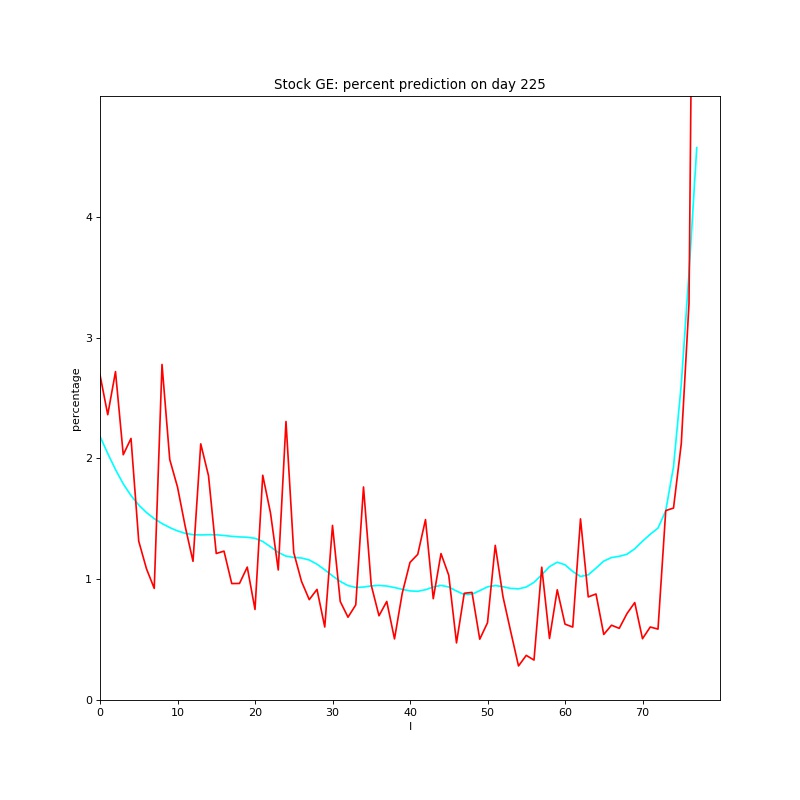}
  \caption{}
  \label{fig:com-GE8}
\end{subfigure}
\caption{comparison of prediction: (a) to (d):baseline models on stock "GE", (e) to (h):our v-state model on stock "GE"}
\label{fig:pre-appen4}
\end{figure}

\begin{figure}[!ht]
\begin{subfigure}{0.25\textwidth}
  \includegraphics[width=\linewidth]{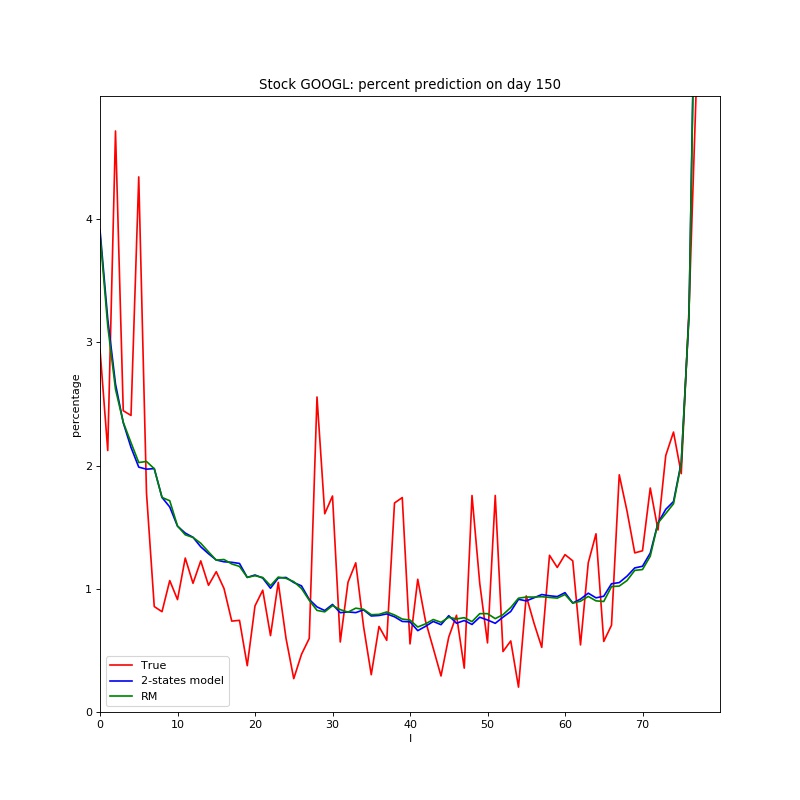}
  \caption{}
  \label{fig:com-GOOGL1}
\end{subfigure}\hfill
\begin{subfigure}{0.25\textwidth}
  \includegraphics[width=\linewidth]{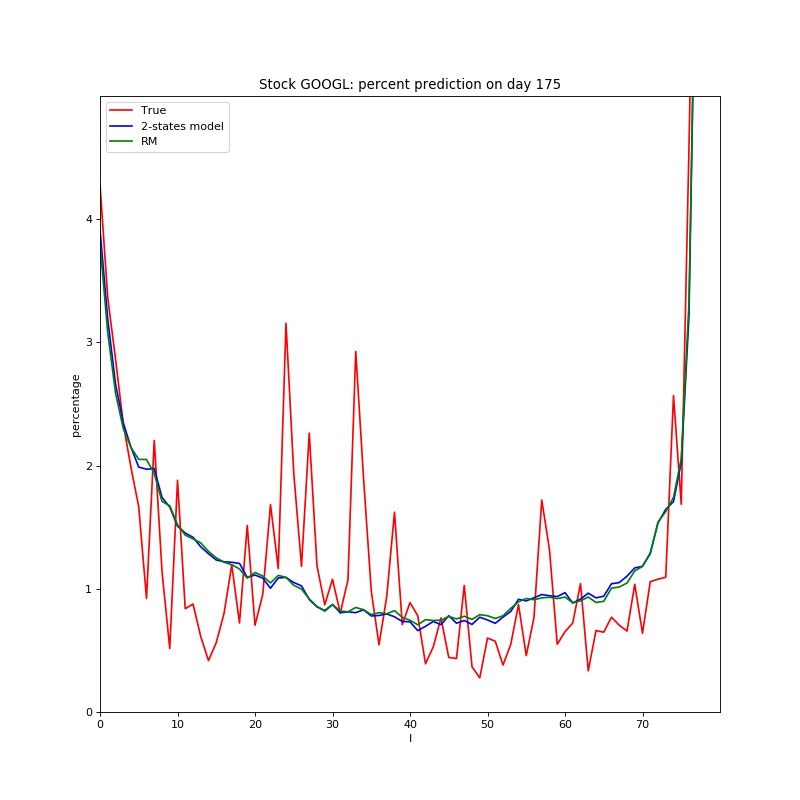}
  \caption{}
  \label{fig:com-GOOGL2}
\end{subfigure}\hfill
\begin{subfigure}{0.25\textwidth}
  \includegraphics[width=\linewidth]{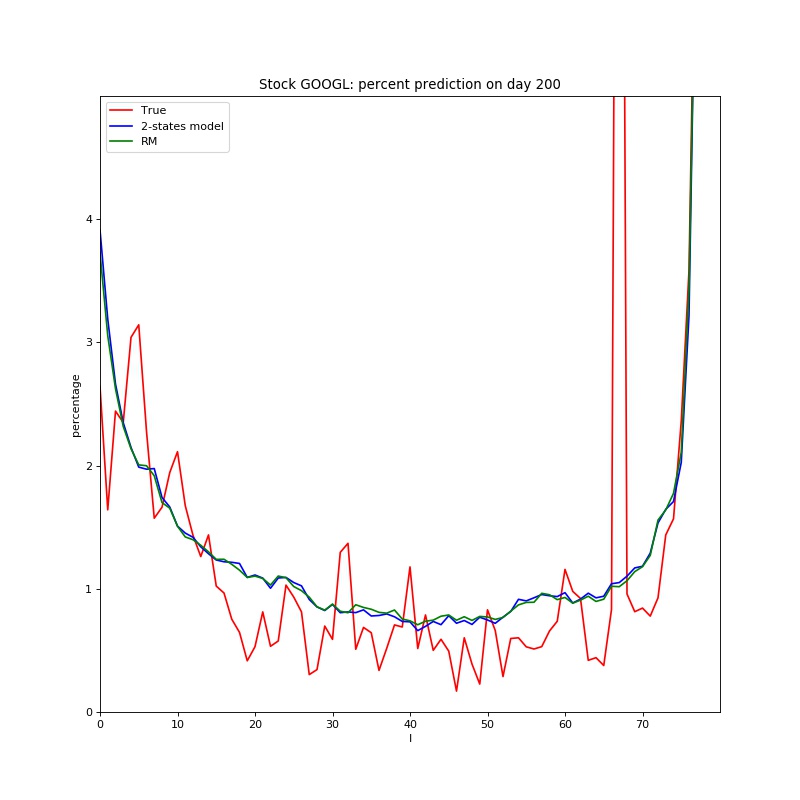}
  \caption{}
  \label{fig:com-GOOGL3}
\end{subfigure}\hfill
\begin{subfigure}{0.25\textwidth}
  \includegraphics[width=\linewidth]{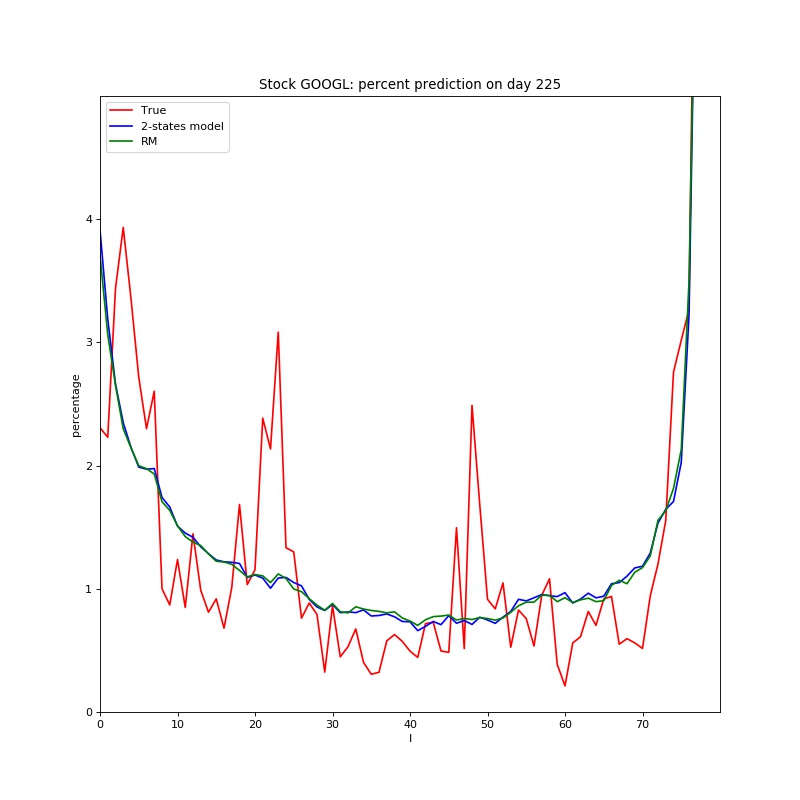}
  \caption{}
  \label{fig:com-GOOGL4}
\end{subfigure}\\
\begin{subfigure}{0.25\textwidth}
  \includegraphics[width=\linewidth]{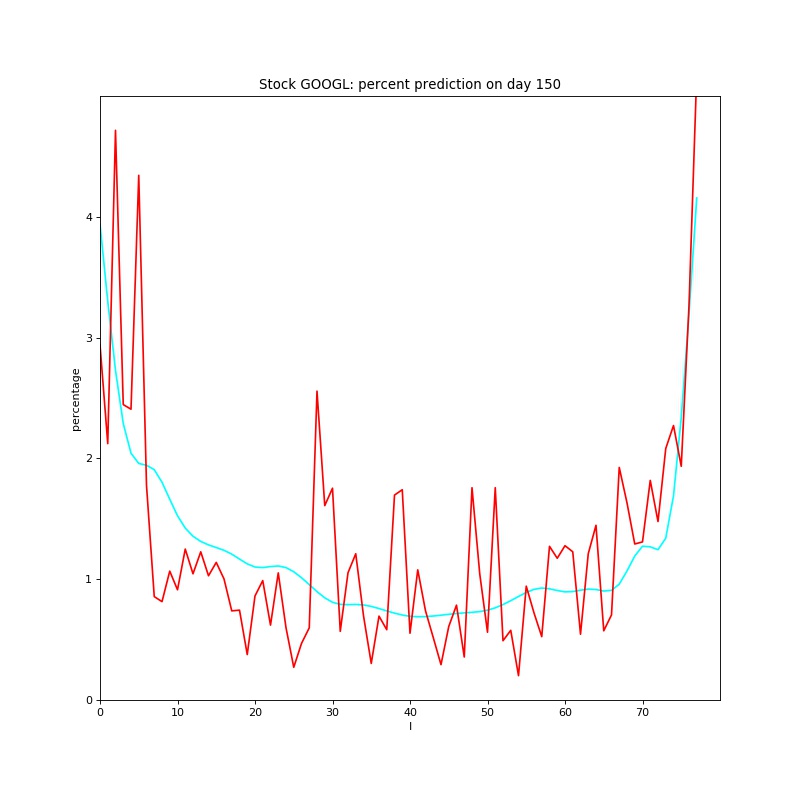}  
  \caption{}
  \label{fig:com-GOOGL5}
\end{subfigure}\hfill
\begin{subfigure}{0.25\textwidth}
  \includegraphics[width=\linewidth]{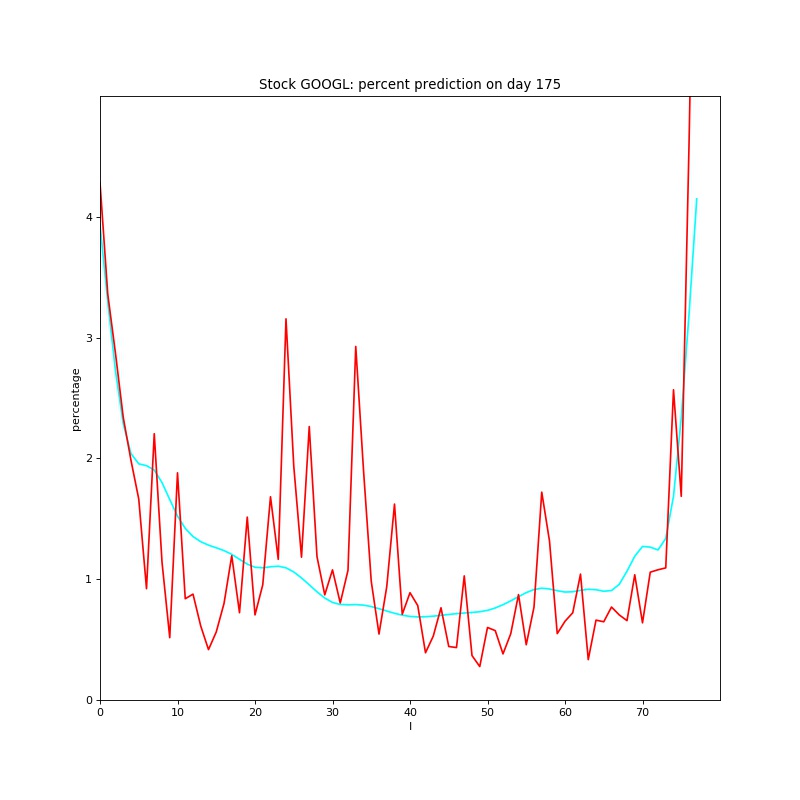}
  \caption{}
  \label{fig:com-GOOGL6}
\end{subfigure}\hfill
\begin{subfigure}{0.25\textwidth}
  \includegraphics[width=\linewidth]{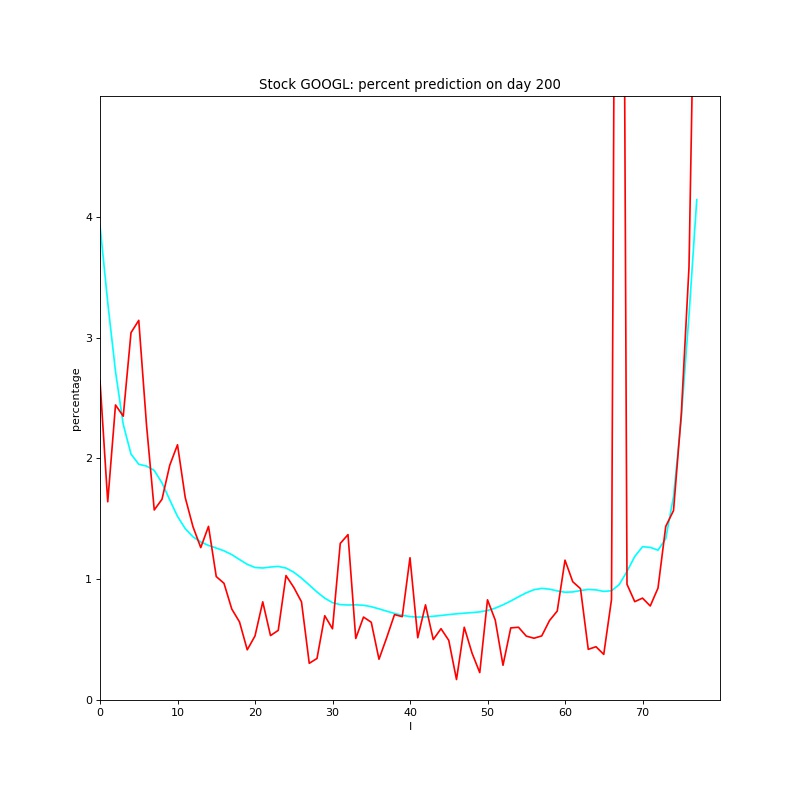}
  \caption{}
  \label{fig:com-GOOGL7}
\end{subfigure}\hfill
\begin{subfigure}{0.25\textwidth}
  \includegraphics[width=\linewidth]{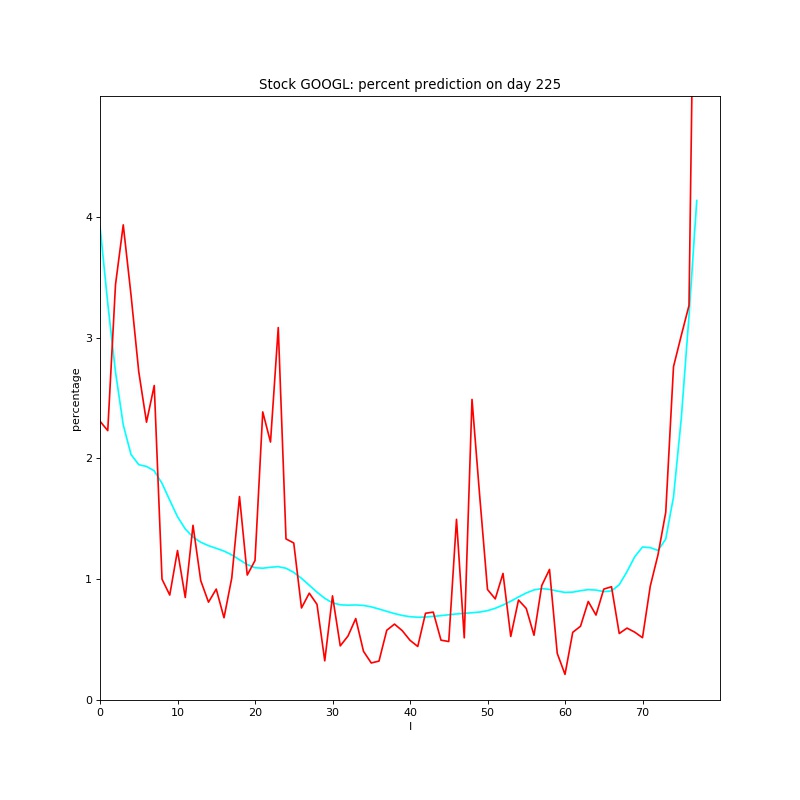}
  \caption{}
  \label{fig:com-GOOGL8}
\end{subfigure}
\caption{comparison of prediction: (a) to (d):baseline models on stock "GOOGL", (e) to (h):our v-state model on stock "GOOGL"}
\label{fig:pre-appen5}
\end{figure}

\begin{figure}[!ht]
\begin{subfigure}{0.25\textwidth}
  \includegraphics[width=\linewidth]{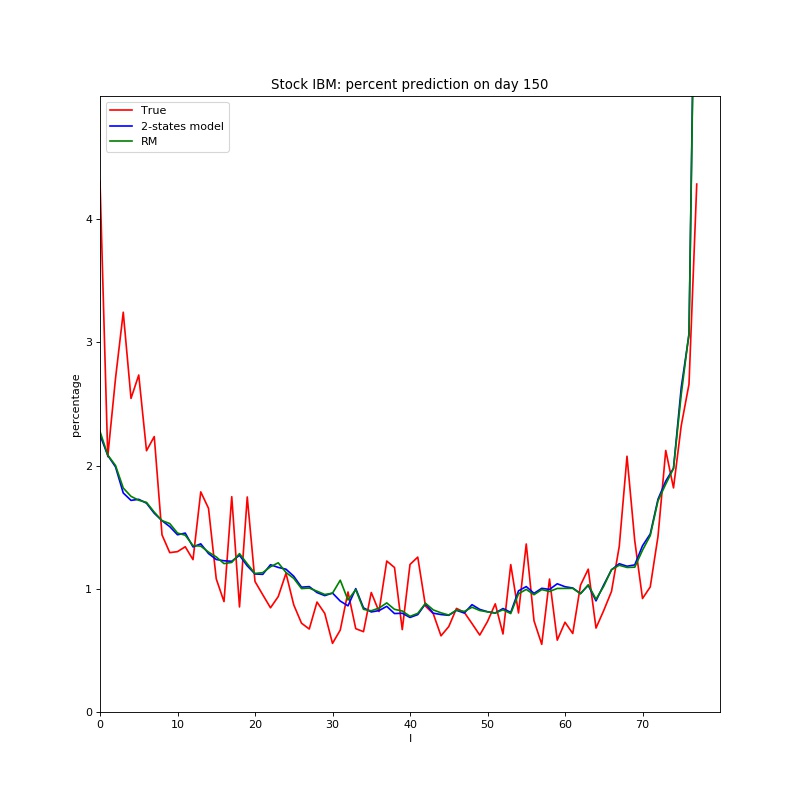}
  \caption{}
  \label{fig:com-IBM1}
\end{subfigure}\hfill
\begin{subfigure}{0.25\textwidth}
  \includegraphics[width=\linewidth]{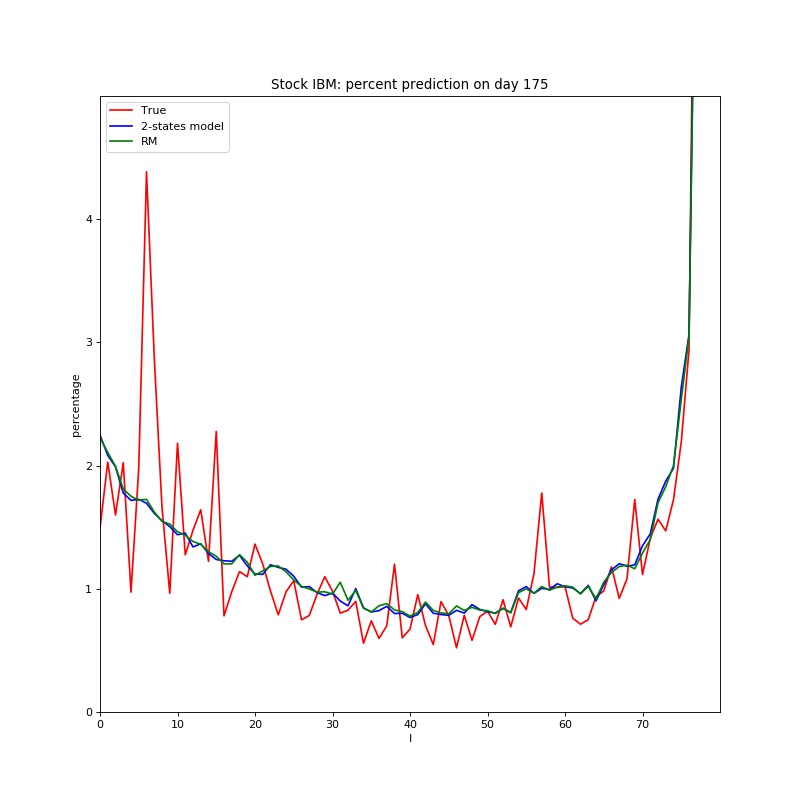}
  \caption{}
  \label{fig:com-IBM2}
\end{subfigure}\hfill
\begin{subfigure}{0.25\textwidth}
  \includegraphics[width=\linewidth]{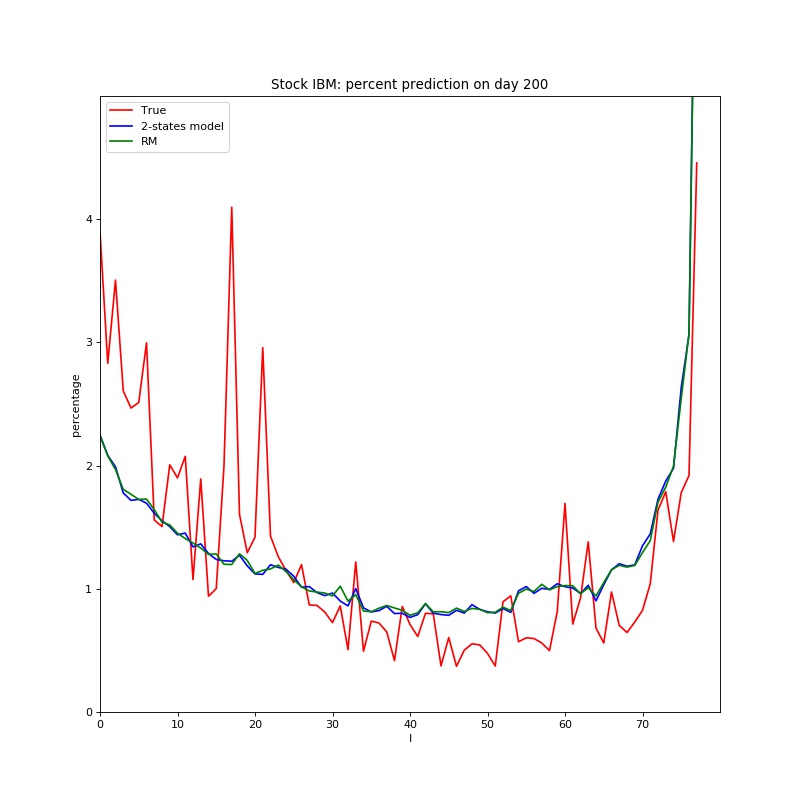}
  \caption{}
  \label{fig:com-IBM3}
\end{subfigure}\hfill
\begin{subfigure}{0.25\textwidth}
  \includegraphics[width=\linewidth]{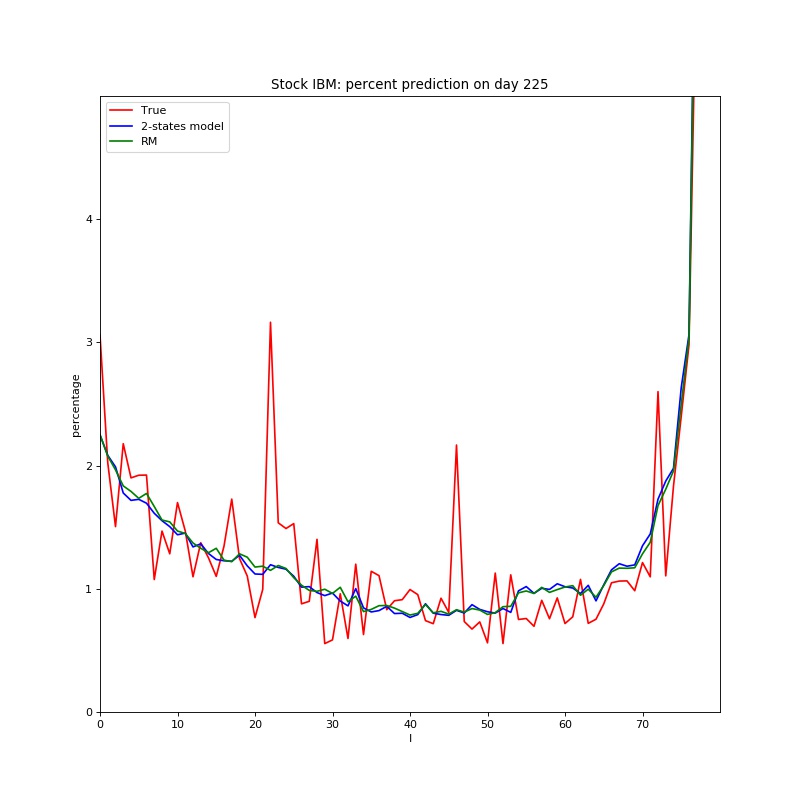}
  \caption{}
  \label{fig:com-IBM4}
\end{subfigure}\\
\begin{subfigure}{0.25\textwidth}
  \includegraphics[width=\linewidth]{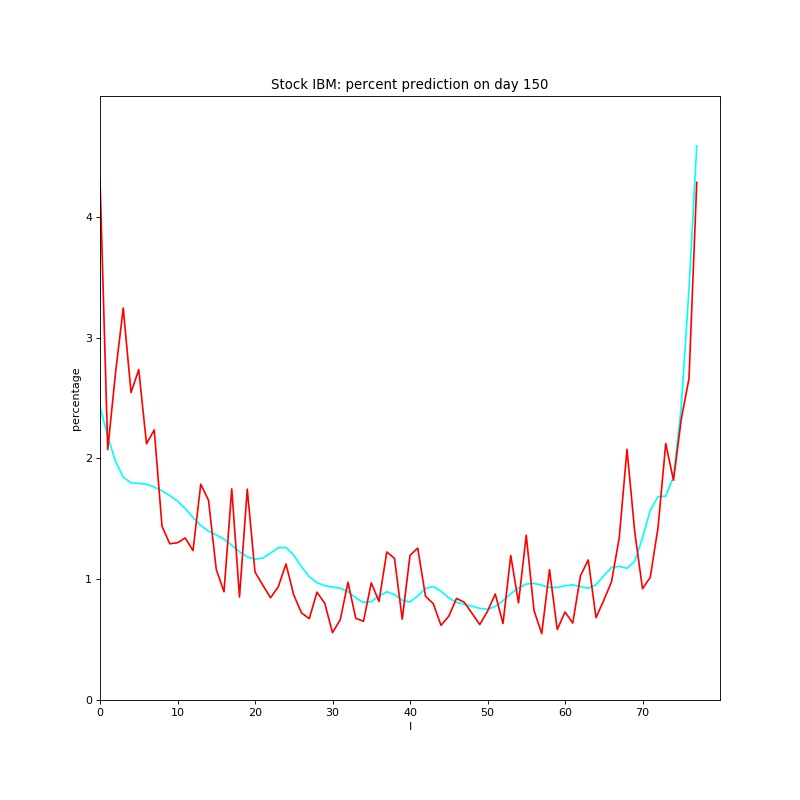}  
  \caption{}
  \label{fig:com-IBM5}
\end{subfigure}\hfill
\begin{subfigure}{0.25\textwidth}
  \includegraphics[width=\linewidth]{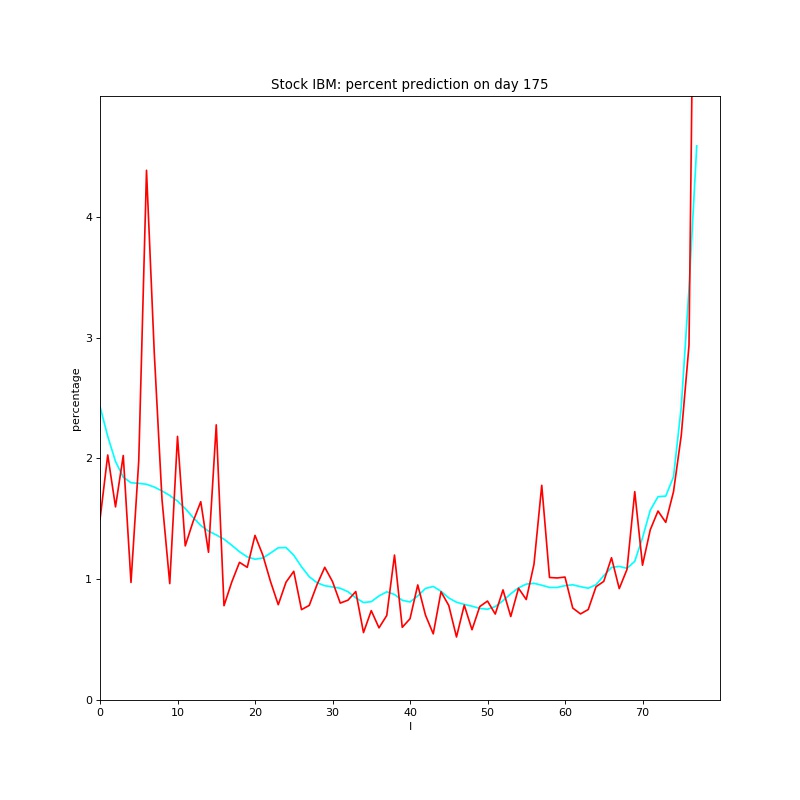}
  \caption{}
  \label{fig:com-IBM6}
\end{subfigure}\hfill
\begin{subfigure}{0.25\textwidth}
  \includegraphics[width=\linewidth]{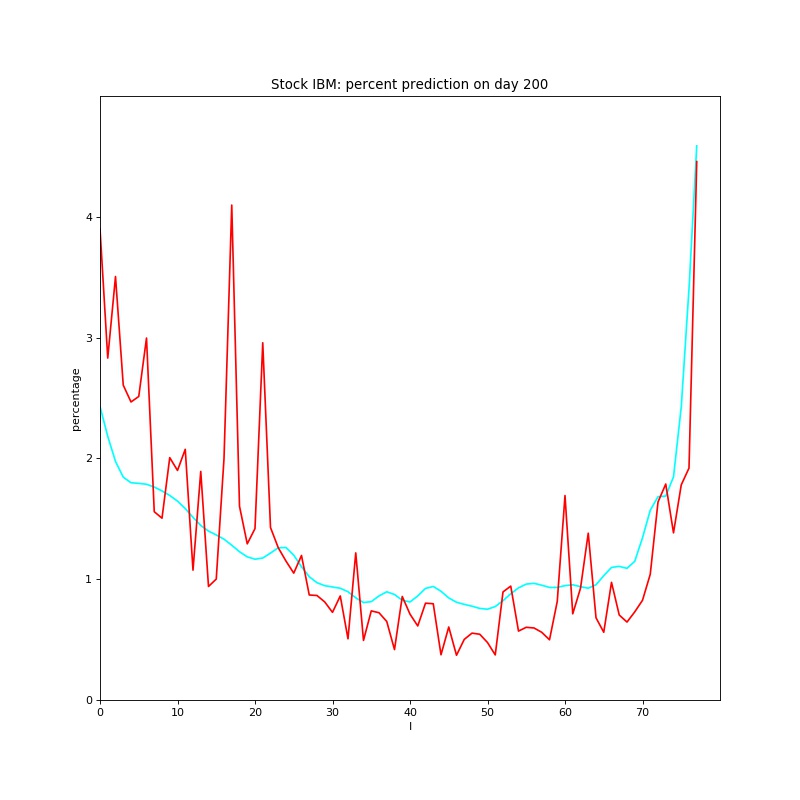}
  \caption{}
  \label{fig:com-IBM7}
\end{subfigure}\hfill
\begin{subfigure}{0.25\textwidth}
  \includegraphics[width=\linewidth]{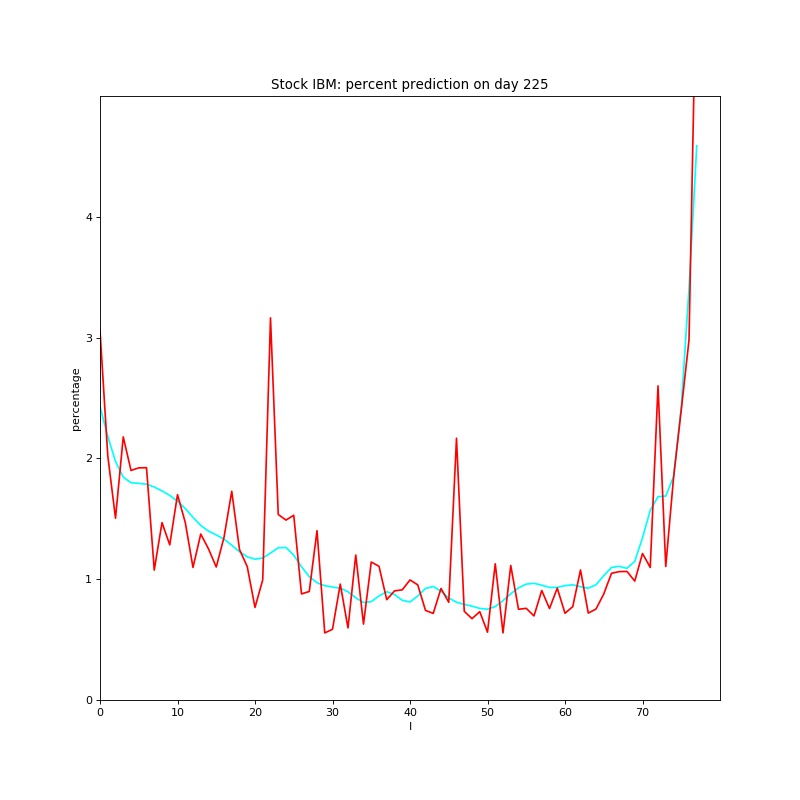}
  \caption{}
  \label{fig:com-IBM8}
\end{subfigure}
\caption{comparison of prediction: (a) to (d):baseline models on stock "IBM", (e) to (h):our v-state model on stock "IBM"}
\label{fig:pre-appen6}
\end{figure}

\begin{figure}[!ht]
\begin{subfigure}{0.25\textwidth}
  \includegraphics[width=\linewidth]{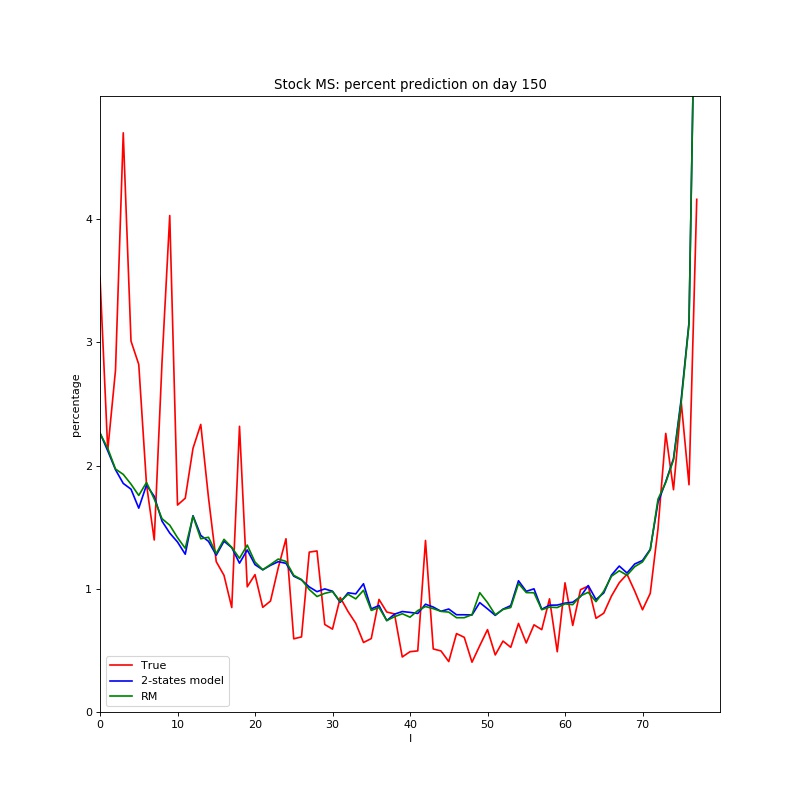}
  \caption{}
  \label{fig:com-MS1}
\end{subfigure}\hfill
\begin{subfigure}{0.25\textwidth}
  \includegraphics[width=\linewidth]{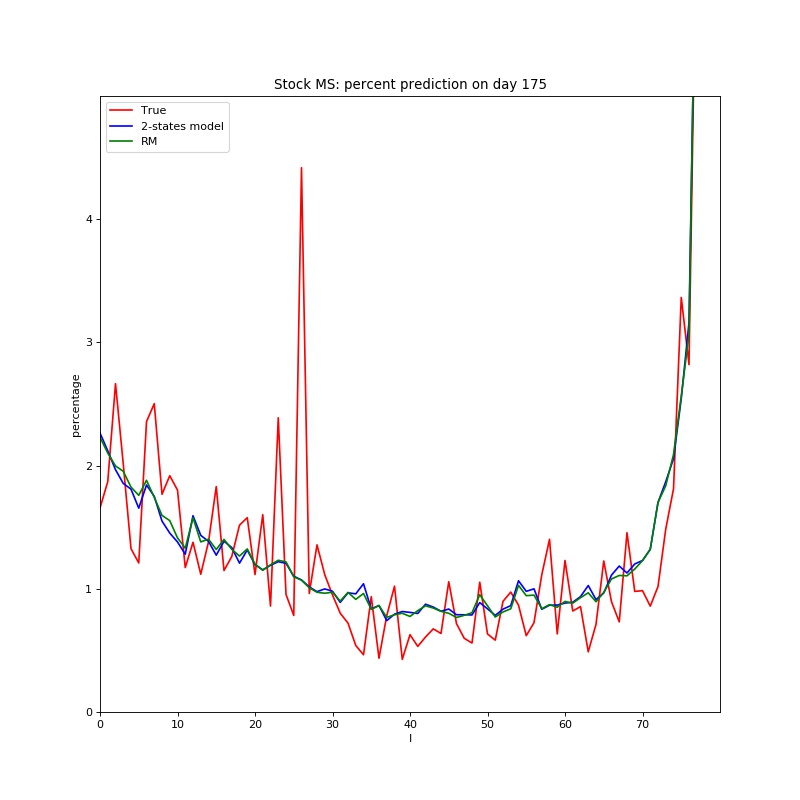}
  \caption{}
  \label{fig:com-MS2}
\end{subfigure}\hfill
\begin{subfigure}{0.25\textwidth}
  \includegraphics[width=\linewidth]{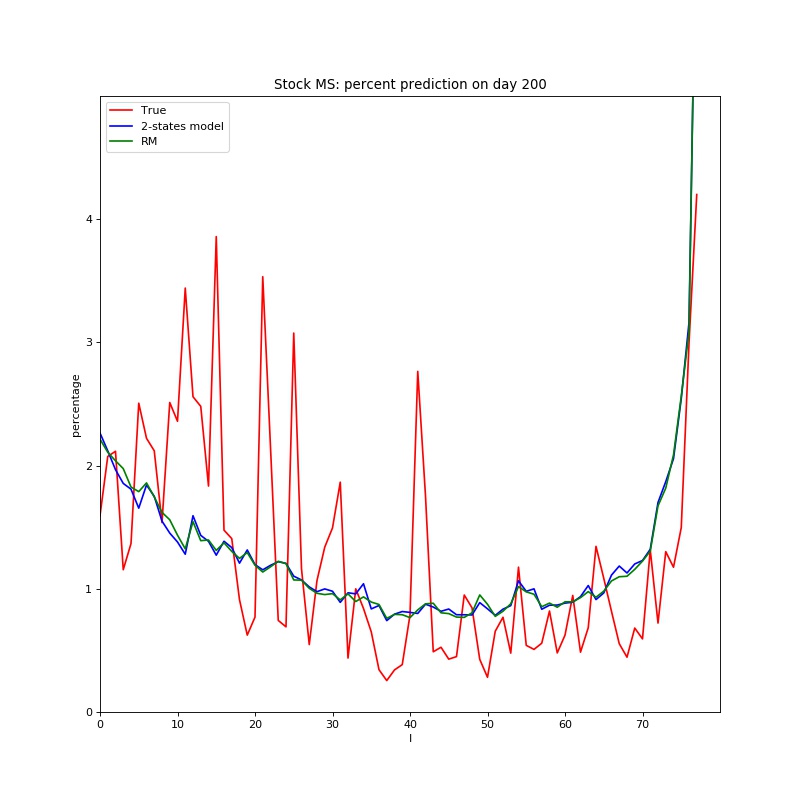}
  \caption{}
  \label{fig:com-MS3}
\end{subfigure}\hfill
\begin{subfigure}{0.25\textwidth}
  \includegraphics[width=\linewidth]{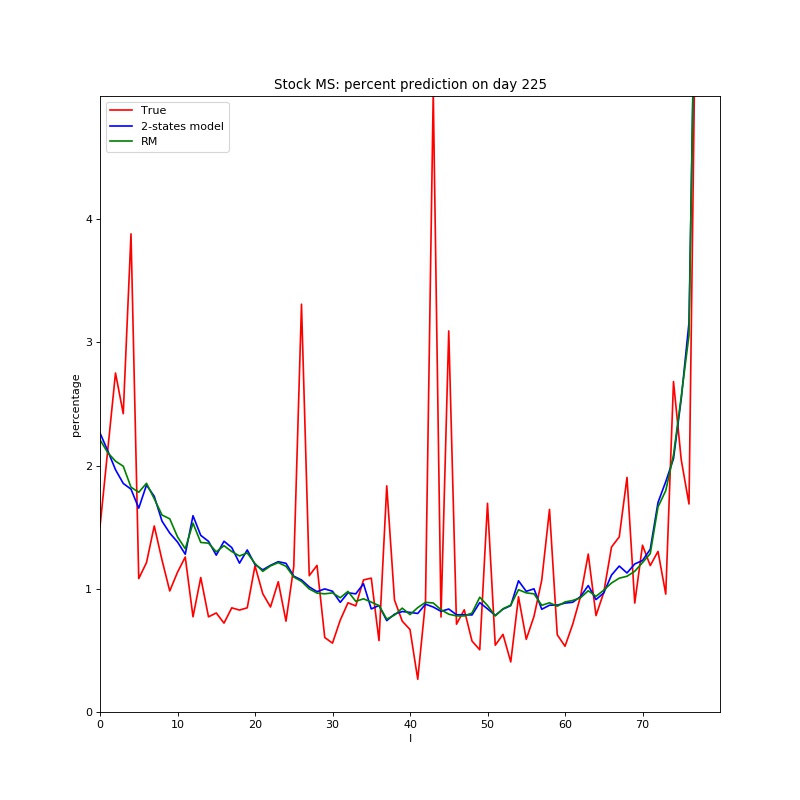}
  \caption{}
  \label{fig:com-MS4}
\end{subfigure}\\
\begin{subfigure}{0.25\textwidth}
  \includegraphics[width=\linewidth]{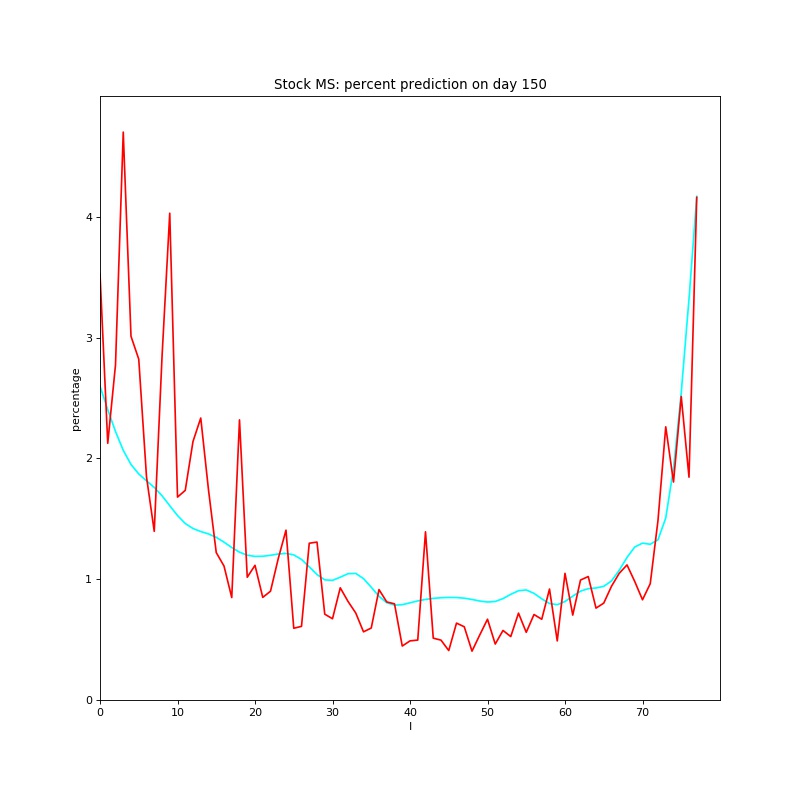}  
  \caption{}
  \label{fig:com-MS5}
\end{subfigure}\hfill
\begin{subfigure}{0.25\textwidth}
  \includegraphics[width=\linewidth]{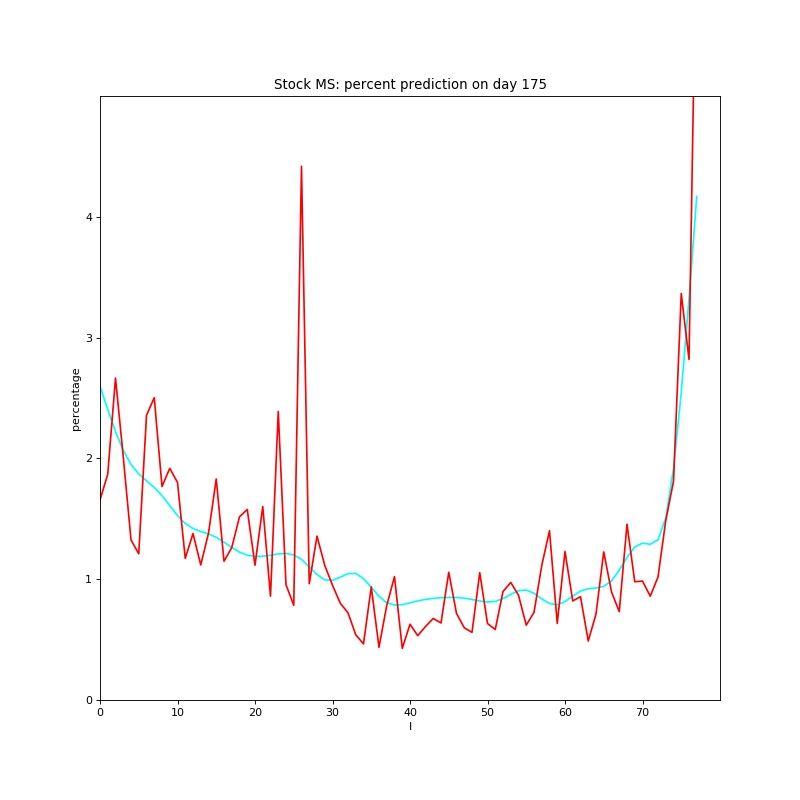}
  \caption{}
  \label{fig:com-MS6}
\end{subfigure}\hfill
\begin{subfigure}{0.25\textwidth}
  \includegraphics[width=\linewidth]{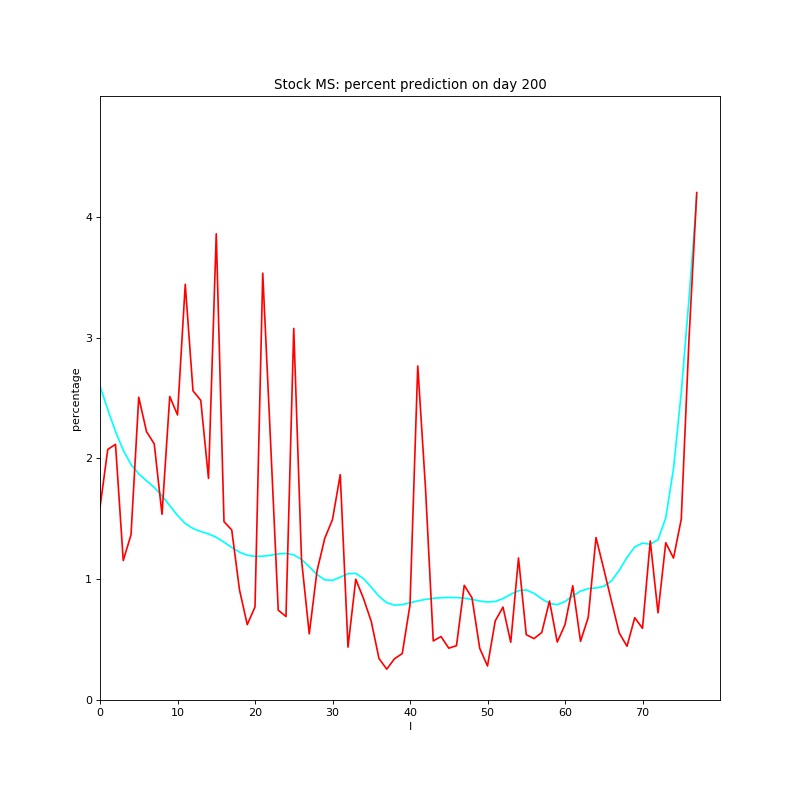}
  \caption{}
  \label{fig:com-MS7}
\end{subfigure}\hfill
\begin{subfigure}{0.25\textwidth}
  \includegraphics[width=\linewidth]{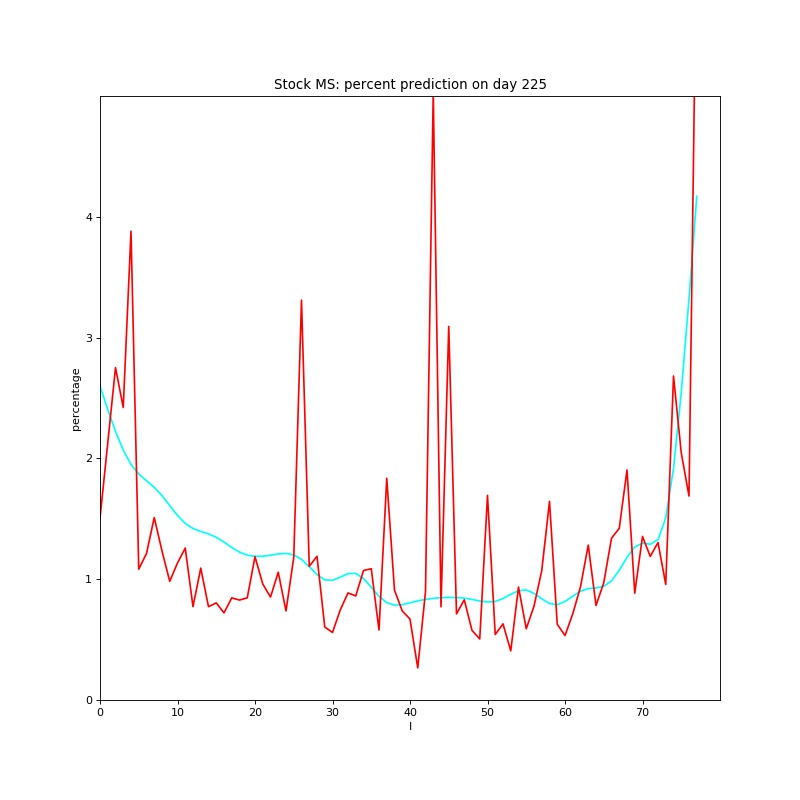}
  \caption{}
  \label{fig:com-MS8}
\end{subfigure}
\caption{comparison of prediction: (a) to (d):baseline models on stock "MS", (e) to (h):our v-state model on stock "MS"}
\label{fig:pre-appen7}
\end{figure}

\begin{figure}[!ht]
\begin{subfigure}{0.25\textwidth}
  \includegraphics[width=\linewidth]{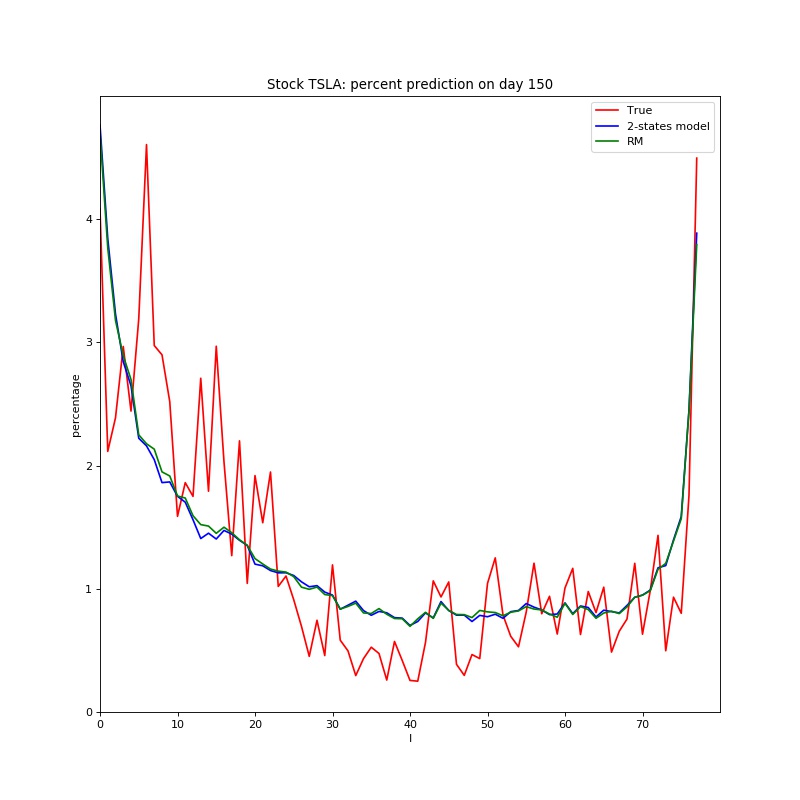}
  \caption{}
  \label{fig:com-TSLA1}
\end{subfigure}\hfill
\begin{subfigure}{0.25\textwidth}
  \includegraphics[width=\linewidth]{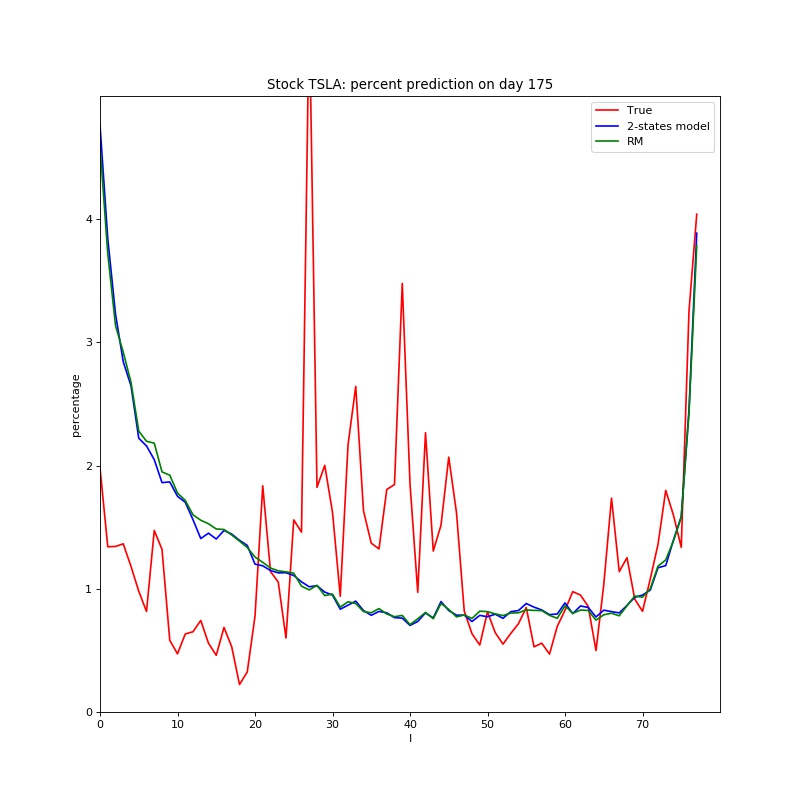}
  \caption{}
  \label{fig:com-TSLA2}
\end{subfigure}\hfill
\begin{subfigure}{0.25\textwidth}
  \includegraphics[width=\linewidth]{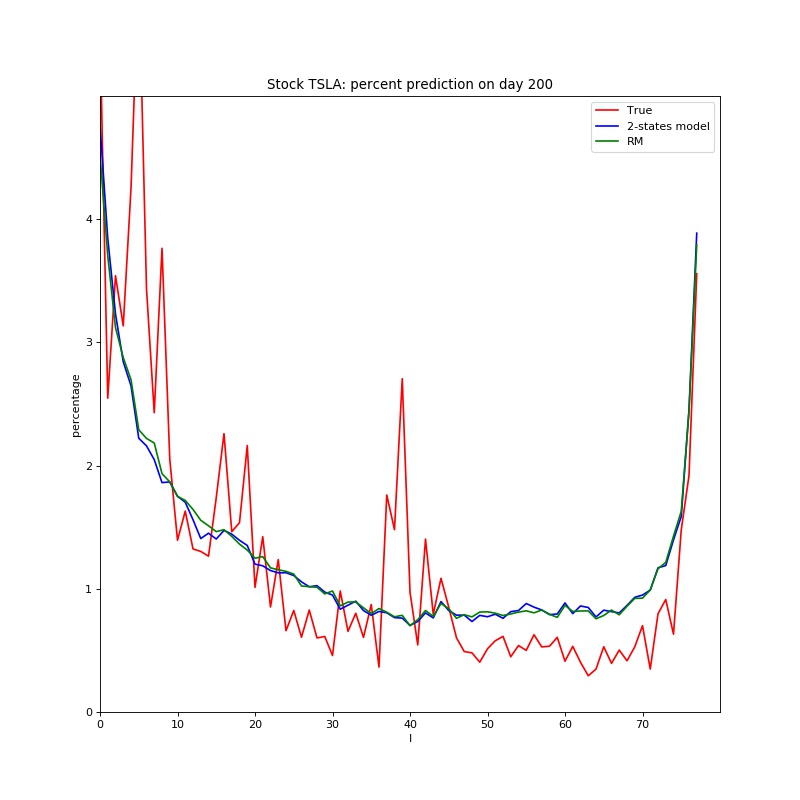}
  \caption{}
  \label{fig:com-TSLA3}
\end{subfigure}\hfill
\begin{subfigure}{0.25\textwidth}
  \includegraphics[width=\linewidth]{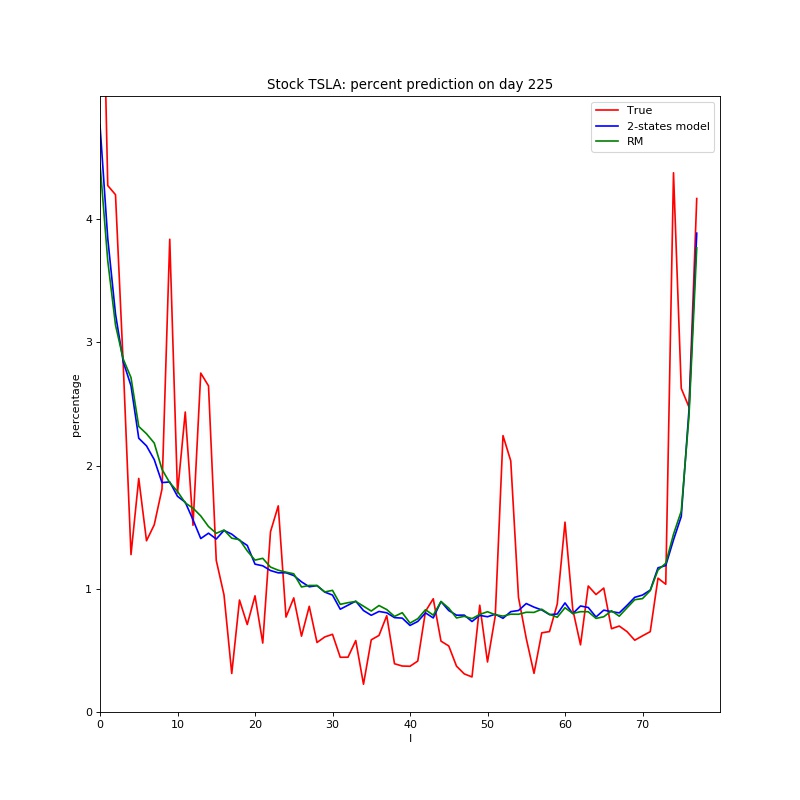}
  \caption{}
  \label{fig:com-TSLA4}
\end{subfigure}\\
\begin{subfigure}{0.25\textwidth}
  \includegraphics[width=\linewidth]{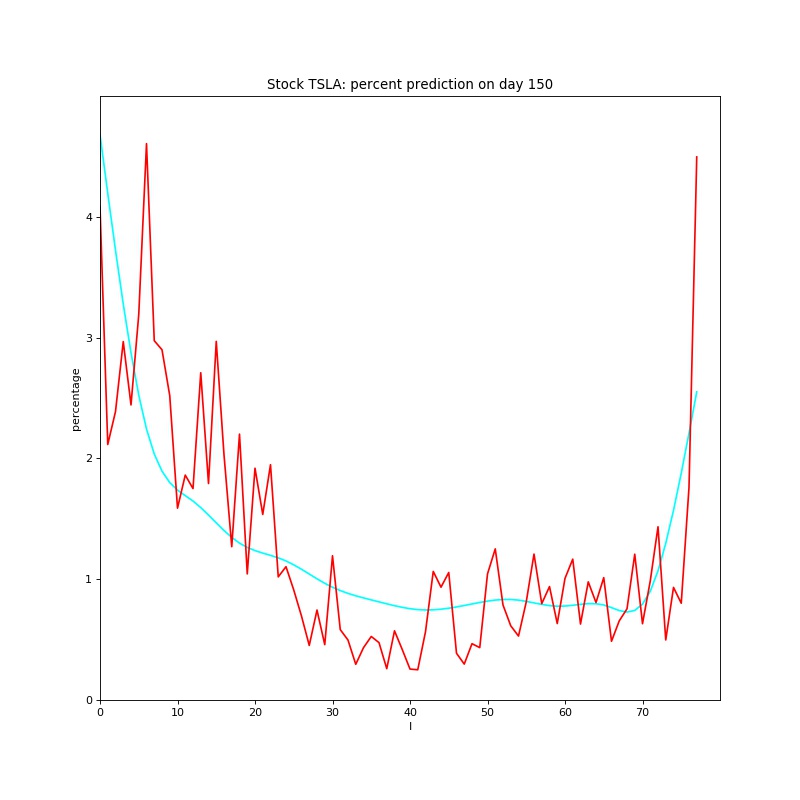}  
  \caption{}
  \label{fig:com-TSLA5}
\end{subfigure}\hfill
\begin{subfigure}{0.25\textwidth}
  \includegraphics[width=\linewidth]{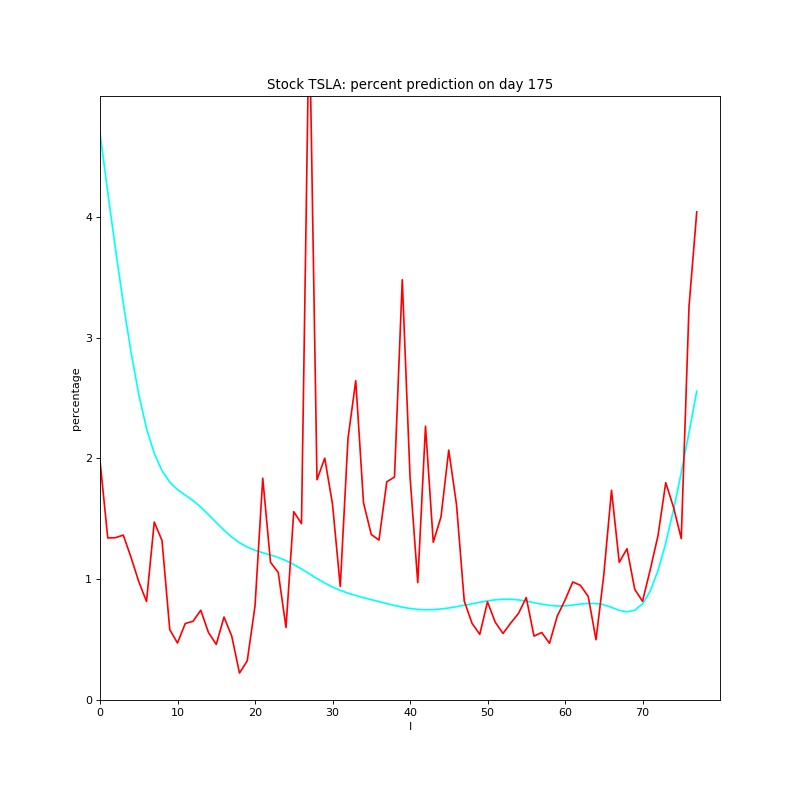}
  \caption{}
  \label{fig:com-TSLA6}
\end{subfigure}\hfill
\begin{subfigure}{0.25\textwidth}
  \includegraphics[width=\linewidth]{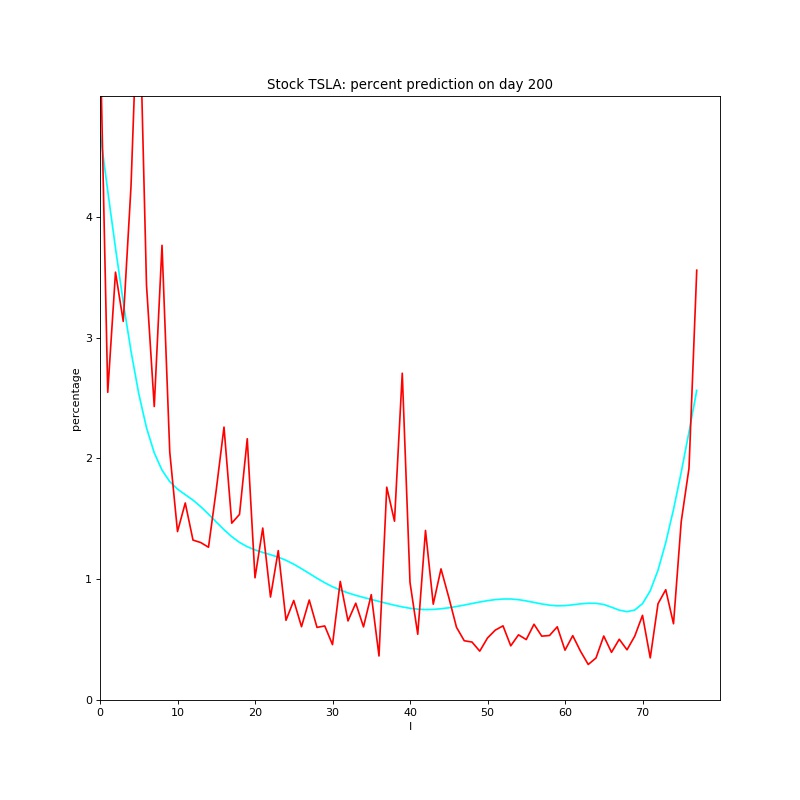}
  \caption{}
  \label{fig:com-TSLA7}
\end{subfigure}\hfill
\begin{subfigure}{0.25\textwidth}
  \includegraphics[width=\linewidth]{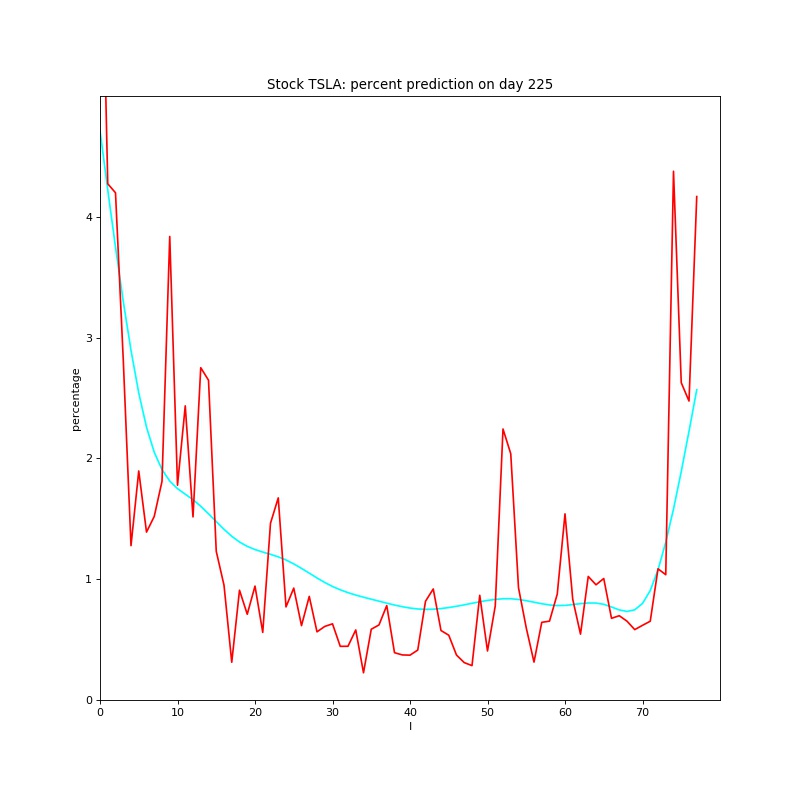}
  \caption{}
  \label{fig:com-TSLA8}
\end{subfigure}
\caption{comparison of prediction: (a) to (d):baseline models on stock "TSLA", (e) to (h):our v-state model on stock "TSLA"}
\label{fig:pre-appen8}
\end{figure}

\begin{figure}[!ht]
\begin{subfigure}{0.23\textwidth}
  \includegraphics[width=\linewidth]{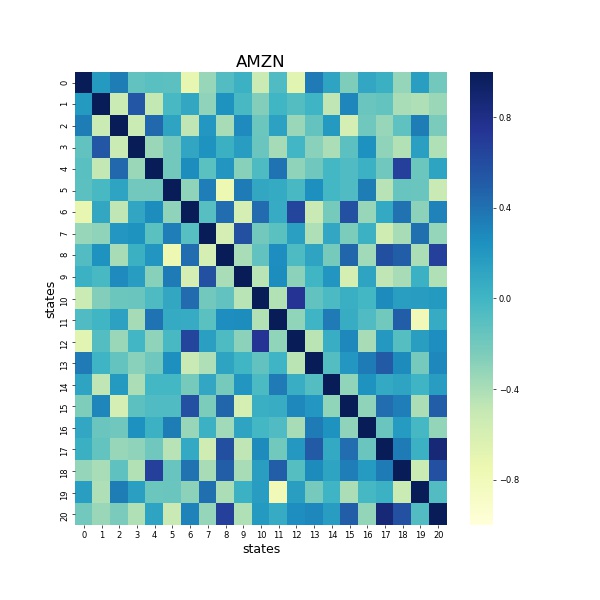}
  \caption{AMZN}
  \label{fig:corr-AMZN}
\end{subfigure} \hfill
\begin{subfigure}{0.23\textwidth}
  \includegraphics[width=\linewidth]{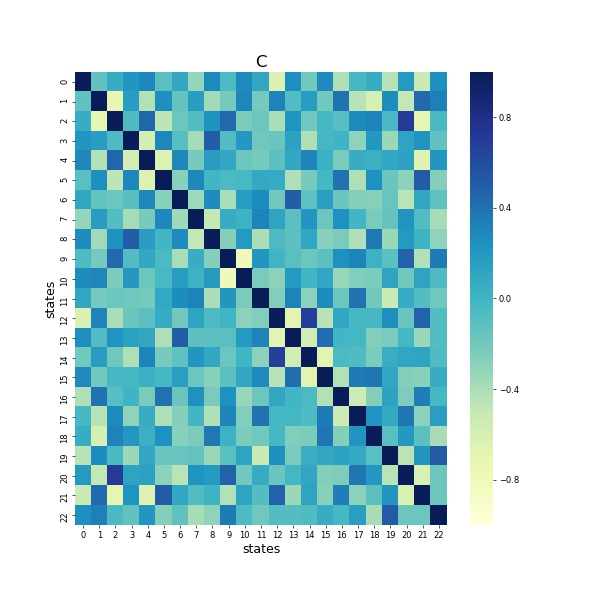}
  \caption{C}
  \label{fig:corr-C}
\end{subfigure} \hfill
\begin{subfigure}{0.23\textwidth}
  \includegraphics[width=\linewidth]{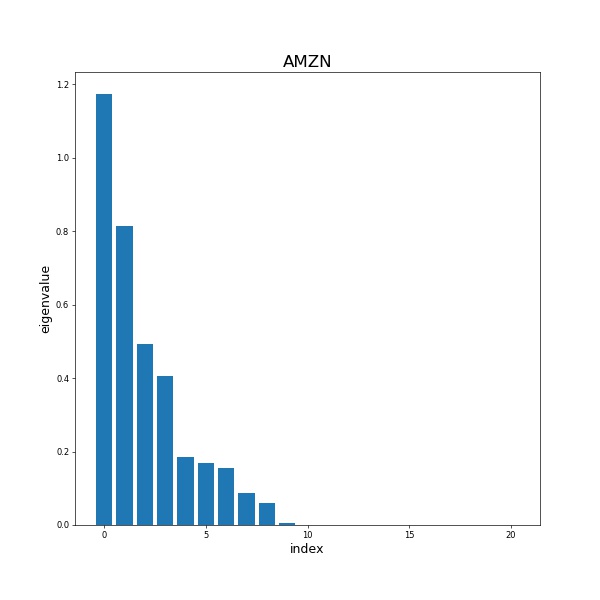}
  \caption{AMZN}
  \label{fig:eig-AMZN}
\end{subfigure} \hfill
\begin{subfigure}{0.23\textwidth}
  \includegraphics[width=\linewidth]{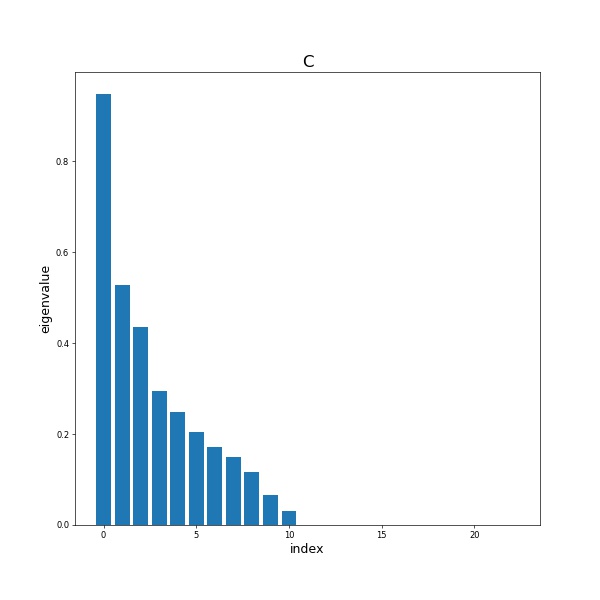}
  \caption{C}
  \label{fig:eig-C}
\end{subfigure}\\
\begin{subfigure}{0.23\textwidth}
  \includegraphics[width=\linewidth]{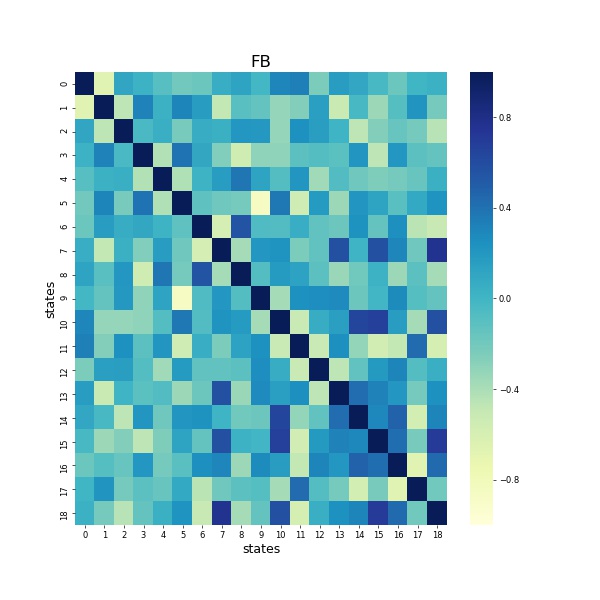}
  \caption{FB}
  \label{fig:corr-FB}
\end{subfigure} \hfill
\begin{subfigure}{0.23\textwidth}
  \includegraphics[width=\linewidth]{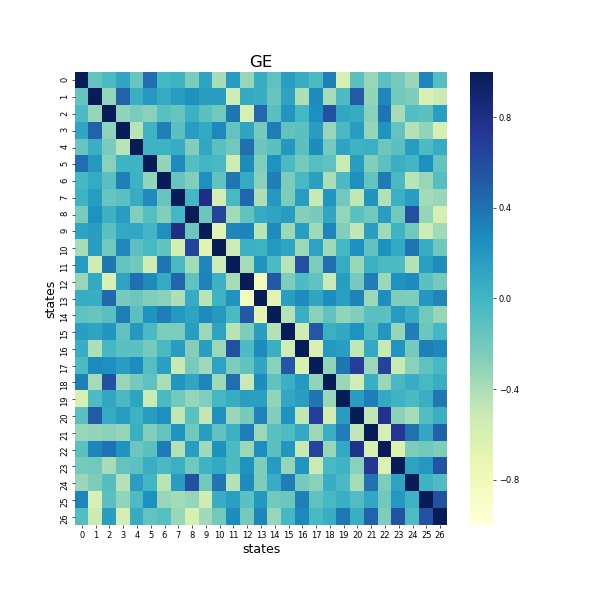}
  \caption{GE}
  \label{fig:corr-GE}
\end{subfigure} \hfill
\begin{subfigure}{0.23\textwidth}
  \includegraphics[width=\linewidth]{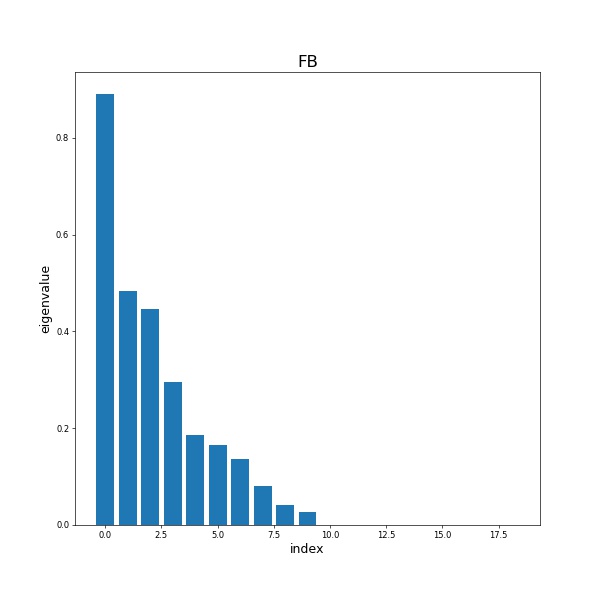}
  \caption{FB}
  \label{fig:eig-FB}
\end{subfigure} \hfill
\begin{subfigure}{0.23\textwidth}
  \includegraphics[width=\linewidth]{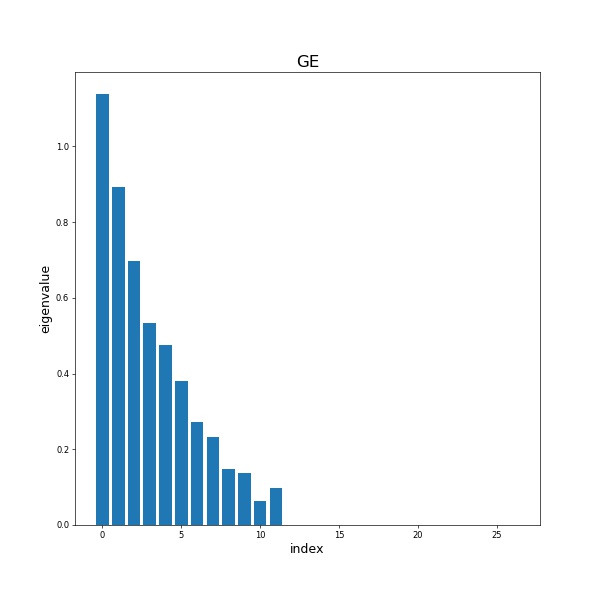}
  \caption{GE}
  \label{fig:eig-GE}
\end{subfigure}
\caption{Correlation matrix of hidden states, eigenvalues of transition covariance matrix, "AMZN", "C", "FB" and "GE"}
\label{fig:transcorr-appen1}
\end{figure}

\begin{figure}[!ht]
\begin{subfigure}{0.23\textwidth}
  \includegraphics[width=\linewidth]{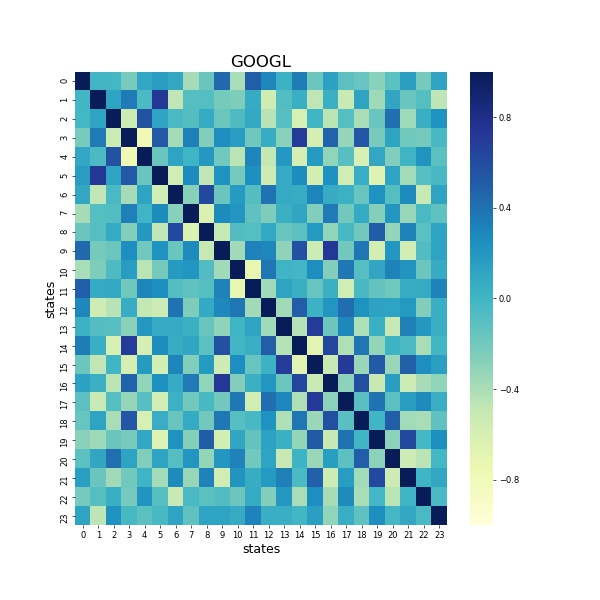}
  \caption{GOOGL}
  \label{fig:corr-GOOGL}
\end{subfigure} \hfill
\begin{subfigure}{0.23\textwidth}
  \includegraphics[width=\linewidth]{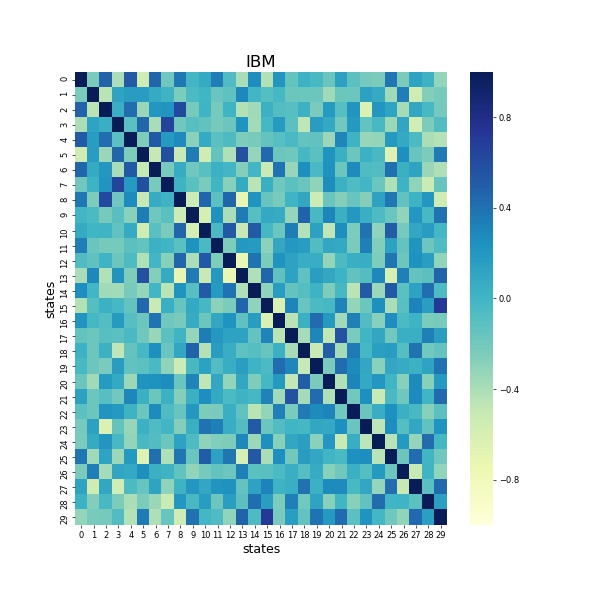}
  \caption{IBM}
  \label{fig:corr-IBM}
\end{subfigure} \hfill
\begin{subfigure}{0.23\textwidth}
  \includegraphics[width=\linewidth]{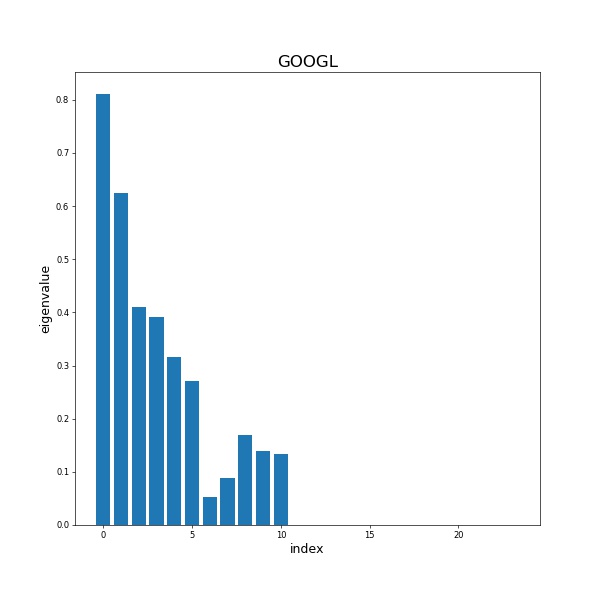}
  \caption{GOOGL}
  \label{fig:eig-GOOGL}
\end{subfigure} \hfill
\begin{subfigure}{0.23\textwidth}
  \includegraphics[width=\linewidth]{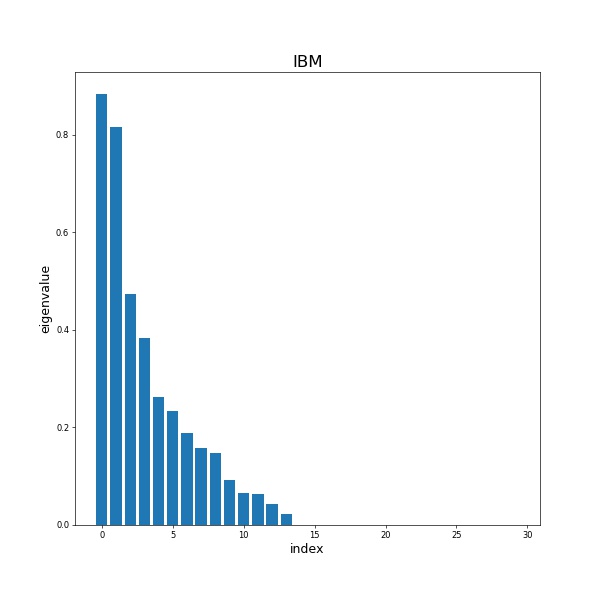}
  \caption{IBM}
  \label{fig:eig-IBM}
\end{subfigure}\\
\begin{subfigure}{0.23\textwidth}
  \includegraphics[width=\linewidth]{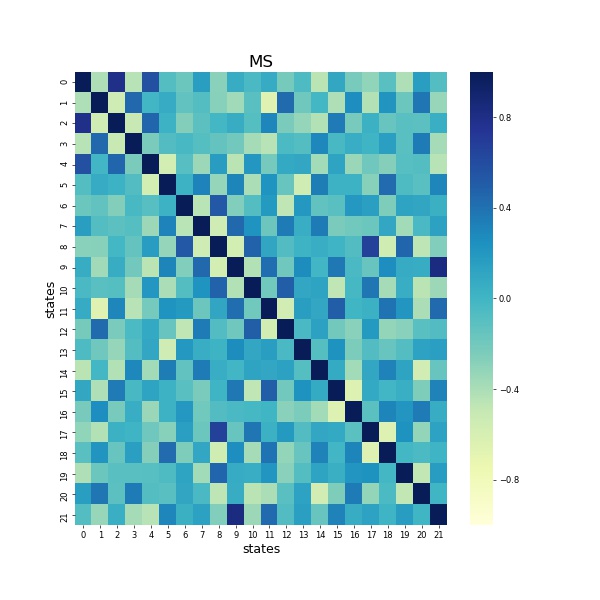}
  \caption{MS}
  \label{fig:corr-MS}
\end{subfigure} \hfill
\begin{subfigure}{0.23\textwidth}
  \includegraphics[width=\linewidth]{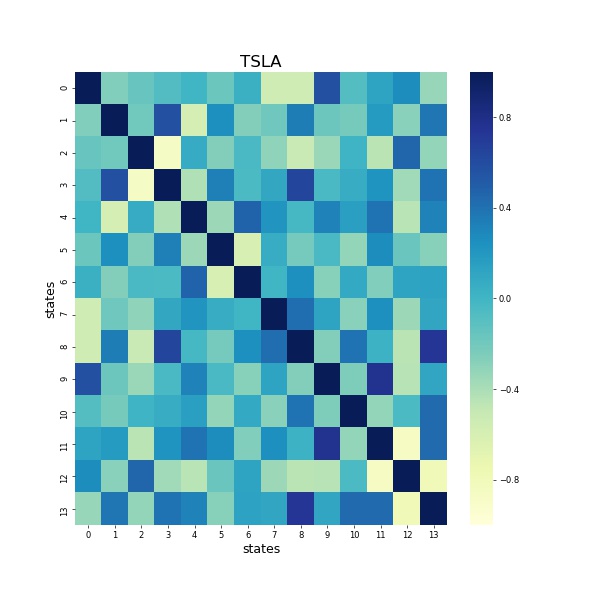}
  \caption{TSLA}
  \label{fig:corr-TSLA}
\end{subfigure} \hfill
\begin{subfigure}{0.23\textwidth}
  \includegraphics[width=\linewidth]{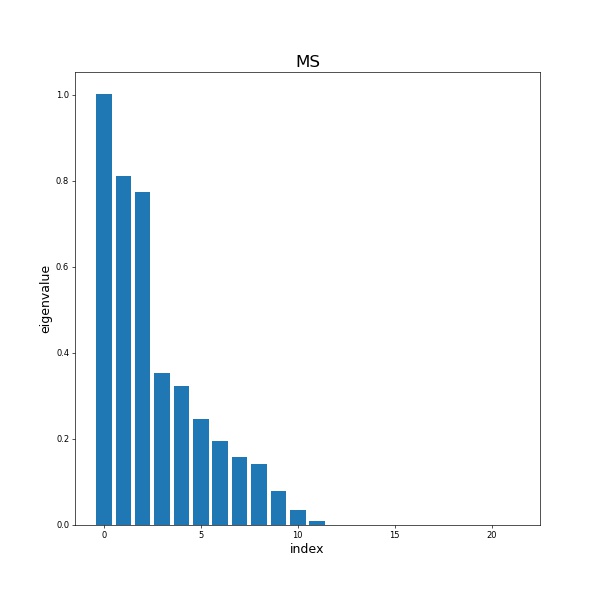}
  \caption{MS}
  \label{fig:eig-MS}
\end{subfigure} \hfill
\begin{subfigure}{0.23\textwidth}
  \includegraphics[width=\linewidth]{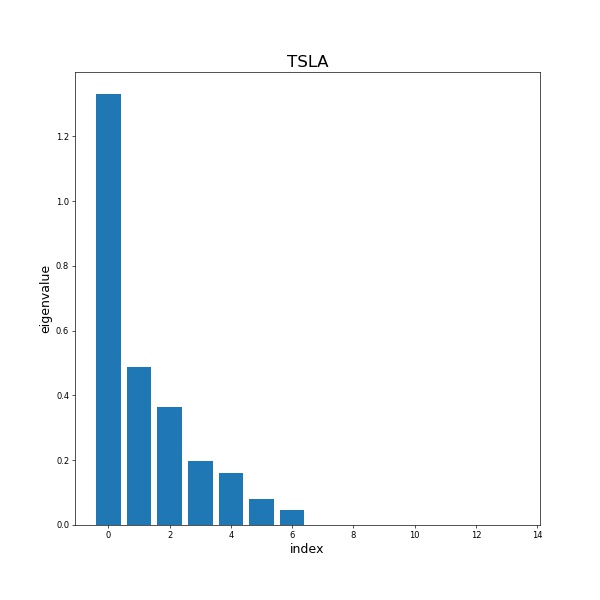}
  \caption{TSLA}
  \label{fig:eig-TSLA}
\end{subfigure}
\caption{Correlation matrix of hidden states, eigenvalues of transition covariance matrix, "GOOGL", "IBM", "MS" and "TSLA"}
\label{fig:transcorr-appen2}
\end{figure}

\end{document}